\documentclass{article}

\usepackage[margin=0.5in]{geometry}
\usepackage[utf8]{inputenc}
\usepackage{csquotes}
\usepackage{lipsum}
\usepackage{amsfonts}
\usepackage{amsthm,amsmath,amssymb,accents,empheq} 
\usepackage{enumitem}
\usepackage{subcaption}
\usepackage{graphicx}
\usepackage{float}
\usepackage{xcolor}
\usepackage{algorithm}
\usepackage[noend]{algpseudocode}
\usepackage{hyperref}

\hypersetup{
    colorlinks=true,
    linkcolor=blue,
    citecolor=blue,
    filecolor=blue,      
    urlcolor=blue,
}

\usepackage[
backend=biber,
style=ieee,
citestyle=numeric,
sorting=none
]{biblatex}

\theoremstyle{remark}

\usepackage{listings}
 
\definecolor{codegreen}{rgb}{0,0.6,0}
\definecolor{codegray}{rgb}{0.5,0.5,0.5}
\definecolor{codepurple}{rgb}{0.58,0,0.82}
\definecolor{backcolour}{rgb}{0.95,0.95,0.92}
 
\lstdefinestyle{mystyle}{
    backgroundcolor=\color{backcolour},   
    commentstyle=\color{codegreen},
    keywordstyle=\color{magenta},
    numberstyle=\tiny\color{codegray},
    stringstyle=\color{codepurple},
    basicstyle=\ttfamily\footnotesize,
    breakatwhitespace=true,         
    breaklines=true,                 
    captionpos=b,                    
    keepspaces=true,                 
    numbers=left,                    
    numbersep=5pt,                  
    showspaces=false,                
    showstringspaces=false,
    showtabs=false,                  
    tabsize=2
}

\usepackage{titlesec}
\titleformat{\paragraph}
{\normalfont\normalsize\bfseries}{\theparagraph}{1em}{}
\titlespacing*{\paragraph}
{0pt}{3.25ex plus 1ex minus .2ex}{1.5ex plus .2ex}

\title{\textbf{A Particle-in-cell Method for Plasmas with a Generalized Momentum Formulation, \\ Part I: Model Formulation} \thanks{
The research of the authors was supported by AFOSR grants FA9550-19-1-0281 and  FA9550-17-1-0394, NSF grant DMS-1912183, and DOE grant DE-SC0023164.}}

\author{Andrew J. Christlieb \thanks{Department of Computational Mathematics, Science and Engineering, Michigan State University, East Lansing, MI, 48824, United States; \href{mailto:christli@msu.edu}{christli@msu.edu}.}
\and William A. Sands \thanks{Department of Mathematical Sciences, University of Delaware, Newark, DE, 19716, United States;
\href{mailto:wsands@udel.edu}{wsands@udel.edu} (corresponding author).}
\and Stephen White \thanks{Department of Computational Mathematics, Science and Engineering, Michigan State University, East Lansing, MI, 48824, United States; \href{mailto:whites73@msu.edu}{whites73@msu.edu}.}
}

\date{}

\begin{document}

\maketitle

\begin{abstract}
     This paper formulates a new particle-in-cell method for the Vlasov-Maxwell system. Under the Lorenz gauge condition, Maxwell's equations for the  electromagnetic fields can be written as a collection of scalar and vector wave equations. The use of potentials for the fields motivates the adoption of a Hamiltonian formulation for particles that employs the generalized (conjugate) momentum. A notable advantage offered by the Hamiltonian formulation is the elimination of time derivatives in the Lorenz gauge formulation that are required by the standard Newton-Lorenz treatment of the particles. This allows the fields to retain the full time-accuracy guaranteed by the field solver. The resulting updates for particles require only knowledge of the fields and their spatial derivatives. An analytical method for constructing these spatial derivatives is presented that exploits the underlying integral solution used in the field solver for the wave equations. Moreover, these derivatives are shown to converge at the same rate as the fields in the both time and space. The Method of Lines Transpose (MOL$^T$) field solver we consider in this work is globally first-order accurate in time and fifth-order accurate in space and belongs to a larger class of methods which are unconditionally stable, can address geometry, and leverage $\mathcal{O}(N)$ fast summation methods for efficiency. We demonstrate the method on several well-established benchmark problems, including a plasma sheath as well as a relativistic particle beam. The efficacy of the proposed formulation is demonstrated through a comparison with standard methods presented in the literature, with one example being the popular finite-difference time-domain (FDTD) method. The new method shows mesh-independent numerical heating properties even in cases where the plasma Debye length is close to the grid spacing. This is an important feature of the new method because it permits the use of coarser grids in space in the representation of the fields. The use of high-order spatial approximations in the new method means that fewer grid points are required in order to achieve a fixed accuracy. Our results also suggest that the new method can be used with fewer simulation particles per cell compared to standard explicit methods, which permits further computational savings.
\end{abstract}

\noindent
{ \footnotesize{\textbf{Keywords}: Vlasov-Maxwell system, generalized momentum, particle-in-cell, method-of-lines-transpose, integral solution} }
% Guidelines for the section

% Introduction of the topic
% and summary of the contributions
%
% Discuss the purpose of the report
%
\section{Introduction}
\label{sec:1 Introduction}

% % Be direct about the content. What is the point to the paper.
% This paper establishes the foundation a new particle-in-cell method for the Vlasov-Maxwell system. Using the Lorenz gauge condition, Maxwell's equations for the  electromagnetic fields are written as a collection of scalar and vector wave equations. The use of potentials for the fields motivates the adoption of a Hamiltonian formulation for particles that employs the generalized (conjugate) momentum. A notable advantage offered by the generalized formulation is the elimination of time derivatives in the Lorenz gauge formulation required by the standard Newton-Lorenz treatment of the particles and allows the fields to retain the full time-accuracy guaranteed by the field solver. 

This paper introduces a new particle-in-cell (PIC) method for the Vlasov-Maxwell (VM) system that is based on a potential formulation in which the particle updates are advanced through a Hamiltonian formulation. The new method shows several \mbox{advantages} over traditional PIC methods, namely: mesh-independent numerical heating; stability and refinement of solutions even when the mesh spacing is comparable to the Debye length (i.e., $ \lambda_{D} / \Delta x \approx 1$);  enhanced ability to capture symmetries that pose difficulties for standard PIC methods; and can be used in problems with complex geometry without resorting to cut-cells and staircase approximations. The method is stable even when the mesh spacing exceeds the Debye length  (i.e., $ \lambda_{D} / \Delta x > 1$), although accuracy may become an issue.  The ability to use $ \lambda_{D} / \Delta x \geq 1$ is important when simulating dense plasmas and implies that the method requires fewer simulation particles than comparable explicit PIC methods based on finite-differences for a given accuracy. Indeed, for benchmark test problems, which simulate plasmas in both periodic and bounded domains, the new method outperforms conventional explicit PIC methods that use staggered meshes. The field solver used in the new method is non-staggered and is fifth-order accurate in space and globally first-order accurate in time. Because we adopt the generalized momentum formulation for the particle push, we do not require temporal derivatives of the fields. For the problems considered in this work, the first-order time accuracy of the fields has been adequate, especially given that the spatial derivatives are computed to high-order accuracy using ``analytical" expressions. As our goal is to eventually move to complex domains, the new method makes use of bilinear (area) weightings for mapping the charge and current density to the mesh. It is well-known that this approach leads to charge conservation issues \cite{VillasenorChargeConservation92}. However, in benchmark problems, we assess violations in the Lorenz gauge condition and find that the new method conserves charge at an acceptable level despite the fact that this is not strictly enforced.

% Basic PIC stuff
PIC methods \cite{BirdsallLangdon,HockneyEastwood} have been extensively applied in numerical simulations of kinetic plasmas and are an important class of techniques used in the design of experimental devices including lasers, pulsed power systems, particle accelerators, among others \cite{Godfrey2016}. The earliest work involving these methods began in the 1950s and 1960s, and it remains an active area of research to this day. At its core, a PIC method combines an Eulerian approach for the electromagnetic fields with a Lagrangian method that evolves collections of samples taken from general distribution functions in phase space. In other words, the fields are evolved using a mesh, while the distribution function is evolved using particles whose equations of motion are set according to characteristics of the partial differential equations (PDEs) that govern the evolution of the distribution functions. Lastly, to combine the two approaches, an interpolation method is used to map data between the mesh and the particles. A typical selection for this map involves some combination of piece-wise constant or linear splines, with tensor products being used to address multi-dimensional problems.

The popularity of the PIC method in engineering applications can be largely attributed to its simplicity, efficiency, and capabilities in simulating complex nonlinear processes in plasmas. Early renditions of these methods were specifically constructed to circumvent the prohibitively expensive force calculation used to calculate the interactions between particles in electrostatic problems. The introduction of a mesh greatly simplified the force calculation because the number of mesh points is typically smaller than the number of particles and fast algorithms could be used to compute the potential. Simulation particles in these methods, which represent an ensemble of physical particles, are initialized by sampling from a prescribed probability distribution. A consequence of this sampling is that bulk processes in plasmas will be well represented, while the tails of the distribution will be largely underresolved even with good sampling methods. This, in turn, necessitates a large number of simulation particles for more systematic refinement studies to prevent certain numerical fluctuations. Realistic simulations of plasma devices have created a demand for new algorithms that simultaneously address the various challenges posed by accuracy constraints in modeling as well as scalability with new computational hardware \cite{BiedronBeam2022}. The goal of this work is to supply new algorithms which aim to enhance the capabilities of existing PIC methods for plasmas. The prevalence of PIC, as a simulation tool, in the plasma physics community, has resulted in the introduction of numerous production codes with different capabilities. Despite these developments, however, comparisons that benchmark the performance of these algorithms have only recently been explored \cite{Smith_2021}. There is a clearly a need for more systematic benchmarking of existing PIC algorithms in the community, and we hope this work contributes significantly in this respect.

A comprehensive review of the literature for PIC methods up to 2005 can be found in the review article \cite{JPVerboncoeur2005review}. Much of the work highlighted by this reference is now largely considered standard, so, we shall emphasize more recent articles that are more aligned with the developments presented in this paper. Most PIC methods evolve the simulation particles explicitly using some form of leapfrog time integration along with the Boris rotation method \cite{Boris1970} in the case of electromagnetic plasmas. The exploration of implicit PIC methods began in the 1980s \cite{Friedman81,BrackbillImplicitPIC82,MasonImplicitPIC1987}. These approaches suffered from a number of unattractive features, including issues with numerical heating and cooling \cite{DirectImplicitPIC1989}, slow nonlinear convergence, and inconsistencies between the fluid moments and particle data. These approaches were later abandoned in favor of explicit treatments. Recent years have shown a resurgence of interest in implicit PIC methods \cite{Lapenta-ImplicitPIC2006,Lapenta2011energy-conserving,Chacon2011es-implicit-pic}. In particular, the implicit PIC method proposed in \cite{Chacon2011es-implicit-pic} addressed many of these issues in the case of the Vlasov-Poisson system. Nonlinear convergence and self-consistency were enforced using a Jacobian-free Newton-Krylov method \cite{JFNKsurvey2004} with a novel fluid preconditioner \cite{Chen-Preconditioning-ImplicitPIC2014} to enforce the continuity equation. This approach eliminated the need to resolve the charge separation in the plasma, which led to remarkable computational savings over explicit methods. These techniques were later extended to curved geometries through the use of smooth grid mappings \cite{Chacon2012mapped}. Recently, an effort has been made to extend these techniques to the full Vlasov-Maxwell system \cite{ChaconVM2020} to avoid the highly restrictive CFL condition posed by the gyrofrequency, as well as the consideration of asymptotic-preserving treatments for particles \cite{Koshkarov-APintegrator2022,ChenAPintegratorESPIC2022}. While these contributions are significant in their own right, there are many opportunities for improvement. Many of these methods are limited to second-order accuracy in both space and time and may greatly benefit from more accurate field solvers. Additionally, applications of interest involve complex geometries which introduce additional complications with stability and are often poorly resolved with uniform Cartesian meshes. Lastly, there is the concern of scalability. Krylov subspace methods present a considerable challenge for scalability on large machines due to the various collective operations used in the algorithms. For this reason, algorithms with explicit structure are more common in applications. It seems that the scalability of these methods could be significantly improved if similar implicit methods could be developed which eliminate these Krylov solves altogether, though this is beyond the scope of the present work.

% FTDT stuff
%
% Would this be better for the second paper???
A challenge associated with developing any solver for Maxwell's equations is the enforcement of the involutions for the fields, namely $\nabla \cdot \mathbf{E} = \rho / \epsilon_0$ and $\nabla \cdot \mathbf{B} = 0$. In the case of a structured Cartesian grid, Maxwell's equations can be discretized using a staggered grid technique introduced by Yee \cite{Yee1966}. The use of a staggered mesh yields a structure-preserving discrete analogue of Maxwell's equations in integral form that automatically enforces the involutions for $\mathbf{E}$ and $\mathbf{B}$ without additional treatment. This is the basis of the well-known finite-difference time-domain method (FDTD) \cite{TafloveHagnessFDTDbook}. It is important to note that this is only true in the absence of moving charge. In the presence of moving charge, a standard linear current weighting will not satisfy the continuity equation $\partial_{t} \rho + \nabla \cdot \mathbf{J} = 0$. In \cite{VillasenorChargeConservation92}, maps for the current were constructed to properly ensure that Gauss' law is satisfied.   Divergence cleaning methods were also analyzed in \cite{Mardahl97conservation} and were shown to be quite effective in simulations of a relativistic beam. Later, higher order versions of particle weighting were proposed in \cite{esirkepov2001exact}. Then in \cite{umeda2003new}, the authors generalized this approach to remove the assumption that particles move in straight lines. Such particle weightings can be useful for dense plasmas because they reduce numerical heating. However, they introduce complications in bounded domains, specifically when the plasma interacts with the boundary. In such cases, the approach of Villasenor and Buneman \cite{VillasenorChargeConservation92} is the preferred weighting for the current density. While the staggering in both space and time used in the original FDTD method is second-order accurate, a fourth-order extension of the spatial discretization was developed as a way of dealing with certain dispersion errors known as numerical Cerenkov radiation \cite{LuginslandCerenkovRadiation}. Pseudo-spectral type discretizations, free of numerical dispersion, have also been considered in simulations of relativistic plasmas  \cite{VayFullPICmethod2016,LeheSpectralPIC2016}. Recent work on relativistic plasmas has used both high-order weighting and high-order methods to study the long-term evolution in problems that require control of energy and momentum conservation \cite{shalaby2017sharp}. In \cite{bacchini2023relsim}, a semi-implicit method was introduced to model systems in which the Debye length cannot be resolved for purposes of practicality. The method in \cite{bacchini2023relsim} shares many of the same properties as the one presented in this work.  While the use of a staggered mesh with finite-differences is quite effective for Cartesian grids, issues arise in problems defined with geometry, such as curved surfaces, in which one resorts to stair step boundaries \cite{JPVerboncoeur2004stairstep}. To mitigate the effect of stair step boundaries in explicit methods, the mesh resolution is increased, resulting in a highly restrictive time step to meet the CFL stability criterion. Conformal PIC methods (see the review \cite{reviewConformalPIC2015,WangConformalPIC2016}), which use smooth grid transformations to address curved boundaries have been developed to address some of these concerns, but there may be certain geometries for which (uniformly) small cells may be required to properly resolve the features. In the case of cut-cell meshes, the time step will be limited by the size of the smallest cell, which can be prohibitively restrictive. Another interesting approach for dealing with geometry in the Yee scheme, which avoids the stair stepping along the boundary, was developed for two-dimensional problems \cite{EngquistCutCell2D}. The grid cells along the boundary in the method are replaced with cut-cells that use generalized finite-difference updates to account for different intersections with the boundary. While this scheme was shown to be energy conserving, a more remarkable feature is that it eliminated the highly restrictive condition on the time step introduced by the cut-cells along the boundary. The theory in this article established half-order accuracy, yet demonstrated first-order accuracy in numerical experiments. While, these schemes have not yet been combined with PIC, they might eliminate some of the commonly encountered stability issues associated with cut-cells.

% FEM-PIC papers such as DG and continuous FEM and methods
While many electromagnetic PIC methods solve Maxwell's equations on Cartesian meshes through the FDTD method, other methods have been developed specifically for addressing issues posed by geometries through the use of unstructured meshes. In \cite{SonnendruckerFEM-PIC95}, a finite-element method (FEM) was coupled with PIC to model plasmas using the Darwin approximation, in which there is considerable (time) scale separation between the fields and the plasma. Truly conformal PIC methods based on the FEM have also been considered to address problems concerning general geometries and parallel scalability \cite{EastwoodConformalFEM-PIC1995}. Explicit finite-volume methods (FVM), which can address geometry, were considered in \cite{munz2000divergence}, which also developed divergence cleaning methods suitable for applications to PIC simulations of the Vlasov-Maxwell system. Discontinuous Galerkin (DG) methods have also been used to develop high-order PIC methods with elliptic \cite{JacobsHesthavenPIC06} and hyperbolic \cite{JacobsHesthavenPIC09} divergence cleaning methods being employed to enforce Gauss' law. Other work in this area has explored more generalized FEM discretizations in order to enforce charge conservation on arbitrary grids \cite{SonnendruckerFEM-PIC14}. Structure-preserving discretizations \cite{Sonnendrucker-Structure-Preserving21,OconnorPIC-Benchmark,OconnorFEM-PIC-ChargeMap,CrawfordFEM-PIC,PintoSemi-Implicit-FEM-PIC22,Kormann-VM-Structure-Preserving21}, which use exact sequence basis functions that follow the de Rham complex at a discrete level and automatically enforce involutions for the electric and magnetic fields, have also been proposed. Another method in this category was presented for the Vlasov-Maxwell system in \cite{KrausGEM-PIC2017}, which exploited the Hamiltonian structure of the system to generate methods with numerous conservative properties that are independent of the basis functions. While the flexibility of these methods is quite appealing, such solvers rely on the solution of large systems of equations. Even with preconditioning, such methods can be slow and difficult implement in a scalable manner. In the case of explicit methods, such as FV and DG approaches, other challenges exist. The basic FVM, without additional reconstructions, is first-order accurate in space. These methods can, of course, be improved to second-order accuracy by performing reconstructions based on a collection of cells. Beyond second-order accuracy, the reconstruction process becomes quite complicated due to the growth in the size of the interpolation patches. DG methods, on the other hand, store cell-wise expansions in a basis, which eliminates the issue encountered in the FVM, typically at the cost of a highly restrictive condition on the size of a time step. Additionally, the significant amount of local work in DG methods makes them appealing for newer hardware, yet the restriction on the time step size is often left unaddressed. However, notable exceptions to this restriction exist for the two-way wave equation including staggered formulations \cite{staggeredDGwave} and Hermite methods \cite{HermiteDGwave}, which allow for a much larger time step. It will be interesting to see the performance of such methods in plasma problems, especially in problems with intricate geometric features.

% Work on ADI and MOLT, as well as the use of gauges
Other methods for Maxwell's equations have been developed with unconditional stability for the time discretization. The first of these methods is the ADI-FDTD method \cite{ZhangADI-FDTD1999,ZhangADI-FDTD2000}, which combined an ADI approach with a two-stage splitting to achieve an unconditionally stable solver. Time stepping in these methods was later generalized using a Crank-Nicolson splitting and several techniques for enhancing the temporal accuracy were proposed \cite{LeeFornberg04}. Of particular significance to this work are methods based on the method-of-lines-transpose (MOL$^T$) \cite{causley2013method,causley2014method,causley2014higher,causley2017wave-propagation}. These methods are unconditionally stable in time and can be obtained by reversing the typical order in which discretization is performed. By first discretizing in time, one can solve a resulting boundary-value problem by formally inverting a dimensionally-split differential operator using a Green's function in conjunction with a fast summation method. Mesh-free methods for plasmas \cite{Grid-free-Christlieb06} have also been developed, which have been extended to Maxwell's equations in the Darwin limit, under the Coulomb gauge \cite{Gibbon2010,Gibbon2017Hamiltonian}. These formulations are in some ways similar to PIC in that they evolve particles with shapes with the exception that no mapping to a mesh is used in the simulation. The elliptic equations are solved using a Green's function and a fast summation method is used for efficiency. Green's function methods have also been used to develop asymptotic preserving schemes. In \cite{cheng2017asymptotic}, a boundary integral formulation with a multi-dimensional Green's function was used to construct a method that recovers the Darwin limit under appropriate conditions. The methods considered in the present work utilize dimensional splittings, which has been used to construct algorithms that are unconditionally stable, permit high-order time accuracy \cite{causley2014higher}, show parallel scalability  \cite{christlieb2020parallel}, and are geometrically flexible \cite{MOLT-EB-2020}. 

In \cite{wolf2016particle}, a PIC method was developed based on the MOL$^T$ discretization. This work leveraged a staggered grid formulation in which Maxwell's equations were cast in terms of the Lorenz gauge. The wave equation for the scalar potential was replaced with an elliptic equation to control errors in the gauge condition. Since the particle equations were written in terms of $\mathbf{E}$ and $\mathbf{B}$, additional finite-difference derivatives were required to compute the electric and magnetic fields from the potentials. In contrast, the formulation developed in this paper eliminates the use of a \mbox{staggered} mesh, so that data for the fields and particles are co-located. Furthermore, spatial derivatives of the potentials, which were originally computed using (low-order) finite-differences, are now evaluated analytically using the integral solution and converge at the same rate as the fields. In this work, we use quadrature that is fifth-order accurate in space. However, this approach naturally extends to arbitrary order and retains the geometric flexibility of the field solver. Lastly, the time integration method used for the particles in this work is fundamentally different from \cite{wolf2016particle}, which considered traditional leapfrog methods. The method we propose is based on a simple modification of a method presented in \cite{Gibbon2017Hamiltonian} and is compatible with the structure of the Hamiltonian formulation. 

The contents of this paper are organized as follows. The relativistic formulation for the Vlasov-Maxwell system, which is the basis of the new PIC method developed in this work, is presented in section \ref{sec:2 Problem formulation}. Details concerning the treatment of the fields are presented in section \ref{sec:3 Methods for field equations}, where we establish stability and time-consistency properties of the field solver at the semi-discrete level. In section \ref{sec:4 new PIC method}, we describe the new PIC method that is based on the field solver of the previous section along with details of the time integration method. Section \ref{sec:5 Numerical results} presents time and space refinement results for the field solver along with an extensive collection of numerical results for the new PIC method. We conclude with a summary of the results and a discussion of future work in section \ref{sec:Conclusion and future work}.

% Introduce the section
%
% Section on the models and non-dimensionalization
%
\section{Problem Formulation}
\label{sec:2 Problem formulation}

In this section, we provide relevant details of the problem formulation used by the plasma applications considered in this work. We begin with a discussion of the relativistic Vlasov-Maxwell system, which is the most general model used in this work, in section \ref{subsec:2 VM-system}. Then, once we have introduced the model, we discuss the treatment of the fields in section \ref{subsec:Maxwell Lorenz}, which expresses Maxwell's equations in terms of potentials. In this work, the fields are cast as wave equations through the Lorenz gauge condition. The generalized momentum formulation used for the particles is presented in section \ref{subsec:2 Particle formulation}. We introduce components of the system in their dimensional form and provide their corresponding non-dimensionalizations in Appendix \ref{app:Non-dim}, with the latter being used in the implementation. We then conclude the section with a brief summary to emphasize the key aspects of the proposed formulation.

% Section on the model
%%% V-M system
\subsection{Relativistic Vlasov-Maxwell System}
\label{subsec:2 VM-system}

In this work, we develop numerical algorithms for plasmas described by the relativistic Vlasov-Maxwell (VM) system, which in SI units, reads as
\begin{empheq}[left=\empheqlbrace]{align}
    &\partial_t f_{s} + \frac{\mathbf{p}}{m_{s}\gamma_{s}} \cdot \nabla_{x} f_{s} + q_{s} \left( \mathbf{E} + \frac{\mathbf{p} \times \mathbf{B}}{m_{s}\gamma_{s}} \right)\cdot \nabla_{p} f_{s} = 0, \label{eq:Vlasov species}\\
    &\nabla \times \mathbf{E} = -\partial_t \mathbf{B}, \label{eq:Farady} \\
    &\nabla \times \mathbf{B} = \mu_{0}\left( \mathbf{J} + \epsilon_{0} \partial_t \mathbf{E} \right), \label{eq:Ampere} \\
    &\nabla \cdot \mathbf{E} = \frac{\rho}{\epsilon_0}, \label{eq:Gauss-E} \\
    &\nabla \cdot \mathbf{B} = 0. \label{eq:Gauss-B}
\end{empheq}
The first equation \eqref{eq:Vlasov species} is the relativistic Vlasov equation which describes the evolution of a probability distribution function $f_{s}\left( \mathbf{x}, \mathbf{p}, t\right)$ for particles of species $s$ in phase space which have mass $m_{s}$ and charge $q_{s}$. Here, we define $\gamma_{s} = 1/\sqrt{1 + \mathbf{p}^2/(m_{s} c)^2} =  1/\sqrt{1 - \mathbf{v}^2/c^2}$, which makes equation \eqref{eq:Vlasov species} Lorentz invariant. Physically, equation \eqref{eq:Vlasov species} describes the time evolution of a distribution function that represents the probability of finding a particle of species $s$ at the position $\mathbf{x}$, with linear momentum $\mathbf{p} = m_{s} \gamma_{s} \mathbf{v}$, at any given time $t$. Since the position and velocity data are vectors with 3 components, the distribution function is a scalar function of 6 dimensions plus time. While the equation itself has fairly simple structure, the primary challenge in numerically solving this equation is its high dimensionality. This growth in the dimensionality has posed tremendous difficulties for grid-based discretization methods, where one often needs to use many grid points to resolve relevant space and time scales in the problem. This difficulty is compounded by the fact that many plasmas of interest contain multiple species. Despite the lack of a collision operator on the right-hand side of \eqref{eq:Vlasov species}, collisions occur in a mean-field sense through the electric and magnetic fields, which appear as coefficients of the gradient in momentum.

Equations \eqref{eq:Farady} - \eqref{eq:Gauss-B} are Maxwell's equations, which describe the evolution of the background electric and magnetic fields. Since the plasma is a collection of moving charges, any changes in the distribution function for each species will be reflected in the charge density $\rho(\mathbf{x},t)$, as well as the current density $\mathbf{J}(\mathbf{x},t)$, which, respectively, are the source terms for Gauss' law \eqref{eq:Gauss-E} and Amp\`ere's law \eqref{eq:Ampere}. For $N_{s}$ species, the total charge density and current density are defined by summing over the species
\begin{equation}
    \rho(\mathbf{x},t) = \sum_{s=1}^{N_{s}} \rho_{s}(\mathbf{x},t), \quad \mathbf{J}(\mathbf{x},t) = \sum_{s=1}^{N_{s}} \mathbf{J}_{s}(\mathbf{x},t), \label{eq:total charge + current densities}
\end{equation}
where the species charge and current densities are defined through moments of the distribution function $f_s$:
\begin{equation}
    \rho_{s}(\mathbf{x},t) = q_{s} \int_{\Omega_{p}} f_{s}(\mathbf{x}, \mathbf{p}, t) \, d\mathbf{p}, \quad \mathbf{J}_{s}(\mathbf{x},t) = q_{s} \int_{\Omega_{p}} \frac{\mathbf{p}}{m_{s} \gamma_{s}} f_{s}(\mathbf{x}, \mathbf{p}, t) \, d\mathbf{p}. \label{eq:species charge + current densities integrals}
\end{equation}
Here, the integrals are taken over the momentum components of phase space, which we have denoted by $\Omega_{p}$. The remaining parameters $\epsilon_{0}$ and $\mu_{0}$ are the permittivity and permeability of free-space. We also have the useful relation $c^{2} = 1/\left( \mu_0 \epsilon_0 \right)$, where $c$ denotes the speed of light. Equations \eqref{eq:Gauss-E} and \eqref{eq:Gauss-B} enforce charge conservation and prevent the appearance of so-called ``magnetic monopoles." It is imperative that numerical schemes for Maxwell's equations satisfy these conditions. This is one of the reasons we adopt a gauge formulation for Maxwell's equations, which is presented in the next section.  

% Sections on the formulation, which are
% separated into fields and particles for clarity
%
% Reformulation of the V-M system with gauges
% We consider Lorenz gauges
%
\subsection{Maxwell's Equations with the Lorenz Gauge}
\label{subsec:Maxwell Lorenz}

Under the potential formulation, with the selection of the Lorenz gauge, Maxwell's equations transform to a system of wave equations of the form
\begin{empheq}[left=\empheqlbrace]{align}
&\frac{1}{c^2} \partial_{tt} \phi -\Delta \phi =  \frac{1}{\epsilon_{0}} \rho, \label{eq:scalar potential eqn lorenz} \\ 
&\frac{1}{c^2} \partial_{tt} \mathbf{A} -\Delta \mathbf{A} = \mu_{0} \mathbf{J}, \label{eq:vector potential eqn lorenz} \\ 
&\frac{1}{c^2} \partial_{t} \phi + \nabla \cdot \mathbf{A} = 0 \label{eq:Lorenz gauge condition},
\end{empheq}
where $c$ is the speed of light, $\epsilon_{0}$ and $\mu_{0}$ represent, respectively, the permittivity and permeability of free-space. Further, we have used $\phi$ to denote the scalar potential and $\mathbf{A}$ is the vector potential. In fact, under any choice of gauge condition, given $\phi$ and $\mathbf{A}$, one can recover $\mathbf{E}$ and $\mathbf{B}$ via the relations
\begin{equation}
    \label{eq:Convert potentials to EB}
    \mathbf{E} = -\nabla \phi - \partial_{t} \mathbf{A}, \quad \mathbf{B} = \nabla \times \mathbf{A},
\end{equation}
where $``\times"$ denotes the vector cross product. The structure of equations \eqref{eq:scalar potential eqn lorenz} and \eqref{eq:vector potential eqn lorenz} is appealing because the system, modulo the gauge condition \eqref{eq:Lorenz gauge condition}, is essentially a system of four ``decoupled" scalar wave equations. Maxwell's equations \eqref{eq:Farady} - \eqref{eq:Gauss-B} are equivalent to \eqref{eq:scalar potential eqn lorenz} and \eqref{eq:vector potential eqn lorenz} as long as the Lorenz gauge condition \eqref{eq:Lorenz gauge condition} is satisfied by $\phi$ and $\mathbf{A}$. This formulation is appealing for several reasons. This form of the system is purely hyperbolic, so it evolves in a local sense. Computationally, this means that a localized method can be used to evolve the system, which will likely be more efficient for parallel computers. Another attractive feature is that many of the methods developed for scalar wave equations, e.g., \cite{causley2014higher,causley2017wave-propagation} can be applied to the system in a straightforward manner.

\subsection{Hamiltonian Formulation for Relativistic Particles}
\label{subsec:2 Particle formulation}

To obtain the Hamiltonian formulation for the relativistic Vlasov-Maxwell system, we first introduce the Lagrangian for a single relativistic particle moving in a potential field, which can be shown to be
\begin{equation}
    \label{eq:Relativistic Lagrangian particle potential field}
    \mathcal{L} = -\frac{m c^2}{\gamma} - q \left( \phi - \mathbf{A} \cdot \mathbf{v} \right).
\end{equation}
Here, we have used $m$ to denote the mass of the particle, $q$ is its charge, $\mathbf{v}$ its velocity, $c$ is the speed of light, $\gamma = 1/\sqrt{1 - \mathbf{v}^2 / c^2}$ is the Lorentz factor, $\phi$ is the scalar potential, and $\mathbf{A}$ is the vector potential. Next, we define the generalized (conjugate) momentum from the relativistic Lagrangian \eqref{eq:Relativistic Lagrangian particle potential field} as
\begin{equation}
    \label{eq:generalized momentum}
    \mathbf{P} = \frac{\partial \mathcal{L}}{\partial \mathbf{v}} = \gamma m \mathbf{v} + q \mathbf{A}.
\end{equation}
In addition, we have the following identities, which can be derived from \eqref{eq:generalized momentum}:
\begin{align}
    \mathbf{v} &= \frac{1}{m\gamma} \left( \mathbf{P} - q \mathbf{A}\right) = \frac{c^2 \left( \mathbf{P} - q \mathbf{A}\right)}{\sqrt{ c^2\left( \mathbf{P} - q \mathbf{A}\right)^2 + \left(mc^2\right)^2}}, \label{eq:velocity from generalized momentum} \\
    \mathbf{v}^2 &= \mathbf{v} \cdot \mathbf{v} = \frac{c^2\left( \mathbf{P} - q \mathbf{A}\right)^2}{\left( \mathbf{P} - q \mathbf{A}\right)^2 + \left(mc\right)^2 }. \label{eq:velocity squared identity}
\end{align}
The Hamiltonian corresponding to the Lagrangian \eqref{eq:Relativistic Lagrangian particle potential field} can be obtained by means of a Legendre transform
\begin{equation}
    \label{eq:Legendre transform to get H}
    \mathcal{H} = \mathbf{P} \cdot \mathbf{v} - \mathcal{L}.
\end{equation}
Using the identities \eqref{eq:velocity from generalized momentum} and \eqref{eq:velocity squared identity} in the transformation \eqref{eq:Legendre transform to get H}, we obtain the relativistic Hamiltonian
\begin{equation}
    \label{eq:Relativistic Hamiltonian particle potential field}
    \mathcal{H} = \sqrt{c^2 \left( \mathbf{P} - q \mathbf{A}\right)^2 + \left(mc^2\right)^2} + q \phi.
\end{equation}
The Hamiltonian for a system of $N_{p}$ particles can be easily obtained by summing the Hamiltonians over individual particles, each having the form \eqref{eq:Relativistic Hamiltonian particle potential field}. From this, we can calculate the equations of motion for each particle using Hamilton's equations, which leads to the system 
\begin{empheq}[left=\empheqlbrace]{align}
    \frac{d\mathbf{x}_{i}}{dt} &= \frac{\partial \mathcal{H}}{\partial \mathbf{P}_{i}} = \frac{c^2 \left(\mathbf{P}_{i} - q_{i}\mathbf{A}\right)}{\sqrt{c^2\left(\mathbf{P}_{i} - q_{i} \mathbf{A}\right)^2 + \left(m_{i} c^2\right)^2}}, \label{eq:Position equation relativistic form} \\
    \frac{d\mathbf{P}_{i}}{dt} &= -\frac{\partial \mathcal{H}}{\partial \mathbf{x}_{i}} = -q_{i} \nabla \phi + \frac{q_{i} c^2 \left( \nabla\mathbf{A}\right) \cdot \left(\mathbf{P}_{i} - q_{i}\mathbf{A}\right)}{\sqrt{c^2\left(\mathbf{P}_{i} - q_{i} \mathbf{A}\right)^2 + \left(m_{i} c^2\right)^2}}, \label{eq:Generalized momentum equation relativistic form}
\end{empheq}
where $i = 1, \cdots, N_{p}$. In each of these equations, it is to be assumed that the potentials are evaluated at the corresponding locations of the particle, i.e., $\phi = \phi(\mathbf{x}_{i})$ and $\mathbf{A} = \mathbf{A}(\mathbf{x}_{i})$. Note that in the non-relativistic limit $v^2 \ll c^2 $, so $\gamma \rightarrow 1$. From the identity \eqref{eq:velocity from generalized momentum} we can see that
\begin{equation*}
    \frac{c^2 \left(\mathbf{P}_{i} - q_{i}\mathbf{A}\right)}{\sqrt{c^2\left(\mathbf{P}_{i} - q_{i} \mathbf{A}\right)^2 + \left(m_{i} c^2\right)^2}} \rightarrow \mathbf{v}_{i} = \frac{1}{m_{i}} \left( \mathbf{P}_{i} - q_{i} \mathbf{A} \right)
\end{equation*}
and we obtain the non-relativistic system
\begin{empheq}[left=\empheqlbrace]{align*}
    \frac{d\mathbf{x}_{i}}{dt} &= \frac{1}{m_{i}} \left( \mathbf{P}_{i} - q_{i} \mathbf{A} \right),  \\
    \frac{d\mathbf{P}_{i}}{dt} &= -q_{i} \nabla \phi + \frac{q_{i}}{m_{i}} \left( \nabla \mathbf{A}\right) \cdot \left( \mathbf{P}_{i} - q_{i} \mathbf{A} \right).
\end{empheq}

\subsection{Summary}
\label{subsec:2 Summary}

In this section, we introduced the relativistic VM system, which is the most general mathematical model that will be used in this work. Maxwell's equations were expressed in potential form, under the choice of the Lorenz gauge, yielding a system of four wave equations that are amenable to the class of unconditionally stable wave solvers developed in our earlier work. We showed how time derivatives of the potentials could be eliminated through the adoption of a Hamiltonian formulation. In the next section, we introduce the wave solvers and propose new methods for evaluating spatial derivatives of the fields that are required in the equations for the particles.

% A key feature of the proposed methods for derivatives is that they retain the convergence rates of the base solver. Additionally, these methods are constructed so that they naturally inherit the stability properties of the base method and can be applied to problems with geometry. While this work is largely focused on aspects concerning the formulation, preserving the geometric flexibility of the methods is essential for the applications we plan to explore in future work.

%
% Section introduction
%
\section{Numerical Methods for the Field Equations}
\label{sec:3 Methods for field equations}

In this section, we describe the algorithms used for wave propagation, which are required in the formulations of Maxwell's equations presented in the previous section. We begin with a general discussion of Green's function methods and integral equations in section \ref{subsec:integral eqn methods}, which is helpful for introducing the methods considered in this paper that employ dimensional splitting techniques. The solver considered in this work converges at a rate that is globally first-order accurate in time and is presented in section \ref{subsubsec:deriving bdf schemes} in its semi-discrete form. A fifth-order quadrature rule is used to approximate the integrals in the fully-discrete case, so the method is high-order accurate in space. A short discussion is presented in section \ref{subsubsec:BDF stability analysis} that addresses the stability of the proposed method in its semi-discrete form. A solution is formulated in terms of one-dimensional operators that can be inverted using the methods discussed in section \ref{subsec:inverting 1-D operators}. We then discuss the methods used to obtain derivatives and demonstrate the application of boundary conditions in section \ref{subsec:derivatives}. These derivatives will be shown in section \ref{subsec:Field-solver experiments} to have the same temporal and spatial convergence rates as the fields. We conclude with a brief summary in section \ref{subsec:3 Summary}.

%
% Introduce the method by discussing integral equations
%
%
% Section on the method of layer potentials
%
\subsection{Integral Equation Methods and Green's Functions}
\label{subsec:integral eqn methods}

Integral equation methods or, more generally, Green's function methods, are a powerful class of techniques used in the solution of boundary value problems that occur in a range of applications, including  acoustics, fluid dynamics, and electromagnetism \cite{ChristliebPenningTrap2004,ChristliebVirtualCathode2004,VeerapaneniVesicle2009,VicoEM-Scattering2016,BarnettAcoustics2020,chiu_moore_quaife_transport_2020,quaife2011Rothe,thavappiragasam2017molt}. Such methods allow one to write an explicit solution of an elliptic PDE in terms of a fundamental solution or Green's function. While explicit, this solution can be difficult or impossible to evaluate, so numerical quadrature is used to evaluate these terms. Layer potentials can then be introduced in the form of surface integrals to adjust the solution to satisfy the prescribed boundary data \cite{FollandBook1995}. We illustrate these features with an example that is the basis for the method presented in this work.

Suppose that we are solving the following modified Helmholtz equation
\begin{equation}
    \label{eq:modified Helmholtz eqn R^n}
    \left( \mathcal{I} - \frac{1}{\alpha^2} \Delta \right) u(\mathbf{x}) = S(\mathbf{x}), \quad \mathbf{x} \in \Omega,
\end{equation}
where $\Omega \subset \mathbb{R}^{n}$ and $\mathcal{I}$ is the identity operator, $\Delta$ is the Laplacian operator in $\mathbb{R}^{n}$, $S$ is a source term, and $\alpha \in \mathbb{R}$ is a parameter. While this method can be broadly applied to other elliptic PDEs, equation \eqref{eq:modified Helmholtz eqn R^n} is of interest to us because it can be obtained from the time discretization of a parabolic or hyperbolic PDE. In this case, the source function includes additional time levels of $u$ and the parameter $\alpha = \alpha(\Delta t)$ is connected to the time discretization of this problem. We shall not prescribe boundary conditions for this problem, and instead consider the most general solution.

To apply a Green's function method to equation \eqref{eq:modified Helmholtz eqn R^n}, one first needs to identify a function $G(\mathbf{x},\mathbf{y})$ that solves the equation
\begin{equation}
    \label{eq:Greens function eqn for R^n}
    \left( \mathcal{I} - \frac{1}{\alpha^2} \Delta \right) G(\mathbf{x},\mathbf{y}) = \delta \left( \mathbf{x}-\mathbf{y} \right), \quad \mathbf{x}, \mathbf{y} \in \mathbb{R}^{n}
\end{equation}
over free-space, with $\delta \left( \mathbf{x}-\mathbf{y} \right)$ being the Dirac delta distribution. The construction of fundamental solutions is quite standard and extensively tabulated for many different operators, including the modified Helmholz operator \cite{DuffyBook2001}. Therefore, we shall not elaborate on this further and, instead, assume that the fundamental solution $G(\mathbf{x},\mathbf{y})$ is known for our problem. The fundamental solution $G(\mathbf{x},\mathbf{y})$, which solves \eqref{eq:Greens function eqn for R^n} can be used to build a solution to the original problem \eqref{eq:modified Helmholtz eqn R^n}. First, let $u$ be a solution of the problem \eqref{eq:modified Helmholtz eqn R^n}. Multiplying the equation \eqref{eq:Greens function eqn for R^n} by $u$, integrating over $\Omega$, and applying the divergence theorem (or integration by parts in the one-dimensional case) leads to the integral identity
\begin{equation}
    \label{eq:modified Helmholtz Green's identity}
    u(\mathbf{x}) = \int_{\Omega} G(\mathbf{x},\mathbf{y}) S(\mathbf{y}) \, dV_{\mathbf{y}} + \frac{1}{\alpha^2} \int_{\partial \Omega} \left(  G(\mathbf{x},\mathbf{y}) \frac{ \partial u }{\partial \mathbf{n}} - u(\mathbf{y}) \frac{\partial G}{\partial \mathbf{n}} \right) \, dS_{\mathbf{y}}. 
\end{equation}
Note that the above identity utilizes the assumption that the function $u$ solves the PDE \eqref{eq:modified Helmholtz eqn R^n}. Since the volume integral term does not enforce boundary conditions, the surface integral contributions involving $u$ are replaced with an ansatz of the form
\begin{equation}
    \label{eq:modified helmholtz BIE solution}
    u(\mathbf{x}) = \int_{\Omega} G(\mathbf{x},\mathbf{y}) S(\mathbf{y}) \, dV_{\mathbf{y}} + \int_{\partial \Omega} \left(  \sigma(\mathbf{y}) G(\mathbf{x},\mathbf{y}) + \gamma(\mathbf{y})\frac{\partial G}{\partial \mathbf{n}} \right) \, dS_{\mathbf{y}}, 
\end{equation}
where $\sigma(\mathbf{y})$ is the single-layer potential and $\gamma(\mathbf{y})$ is the double-layer potential, which must now be determined to enforce the boundary conditions. The choice of names is reflected by the behavior of the Green's function associated with each of the terms. The Green's function itself is continuous, but its derivative will have a ``jump." Based on the boundary conditions, one selects either a single or double layer form as the ansatz for the solution. The single layer form is used in the Neumann problem, while the double layer form is chosen for the Dirichlet problem.

The algorithms presented in the subsequent sections are essentially a one-dimensional analogue of these methods. Rather than invert the multi-dimensional operator corresponding to \eqref{eq:modified helmholtz BIE solution}, the methods presented here, instead, factor the Laplacian and invert one-dimensional operators, dimension-by-dimension, using the one-dimensional form of \eqref{eq:modified helmholtz BIE solution}. We will see, later, the resulting methods solve for something that looks like a layer potential, with the key difference being that the linear system is now only a small, $2 \times 2$ matrix, which can be inverted by hand, rather than with an iterative method. Similarly, the particular solution along a given line segment can be rapidly computed with a lightweight, recursive, fast summation method, rather than a more complicated algorithm, such as the FMM. Moreover, these methods retain the geometric flexibility since the domain can be represented using one-dimensional line segments with termination points specified by the geometry. This approach, which was originally introduced in \cite{causley2014method}, was later extended to all orders in time and space and shown to be unconditionally stable \cite{causley2014higher}.

% 
% Talk about the semi-discrete schemes
%
%
% Section on semi-discrete methods for the wave equation
%
\subsection{Description of the Wave Solver}
\label{subsec:wave solver}

Unlike the high-order time accurate methods from our previous work \cite{causley2014higher}, which are based on successive convolution, we discretize the time derivatives of the wave equation using a first-order accurate backwards difference formula (BDF). Again, we wish to emphasize that the spatial discretization in the fully-discrete method is fifth-order accurate, so we retain high-order spatial accuracy. A short section on the stability analysis of the semi-discrete method is presented. Then, we establish a time consistency property that applies when the proposed discretization is used for the potentials in the Lorenz gauge formulation. We briefly discuss the splitting technique that reduces multi-dimensional problems into a sequence of one-dimensional updates.

% BDF method
\subsubsection{The Semi-discrete BDF Scheme}
\label{subsubsec:deriving bdf schemes}

To derive the first-order (time) BDF wave solver, we start with the equation
\begin{equation}
    \label{eq:two-way wave eqn}
    \frac{1}{c^2} \partial_{tt} u - \Delta u = S(\mathbf{x},t),
\end{equation}
where $c$ is the wave speed and $S$ is a source function. Then, using the notation $u(\mathbf{x}, t^{n}) = u^{n}$, we can apply a three-point backwards finite-difference stencil for the second derivative
\begin{equation*}
    \partial_{tt} u \Big\rvert_{t = t^{n+1}} = \frac{u^{n+1} - 2 u^{n} + u^{n-1} }{\Delta t^2} + \mathcal{O}(\Delta t),
\end{equation*}
where $\Delta t = t^{k} - t^{k-1}$, for any $k$, is the grid spacing in time. Evaluating the remaining terms in equation \eqref{eq:two-way wave eqn} at time level $t^{n+1}$, and inserting the above difference approximation, we obtain
\begin{equation*}
    \frac{1}{c^2 \Delta t^2} \left( u^{n+1} - 2 u^{n} + u^{n-1} \right) - \Delta u^{n+1} = S^{n+1}(\mathbf{x}) + \mathcal{O}\left(\Delta t\right),
\end{equation*}
which can be rearranged to obtain the semi-discrete equation
\begin{equation}
    \label{eq:BDF-1 semi-discrete equation}
    \left( \mathcal{I} - \frac{1}{\alpha^2} \Delta \right) u^{n+1} = \left( 2 u^{n} - u^{n-1} \right) + \frac{1}{\alpha^2} S^{n+1}(\mathbf{x}) + \mathcal{O}\left(\frac{1}{\alpha^3}\right), \quad \alpha := \frac{1}{c \Delta t}.
\end{equation}
We note that the source term is treated implicitly in this method, which creates additional complications if the source function $S$ depends on $u$. This necessitates some form of iteration, which increases the cost of the method.

% Stability analysis
\subsubsection{Stability and Dispersion Analysis of the Semi-discrete BDF Scheme}
\label{subsubsec:BDF stability analysis}

We now analyze the stability of the first-order semi-discrete BDF scheme given by equation \eqref{eq:BDF-1 semi-discrete equation}. Suppose that the solution $u(\mathbf{x})$ takes the form of the plane wave given by
\begin{equation*}
    u^{n}(\mathbf{x}) = e^{i \left( \mathbf{k} \cdot \mathbf{x} - \omega t^{n} \right)} \equiv \lambda^{n} e^{i \mathbf{k} \cdot \mathbf{x}}.
\end{equation*}
Substituting this ansatz into the semi-discrete scheme \eqref{eq:BDF-1 semi-discrete equation} and ignoring contributions due to sources, we obtain the polynomial equation
\begin{equation*}
    \left( 1 + z^2 \right) \lambda^{2} - 2 \lambda + 1 = 0.
\end{equation*}
In the above equation, we have defined the real number $z^{2} = |\mathbf{k}|^2/\alpha^2$ for simplicity. The roots of this polynomial are a pair of complex conjugates that can be written as
\begin{align*}
    \lambda_{\pm} = \frac{1}{1 + z^2} \left(1 \pm i \sqrt{z^2} \right),
\end{align*}
which satisfy the condition $\lvert \lambda_{\pm} \rvert \leq 1$ for any $\Delta t$. This shows that the amplitude of the plane wave does not grow in time, so the scheme is unconditionally stable.

The phase error introduced by the semi-discrete scheme can also be determined by first noting that $t^{n} = n \Delta t$, so that
\begin{equation}
    \label{eq:omega as a function of lambda}
    \lambda = e^{-i \omega \Delta t} \iff \omega = -\frac{1}{i \Delta t} \log \left( \lambda \right).
\end{equation}
Then, we insert the factor $\lambda_{+}$ into
equation \eqref{eq:omega as a function of lambda} and expand the resulting expression into a Puiseux series about $z = 0$, which gives
\begin{align*}
    \omega_{+} &= -\frac{1}{i \Delta t} \log \left( \lambda_{+} \right), \\
               &= -\frac{1}{i \Delta t} \left( -iz - \frac{z^{2}}{2} + \frac{iz^{3}}{3} + \frac{z^{4}}{4} + \mathcal{O}(z^{5}) \right), \quad z \ll 1.
\end{align*}
Since $z = |\mathbf{k}|/\alpha = c |\mathbf{k}| \Delta t$, the last expression can be further simplified to
\begin{equation*}
    \omega_{+} = c |\mathbf{k}| - \frac{i c^{2} |\mathbf{k}|^{2} \Delta t}{2} - \frac{c^{3} |\mathbf{k}|^{3} \Delta t^{2}}{3} + \frac{i c^{4} |\mathbf{k}|^{4} \Delta t^{3}}{2} + \mathcal{O}(c^{5} |\mathbf{k}|^{5} \Delta t^{4}).
\end{equation*}
Since the analytical dispersion relation for the plane wave solution is $\omega = c |\mathbf{k}|$, the phase error is
\begin{equation*}
    \omega - \omega_{+} = \frac{i c^{2} |\mathbf{k}|^{2} \Delta t}{2} + \mathcal{O} \left( c^{3} |\mathbf{k}|^{3} \Delta t^{2} \right).
\end{equation*}
Moreover, since the leading order term in the error is imaginary, this mode decays with time, which introduces dissipation into the scheme. Similar behavior is observed with the factor $\lambda_{-}$, so we exclude it from the discussion.

\subsubsection{Splitting Method Used for Multi-dimensional Problems}
\label{subsubsec:splitting for multi-D}

The semi-discrete equation \eqref{eq:BDF-1 semi-discrete equation} is a modified Helmholtz equation of the form \eqref{eq:modified Helmholtz eqn R^n}. Instead of appealing to \eqref{eq:modified helmholtz BIE solution}, which formally inverts the multi-dimensional modified Helmholtz operator, we apply a factorization into a product of one-dimensional operators. For example, in two-spatial dimensions, the factorization is given by
\begin{align*}
     \mathcal{I} - \frac{1}{\alpha^2} \Delta &=  \left( \mathcal{I} - \frac{1}{\alpha^2} \partial_{xx} \right)\left( \mathcal{I} - \frac{1}{\alpha^2} \partial_{yy} \right) + \frac{1}{\alpha^4} \partial_{xx} \partial_{yy}, \\
     &\equiv \mathcal{L}_{x} \mathcal{L}_{y} + \frac{1}{\alpha^4} \partial_{xx} \partial_{yy},
\end{align*}
where $\mathcal{L}_{x}$ and $\mathcal{L}_{y}$ are one-dimensional operators and the last term represents the splitting error associated with the factorization step. Note that the coefficient of the splitting error is $1/\alpha^4 = \mathcal{O}(\Delta t^4)$, which can be ignored for first-order accuracy. Therefore, the semi-discrete equation \eqref{eq:BDF-1 semi-discrete equation} in two-dimensions can be written more compactly (dropping error terms) as
\begin{equation}
    \label{eq:BDF-1 2d semi-discrete}
    \mathcal{L}_{x} \mathcal{L}_{y} u^{n+1}(\mathbf{x}) = 2 u^{n}(\mathbf{x}) - u^{n-1}(\mathbf{x}) + \frac{1}{\alpha^2} S^{n+1}(\mathbf{x}).
\end{equation}

Considerable effort has been made to address issues associated with the splitting error. For example, in \cite{causley2017phase-field}, a technique was developed to remove the splitting error in multi-dimensional applications involving parabolic equations. Successive convolution methods for the wave equation, introduced in the paper \cite{causley2014higher}, can achieve higher-order accuracy in time through more elaborate operator expansions that perform additional sweeps that remove this error. Such approaches are not considered in this paper, as we are primarily concerned with the formulation of new particle methods. Moreover, for first and second-order (time) discretizations, the splitting error can be neglected; however, we point out that in the case of higher-order methods, this term will need to be addressed in a manner that aligns with the proposed approach for computing derivatives on the mesh, which is presented in section \ref{subsec:derivatives}. For this reason, methods with higher-order time accuracy will be explored in future work. Next, we discuss the procedure used to invert the one-dimensional operators used in the factorization.

%
% Process for inverting 1-D operators
%
%
% Inverting operators in 1-D
%
\subsection{Inverting One-dimensional Operators}
\label{subsec:inverting 1-D operators}

The choice of factoring the multi-dimensional modified Helmholtz operator means we now have to solve a sequence of one-dimensional boundary value problems (BVPs) of the form
\begin{equation}
    \label{eq:one-dim BVP MHE}
    \left( \mathcal{I} - \frac{1}{\alpha^2} \partial_{xx} \right) w(x) = f(x), \quad x \in [a,b],
\end{equation}
where $[a,b]$ is a one-dimensional line and $f$ is a new source term that can be used to represent a time history or an intermediate variable constructed from the inversion of an operator along another direction. We also point out that the parameter $\alpha$ depends on the choice of the semi-discrete scheme employed to solve the problem. For the BDF scheme presented in this paper, $\alpha$ is defined by the semi-discrete update \eqref{eq:BDF-1 semi-discrete equation}. We will show the process by which one obtains the general solution to the problem \eqref{eq:one-dim BVP MHE}, deferring the application of boundary conditions to section \ref{subsec:derivatives}. Further, this section also discusses the construction of spatial derivatives.

\subsubsection{Integral Solution}
\label{subsubsec:Definining the inverse}

Since the BVP \eqref{eq:one-dim BVP MHE} is linear, its general solution can be expressed using the one-dimensional analogue of the integral solution \eqref{eq:modified Helmholtz Green's identity}:
\begin{equation}
    \label{eq:1d MHE general solution 1}
    w(x) = \int_{a}^{b} G(x,y) f(y) \,dy + \frac{1}{\alpha^2} \left[ G(x,y) \partial_{y} u(y) - u(y) \partial_{y} G(x,y) \right] \Bigg\rvert_{y=a}^{y=b},
\end{equation}
where the free-space Green's function in one-dimension is
\begin{equation}
    \label{eq:1D free-space greens function}
    G(x,y) = \frac{\alpha}{2} e^{-\alpha \lvert x - y \rvert}.
\end{equation}
In order to use the relation \eqref{eq:1d MHE general solution 1}, we need to evaluate the derivatives of the Green's function near the boundary. We note that
\begin{equation*}
    \partial_{y} G(x,y) =
    \begin{cases}
    ~~\dfrac{\alpha}{2} e^{-\alpha (x-y)}, \quad x \geq y, \\[10pt]
    -\dfrac{\alpha}{2} e^{-\alpha (y-x)}, \quad x < y.
    \end{cases}
\end{equation*}
Taking limits, we find that
\begin{align*}
    \lim_{y \to a} \partial_{y} G(x,y) &= \frac{\alpha}{2} e^{-\alpha (x-a)}, \\
    \lim_{y \to b} \partial_{y} G(x,y) &= -\frac{\alpha}{2} e^{-\alpha (b-x)}.
\end{align*}
Combining these limits with \eqref{eq:1d MHE general solution 1}, we obtain the general solution
\begin{equation}
    \label{eq:1d MHE general solution 2}
    w(x) = \frac{\alpha}{2} \int_{a}^{b} e^{-\alpha \lvert x - y \rvert} f(y) \,dy + A e^{-\alpha (x - a)} + B e^{-\alpha (b - x)},
\end{equation}
where $A$ and $B$ are constants that are determined by boundary conditions. Comparing with \eqref{eq:modified helmholtz BIE solution}, these terms serve the same purpose as the layer potentials. Further, we identify the general solution \eqref{eq:1d MHE general solution 2} as the inverse of the one-dimensional modified Helmholtz operator. In other words, we define $\mathcal{L}_{x}^{-1}$ so that
\begin{align}
    w(x) &= \mathcal{L}_{x}^{-1} \left[f\right](x), \label{eq:general inverse operator 1-D def 1} \\
        &\equiv \frac{\alpha}{2} \int_{a}^{b} e^{-\alpha \lvert x - y \rvert} f(y) \,dy + A e^{-\alpha (x - a)} + B e^{-\alpha (b - x)}, \label{eq:general inverse operator 1-D def 2} \\
        &\equiv \mathcal{I}_{x}[f](x) + A e^{-\alpha (x - a)} + B e^{-\alpha (b - x) }. \label{eq:general inverse operator 1-D def 3}
\end{align}
Section \ref{subsec:derivatives} will make repeated use of definitions \eqref{eq:general inverse operator 1-D def 1}-\eqref{eq:general inverse operator 1-D def 3} in the construction of spatial derivatives and the application of boundary conditions. The integral operator $\mathcal{I}_{x}[f](x)$ is evaluated as
\begin{equation}
    \label{eq:Average of IR and IL}
    \mathcal{I}_{x}[f](x) = \frac{1}{2} \Big( \mathcal{I}_{x}^{R}[f](x) + \mathcal{I}_{x}^{L}[f](x) \Big).
\end{equation}
where the integrals
\begin{align}
    \mathcal{I}_{x}^{R}[f](x) \equiv \alpha \int_{a}^{x} e^{-\alpha (x - y)} f(y) \,dy , \label{eq:right integral of operand} \\
    \mathcal{I}_{x}^{L}[f](x) \equiv \alpha \int_{x}^{b} e^{-\alpha (y - x)} f(y) \,dy, \label{eq:left integral of operand}
\end{align}
are computed with a recursive fast summation method. Details of this evaluation exist in numerous instances of previous work, e.g., \cite{causley2014method,causley2014higher,christlieb2020parallel,causley2017phase-field,christlieb2020nonuniformHJE}. For completeness, the details of the fast summation and approximation of these integrals are included in the Appendices \ref{app:Fast summation for 1-D} and \ref{app:computing local integrals}, respectively.

%
% Derivatives for the 1-D scheme and BCs
%
%
% Construction of derivatives
%
\subsection{Methods for the Construction of Spatial Derivatives}
\label{subsec:derivatives}

To set the stage for the ensuing discussion, note that the semi-discrete update for the first-order BDF method, in one-spatial dimension, can be obtained by combining \eqref{eq:1d MHE general solution 2} with the semi-discrete equation \eqref{eq:BDF-1 semi-discrete equation}. Defining the operand
\begin{equation*}
    R(x) = 2 u^{n}(x) - u^{n-1}(x) + \frac{1}{\alpha^2} S^{n+1}(x),
\end{equation*}
we obtain the update
\begin{align}
    u^{n+1}(x) &= \frac{\alpha}{2} \int_{a}^{b} e^{-\alpha \lvert x - y \rvert} R(y) \,dy + A e^{-\alpha (x - a)} + B e^{-\alpha (b - x)}, \label{eq:BDF-1 update in 1-D} \\
            &\equiv \mathcal{I}_{x}[R](x) + A e^{-\alpha (x - a)} + B e^{-\alpha (b - x)}, \label{eq:BDF-1 update in 1-D compact}
\end{align}
where we have used $\mathcal{I}_{x}[\cdot]$ to denote the term involving the convolution integral which is not to be confused with the identity operator. 

In order to enforce conditions on the derivatives of the solution, we will also need to compute a derivative of the update \eqref{eq:BDF-1 update in 1-D} (equivalently \eqref{eq:BDF-1 update in 1-D compact}). For this, we observe that the dependency for $x$ appears only on analytical functions, i.e., the Green's function (kernel) and the exponential functions in the boundary terms. To differentiate \eqref{eq:BDF-1 update in 1-D} we start with the definition \eqref{eq:Average of IR and IL}, which splits the integral at the point $y = x$ and makes the kernel easier to manipulate. Then, using the fundamental theorem of calculus, we can calculate derivatives of
\eqref{eq:right integral of operand} and \eqref{eq:left integral of operand} to find that
\begin{align}
    \frac{d}{dx} \left( \mathcal{I}_{x}^{R}[f](x) \right) =  \frac{d}{dx} \left( \alpha \int_{a}^{x} e^{-\alpha (x - y)} f(y) \,dy \right) = -\alpha \mathcal{I}_{x}^{R}[f](x) + \alpha f(x), \label{eq:ddx right integral of operand} \\
    \frac{d}{dx} \left(
    \mathcal{I}_{x}^{L}[f](x) \right) = \frac{d}{dx} \left( \alpha \int_{x}^{b} e^{-\alpha (y - x)} f(y) \,dy \right) =
    \alpha \mathcal{I}_{x}^{L}[f](x) - \alpha f(x), \label{eq:ddx left integral of operand}
\end{align}
These results can be combined according to \eqref{eq:Average of IR and IL}, which provides an expression for the derivative of the convolution term:
\begin{equation}
    \label{eq:ddx of convolution}
    \frac{d}{dx} \Big( \mathcal{I}_{x}[f](x) \Big) = \frac{\alpha}{2} \Big( - \mathcal{I}_{x}^{R}[f](x) +  \mathcal{I}_{x}^{L}[f](x) \Big).
\end{equation}
Additionally, by evaluating this equation at the ends of the interval, we obtain the identities
\begin{align}
    \frac{d}{dx} \Big( \mathcal{I}_{x}[f](a) \Big)  &= \alpha  \mathcal{I}_{x}[f](a), \label{eq:ddx of convolution at a} \\
    \frac{d}{dx} \Big( \mathcal{I}_{x}[f](b) \Big)  &= -\alpha  \mathcal{I}_{x}[f](b), \label{eq:ddx of convolution at b}
\end{align}
which are helpful in enforcing the boundary conditions. The relation \eqref{eq:ddx of convolution} can be used to obtain a derivative for the solution at the new time level. From the update \eqref{eq:BDF-1 update in 1-D compact}, a direct computation reveals that
\begin{equation}
    \label{eq:BDF-1 ddx update in 1-D}
    \frac{d  u^{n+1}}{dx} = \frac{\alpha}{2} \left( - \mathcal{I}_{x}^{R}[R](x) +  \mathcal{I}_{x}^{L}[R](x) \right) -\alpha A e^{-\alpha (x-a)} + \alpha B e^{-\alpha (b-x)}.
\end{equation}
Notice that no additional approximations have been made beyond what is needed to compute $\mathcal{I}_{x}^{R}$ and $\mathcal{I}_{x}^{L}$. These terms are already evaluated as part of the base method. For this reason, we think of equation \eqref{eq:BDF-1 ddx update in 1-D} as an analytical derivative. The boundary coefficients $A$ and $B$ appearing in \eqref{eq:BDF-1 ddx update in 1-D} will be calculated in the same way as the update \eqref{eq:BDF-1 update in 1-D compact}, and are discussed in the remaining subsections. This treatment ensures that the discrete derivative will be consistent with the conditions imposed on the solution variable. 

Applying different boundary conditions amounts to determining the values of $A$ and $B$ used in \eqref{eq:BDF-1 update in 1-D compact}. We shall assume that the boundary conditions at the ends of the one-dimensional domain are the same, though this is not essential. Using slight variations of the cases illustrated below, one can mix the boundary conditions at the ends of the line segments.

\subsubsection{Dirichlet Boundary Conditions}
\label{subsubsec:BDF-1 Dirichlet}

Suppose we are given the function values along the boundary, which are represented by the data
\begin{equation*}
    u^{n+1}(a) = g_{a} \left(t^{n+1}\right), \quad u^{n+1}(b) = g_{b} \left(t^{n+1}\right).
\end{equation*}
If we evaluate the BDF-1 update \eqref{eq:BDF-1 update in 1-D compact} at the ends of the interval, we obtain the conditions
\begin{align*}
    g_{a} \left(t^{n+1}\right) &= \mathcal{I}_{x}[R](a) + A + \mu B, \\
    g_{b} \left(t^{n+1}\right) &= \mathcal{I}_{x}[R](b) + \mu A + B,
\end{align*}
where we have defined $\mu = e^{-\alpha(b-a)}$. This is a simple linear system for the boundary coefficients $A$ and $B$, which can be inverted by hand. Proceeding, we find that
\begin{align*}
    A &= \frac{g_{a} \left(t^{n+1}\right) - \mathcal{I}_{x}[R](a) - \mu \left( g_{b} \left(t^{n+1}\right) - \mathcal{I}_{x}[R](b) \right)}{1 - \mu^2}, \\
    B &= \frac{g_{b} \left(t^{n+1}\right) - \mathcal{I}_{x}[R](b) - \mu \left( g_{a} \left(t^{n+1}\right) - \mathcal{I}_{x}[R](a) \right)}{1 - \mu^2}.
\end{align*}

\subsubsection{Neumann Boundary Conditions}
\label{subsubsec:BDF-1 Neumann}

We can also enforce conditions on the derivatives at the end of the domain. Given the Neumann data
\begin{equation*}
    \frac{d u^{n+1}(a)}{dx} = h_{a} \left(t^{n+1}\right), \quad \frac{d  u^{n+1}(b)}{dx} = h_{b} \left(t^{n+1}\right),
\end{equation*}
we can evaluate the derivative formula for the update \eqref{eq:BDF-1 ddx update in 1-D} and use the identities \eqref{eq:ddx of convolution at a} and \eqref{eq:ddx of convolution at b}. Performing these evaluations, we obtain the system of equations
\begin{align*}
    -A + \mu B &= \frac{1}{\alpha} h_{a} \left(t^{n+1}\right) - \mathcal{I}_{x}[R](a), \\
    -\mu A + B &= \frac{1}{\alpha} h_{b} \left(t^{n+1}\right) + \mathcal{I}_{x}[R](b),
\end{align*}
where, again, $\mu = e^{-\alpha(b-a)}$. Solving this system, we find that
\begin{align*}
    A &= -\frac{\frac{1}{\alpha} h_{a} \left(t^{n+1}\right) - \mathcal{I}_{x}[R](a) - \mu \left( \frac{1}{\alpha} h_{b} \left(t^{n+1}\right) + \mathcal{I}_{x}[R](b) \right)}{1-\mu^2}, \\
    B &= -\frac{ \mu \left( \frac{1}{\alpha} h_{a} \left(t^{n+1}\right) - \mathcal{I}_{x}[R](a) \right) - \left( \frac{1}{\alpha} h_{b} \left(t^{n+1}\right) + \mathcal{I}_{x}[R](b) \right)}{1-\mu^2}. 
\end{align*}
We note that Robin boundary conditions, which combine Dirichlet and Neumann conditions can be enforced in a nearly identical manner.

\subsubsection{Periodic Boundary Conditions}
\label{subsubsec:BDF-1 periodic}

Periodic boundary conditions are enforced by taking
\begin{equation*}
    u^{n+1}(a) = u^{n+1}(b), \quad \partial_{x} u^{n+1}(a) = \partial_{x} u^{n+1}(b).
\end{equation*}
Enforcing these conditions through the update \eqref{eq:BDF-1 update in 1-D compact} and its derivative \eqref{eq:BDF-1 ddx update in 1-D}, using the identities \eqref{eq:ddx of convolution at a}-\eqref{eq:ddx of convolution at b}, leads to the system of equations
\begin{align*}
    (1 - \mu)A + (\mu - 1)B &= \mathcal{I}_{x}[R](b) - \mathcal{I}_{x}[R](a), \\
    (\mu - 1)A + (\mu - 1)B &= -\mathcal{I}_{x}[R](b) - \mathcal{I}_{x}[R](a),
\end{align*}
with $\mu = e^{-\alpha(b-a)}$. The solution of this system, after some simplifications is given by
\begin{align*}
    A &= \frac{\mathcal{I}_{x}[R](b)}{1-\mu}, \\
    B &= \frac{\mathcal{I}_{x}[R](a)}{1-\mu}. 
\end{align*}

\subsection{Summary}
\label{subsec:3 Summary}

In this section we discussed the methods used for the fields. Inspired by the underlying connection to integral equations, we obtained analytical expressions for the spatial derivatives of the scalar fields. The evaluation of the derivatives relies on the same core algorithms used to evolve the scalar fields, allowing the proposed methods for derivatives to naturally inherit the geometric flexibility offered by the base field solver. We discussed the essential components used to solve these one-dimensional problems including the fast summation method, as well as the application of boundary conditions. In the next section, we combine the proposed methods for fields and their derivatives with time integration methods for particles to construct new particle-in-cell methods for plasmas.

% Section introduction
\section{Development of a New PIC Method}
\label{sec:4 new PIC method}

This section describes the construction of a new PIC method that leverages the field solvers introduced in the previous section. We begin by introducing the concept of a macroparticle that is the foundation of all PIC methods in section \ref{subsec:particle weights}. Then, we present several recently developed time integration methods for non-separable Hamiltonian systems in section \ref{subsec:4 Integrators for non-separable Hamiltonians}, which are designed to advance the particles in the generalized momentum formulation. An algorithm which couples the time integration method for particles with the proposed field solvers is also presented. We conclude with a brief summary of the section contents in section \ref{subsec:4 Summary}. 

% PIC stuff
%
% Section on the weight scheme
%
\subsection{Moving from Point-particles to Macroparticles}
\label{subsec:particle weights}

In a particle method, the charge density and current density are defined as linear combinations of Dirac distributions. For example, in the non-relativistic limit, these take the form
\begin{align} 
    \rho(\mathbf{x},t) &= \sum_{p=1}^{N_{p}} q_{p} \delta (\mathbf{x}-\mathbf{x}_{p}(t)), \label{eq:charge density}\\
    \mathbf{J}(\mathbf{x},t) & = \sum_{p=1}^{N_{p}} q_{p} \mathbf{v}_{p}(t) \delta (\mathbf{x}-\mathbf{x}_{p}(t)) \label{eq:current density}.
\end{align}
In the above equations, $q_p$, $\mathbf{x}_p$, and $\mathbf{v}_p$ denote the charge, position, and velocity, respectively, of a particle whose label is $p$. In defining things this way, we have dropped the reference to the species altogether, since each particle can be thought of as its own entity.

An essential feature of PIC methods is that the simulation particles are not physical particles. Instead, they represent a collection of particles sampled from an underlying probability distribution function. For this reason, they are often called \textit{simulation particles} or \textit{macroparticles}. It is important to note that the motion of the physical particles (which comprise a given macroparticle) is not tracked during a simulation. The particular ``size" of this sample is reflected in the weight associated with a given macroparticle $w_{mp}$, which can be calculated as
\begin{equation*}
    w_{mp} = \frac{N_{\text{real}}}{N_{\text{simulation}}}.
\end{equation*}
Here, we use $N_{\text{real}}$ to denote the number of physical particles contained within a simulation domain and $N_{\text{simulation}}$ to be the number of simulation particles. The calculation of $N_{\text{real}}$ is problem dependent, but can be expressed in terms of the average macroscopic number density $\bar{n}$ that describes the plasma and a volume that is associated with either the domain or beam being considered. Once the weight for each particle is calculated, it can be absorbed into properties of the particle species, such as the charge, so that $ w_{mp} q_{i}$ can be shortened to $q_{i}$.

While PIC methods can be developed to work with these point-particle representations (see e.g., \cite{OconnorFEM-PIC-ChargeMap}), most PIC methods, including the ones developed in this work, represent particles using shape functions, which replace equations \eqref{eq:charge density} and \eqref{eq:current density} with
\begin{align}
    \rho \left( \mathbf{x}, t \right) &= \sum_{p=1}^{N_{p}} q_{p} S \left(\mathbf{x} - \mathbf{x}_{p}(t) \right), \label{eq:PIC charge density} \\
    \mathbf{J} \left( \mathbf{x}, t \right) &= \sum_{p=1}^{N_{p}} q_{p} \mathbf{v}_{p}(t) S \left(\mathbf{x} - \mathbf{x}_{p}(t) \right), \label{eq:PIC current density}
\end{align}
where the shape function $S$ is now used to represent a simulation particle. The shape functions most often employed in PIC simulations are B-splines, which are compact (local) and positive. Furthermore, they can be easily extended to include additional dimensions using tensor products. While higher-order splines produce smoother mappings to the mesh and possess higher degrees of continuity, the extended support regions create complications in plasma simulations on bounded domains. For simplicity, the particle methods developed in this work employ linear splines to represent particle shapes. The linear spline function that represents the particle $x_p$ on the mesh with spacing $\Delta x$ is given by
\begin{equation}
\label{eq:Linear shape function}
S(x - x_p) = 
\begin{cases} 
  1 - \dfrac{\lvert x - x_p \rvert}{\Delta x}, & 0\leq \lvert x - x_p \rvert \leq \Delta x, \\
  0, & \lvert x - x_p \rvert > \Delta x. 
\end{cases} 
\end{equation}
The shape function \eqref{eq:Linear shape function} generally serves two purposes: (1) It provides a way to map particle data onto the mesh (scatter operation) and (2) can be used to interpolate mesh based quantities to the particles during the time integration (gather operation). For consistency in a PIC method, it is important that the maps between the mesh and the particle be identical. It is well-known that the use of bilinear maps to approximate the charge and current densities is not consistent with the continuity equation \cite{VillasenorChargeConservation92}. As will be shown experimentally in section \ref{subsec:Plasma test problems}, the methods proposed in this work show adequate accuracy for satisfying the gauge condition, even for problems known to be sensitive to subtle violations in the continuity equation. In the next section, we discuss the time integration method used to evolve the particles.

% Particle integrators
%
% Description of Tao's method
%
\subsection{Time Integration of Non-separable Hamiltonian Systems}
\label{subsec:4 Integrators for non-separable Hamiltonians}

In this section, we describe the time integration methods used to evolve the simulation particles. In contrast to the usual Newton-Lorenz treatment for particles, the adoption of a Hamiltonian formulation results in a non-separable system of equations. A Hamiltonian $\mathcal{H}$ is said to be \textit{separable} if it can be written in the form
\begin{equation*}
    \mathcal{H}(\mathbf{x}, \mathbf{P}) = \mathcal{K}\left(\mathbf{P}\right) + \mathcal{U}\left(\mathbf{x}\right),
\end{equation*}
Where $\mathcal{K}$ and $\mathcal{U}$ denote the kinetic and potential energy of the system. In contrast, the Hamiltonian for the VM system considered in this work is non-separable because it contains a momentum-dependent potential and is of the form
\begin{equation*}
    \mathcal{H}(\mathbf{x}, \mathbf{P}) = \mathcal{K}\left(\mathbf{P}\right) + \mathcal{U}\left(\mathbf{x},\mathbf{P}\right).
\end{equation*}

Symplectic integration methods for this class of problems are generally limited to fully-implicit Runge-Kutta type methods \cite{leimkuhler_reich_2005}, which can become prohibitively expensive for systems with many simulation particles. As an example, the simplest method among this class of algorithms is the second-order implicit midpoint rule. Recently, an explicit, symplectic approach with fractional time steps was presented in \cite{Tao2016particles} that extends phase space by duplicating variables and prescribes a certain mixing operator to keep these copies ``close" together; however, the numerical experiments they presented did not consider problems with self-fields, so over time, these copies can drift apart and can lead to certain instabilities and other non-physical behavior. Additionally, the duplication of phase space variables also applies to the fields associated with each set of particle data. This makes the approach computationally demanding in terms of memory usage. Instead, this work seeks a simpler approach that provides a fair trade-off between accuracy and computational efficiency. We provide an outline of the base time integration method in section \ref{subsubsec:AEM method description} and offer an improvement in section \ref{subsubsec:Improved AEM method description} using a correction from a Taylor expansion.

\subsubsection{The Asymmetrical Euler Method}
\label{subsubsec:AEM method description}

A time integration method suitable for non-separable Hamiltonian systems was recently proposed in \cite{Gibbon2017Hamiltonian}, which developed mesh-free methods for solving the VM system in the Darwin limit. Their adoption of a generalized Hamiltonian model for particles was largely motivated by the numerical instabilities associated with time derivatives of the vector potential in this particular limit, which effectively sends the speed of light $c \rightarrow \infty$. The resulting model, which is essentially identical to the formulation \eqref{eq:Position equation relativistic form}-\eqref{eq:Generalized momentum equation relativistic form}, trades additional coupling of phase space for numerical stability through the elimination of this time derivative. They proposed a semi-implicit method, dubbed the asymmetrical Euler method (AEM), which has the form
\begin{empheq}[left=\empheqlbrace]{align}
    \mathbf{x}_{i}^{n+1} &= \mathbf{x}_{i}^{n} + \mathbf{v}_{i}^{n} \Delta t, \label{eq:AEM x update} \\
    \mathbf{P}_{i}^{n+1} &= \mathbf{P}_{i}^{n} + q_i \Bigg( - \nabla \phi^{n+1} + \nabla \mathbf{A}^{n+1} \cdot \mathbf{v}_{i}^{n} \Bigg)\Delta t, \label{eq:AEM P update} \\
    \mathbf{v}_{i}^{n+1} &\equiv \frac{c^2 \left( \mathbf{P}_{i}^{n+1} - q_{i} \mathbf{A}^{n+1}\right)}{\sqrt{ c^2\left( \mathbf{P}_{i}^{n+1} - q_{i} \mathbf{A}^{n+1}\right)^2 + \left(m_{i}c^2\right)^2}}. \label{eq:AEM v defn}
\end{empheq}
This method, which is globally first-order accurate in time, proceeds by, first, performing an explicit update of the particle positions using \eqref{eq:AEM x update}. Next, with the new positions $\mathbf{x}^{n+1}$ and the old velocity $\mathbf{v}^{n}$, we obtain the charge density $\rho^{n+1}$ and an approximate current density $\tilde{\mathbf{J}}^{n+1}$ which are used to evolve the fields under the BDF-1 discretization. We note that the use of $\mathbf{v}^{n}$ in the construction of $\tilde{\mathbf{J}}^{n+1}$ is consistent with a first-order approximation of the true current density $\mathbf{J}^{n+1}$. Finally, once the fields are updated, the generalized momentum $\mathbf{P}^{n+1}$ and its corresponding velocity $\mathbf{v}^{n+1}$ are updated according to equations \eqref{eq:AEM P update} and \eqref{eq:AEM v defn}, respectively.

\subsubsection{An Improved Asymmetrical Euler Method}
\label{subsubsec:Improved AEM method description}

One of the issues with the AEM, which was discussed in the previous section, concerns the explicit treatment of velocity in the generalized momentum equation for problems with magnetic fields. In such cases, this update resembles the explicit Euler method, which is known to generate artificial energy when applied to Hamiltonian systems. We offer a simple modification for such problems in an effort to increase the accuracy and reduce such energy violations. If the update for the generalized momentum equation \eqref{eq:AEM P update} were treated implicitly with a backward Euler discretization, then we would instead compute
\begin{equation*}
    \mathbf{P}_{i}^{n+1} = \mathbf{P}_{i}^{n} + q_i \Bigg( - \nabla \phi^{n+1} + \nabla \mathbf{A}^{n+1} \cdot \mathbf{v}_{i}^{n+1} \Bigg)\Delta t.
\end{equation*}
Unfortunately, this approach necessitates iteration on $\mathbf{P}_{i}^{n+1}$ (through $\mathbf{v}_{i}^{n+1}$), which we are trying to avoid. Instead, with the aid of a Taylor expansion, we linearize the velocity about time level $t^{n}$ so that
\begin{align*}
    \mathbf{v}_{i}^{n+1} &= \mathbf{v}_{i}^{n} + \frac{d \mathbf{v}_{i}^{n}}{dt} \Delta t + \mathcal{O}(\Delta t^2), \\
    &\approx 2\mathbf{v}_{i}^{n} - \mathbf{v}_{i}^{n-1}, \\
    &\equiv \mathbf{v}_{i}^{*}.
\end{align*}
While this treatment is not symplectic, the numerical results presented in section \ref{subsec:Plasma test problems} for the evolution of a single particle indicate that the improved accuracy from the linear correction manages to tame the otherwise significant energy increase introduced by the explicit Euler discretization. Therefore, in problems with magnetic fields, we shall, instead, use the modified update
\begin{align*}
    \mathbf{P}_{i}^{n+1} &= \mathbf{P}_{i}^{n} + q_i \Bigg( - \nabla \phi^{n+1} + \nabla \mathbf{A}^{n+1} \cdot \mathbf{v}_{i}^{*} \Bigg)\Delta t, \\
    \mathbf{v}_{i}^{*} &= 2\mathbf{v}_{i}^{n} - \mathbf{v}_{i}^{n-1},
\end{align*}
as an improvement to the generalized momentum update \eqref{eq:AEM P update}. Since this approach is used to evolve particles in the electromagnetic examples considered in this work, its integration with the PIC lifecycle is presented in Algorithm \ref{alg:IAEM}. Henceforth, we shall call refer to this as the improved asymmetrical Euler method (IAEM).

\begin{algorithm}[t]
    \caption{Outline of the PIC algorithm with the improved asymmetric Euler method (IAEM)}
    Perform one time step of the PIC cycle using the improved asymmetric Euler method. 
    \label{alg:IAEM}
    \begin{algorithmic}[1]
    \State \textbf{Given}: $(\mathbf{x}_{i}^{0}, \mathbf{P}_{i}^{0}, \mathbf{v}_{i}^{0})$, as well as the fields $ \left( \phi^{0}, \nabla \phi^{0} \right)$ and $\mathbf{A}^{0}, \nabla \mathbf{A}^{0}$
    % \State Initialize $\mathbf{v}_{i}^{-1} = \mathbf{v}_{i}(-\Delta t)$ using the second-order Taylor approximation
    % \begin{equation*}
    %     \mathbf{v}_{i}^{-1} \approx \mathbf{v}_{i}^{0} - \frac{q_i}{m_i} \left(-\nabla \phi^{0} + \nabla \mathbf{A}^{0} \cdot \mathbf{v}^{0} \right)\Delta t.
    % \end{equation*}
    \State Initialize $\mathbf{v}_{i}^{-1} = \mathbf{v}_{i}(-\Delta t)$ using a Taylor approximation.
    \While{stepping}
    
    \vspace{10pt}
    
    \State \label{start time loop} Update the particle positions with $$\mathbf{x}_{i}^{n+1} = \mathbf{x}_{i}^{n} + \mathbf{v}_{i}^{n} \Delta t.$$
    
    \State Using the position data $\mathbf{x}_{i}^{n+1}$ and velocity data $\mathbf{v}_{i}^{n}$, compute the current density $$\tilde{\mathbf{J}}^{n+1} = \mathbf{J}^{n+1} + \mathcal{O}(\Delta t).$$
    
    \State Using the position data $\mathbf{x}_{i}^{n+1}$, compute the charge density $\rho^{n+1}$.
    
    \State Compute the potentials and their derivatives at time level $t^{n+1}$ using the BDF field solver.
    
    \State Evaluate the Taylor corrected particle velocities $$\mathbf{v}_{i}^{*} = 2 \mathbf{v}_{i}^{n} - \mathbf{v}_{i}^{n-1}.$$
    
    \State Calculate the new generalized momentum according to $$\mathbf{P}_{i}^{n+1} = \mathbf{P}_{i}^{n} + q_i \Big( - \nabla \phi^{n+1} + \nabla \mathbf{A}^{n+1} \cdot \mathbf{v}_{i}^{*} \Big)\Delta t.$$
    
    \State Convert the new generalized momenta into new particle velocities with $$ \mathbf{v}_{i}^{n+1} =  \frac{c^2 \left( \mathbf{P}_{i}^{n+1} - q_{i} \mathbf{A}^{n+1}\right)}{\sqrt{ c^2\left( \mathbf{P}_{i}^{n+1} - q_{i} \mathbf{A}^{n+1}\right)^2 + \left(m_{i}c^2\right)^2}}. $$
    
    \State Shift the time history data and return to step \ref{start time loop} to begin the next time step. 
    
    \EndWhile
    \end{algorithmic}
\end{algorithm}

\subsection{Summary}
\label{subsec:4 Summary}

In this section we proposed new PIC methods for the numerical simulation of plasmas. To this end, we combined methods for fields and their derivatives, which were introduced in section \ref{sec:3 Methods for field equations}, with time integration methods for non-separable Hamiltonian systems. A high level description of the particle method was presented. In the next section, we present results from the numerical experiments conducted with the algorithms introduced in this paper. First, we establish the refinement properties of the field solver and methods for derivatives. Then, we demonstrate the performance of the proposed PIC methods in several key test problems involving plasmas with varying complexity.
% Section introduction
%
% Section on numerical examples
%
\section{Numerical Examples}
\label{sec:5 Numerical results}

This section presents numerical results that demonstrate the proposed methods for fields and particles that comprise the formulation adopted in this work. First, we establish the convergence properties of the BDF field solver and methods for evaluating spatial derivatives. The proposed methods are demonstrated using boundary conditions that will be considered in the applications involving plasmas. Once the refinement properties of the field solver are established, we focus on applications to plasmas. We begin with a single particle example involving cyclotron motion before moving to more complex problems involving self-fields. After benchmarking the time integration methods used for the generalized momentum formulation, we apply the proposed PIC methods to a suite of electrostatic and electromagnetic test problems.

% Field solver numerical examples
%
% Numerical examples for field solvers
%
\subsection{Numerical Experiments for Field Solvers}
\label{subsec:Field-solver experiments}

In this section we establish the refinement properties of the BDF field solver and the proposed methods for computing spatial derivatives. Results for space and time refinement experiments are presented from a suite of two-dimensional test problems using boundary conditions that are relevant to the plasma examples considered in this work.

\subsubsection{Periodic Boundary Conditions}
\label{subsubsec:field-solver periodic results}

We first consider the two-dimensional inhomogeneous scalar wave equation
\begin{equation}
    \label{eq:field-solver 2-D periodic}
    \partial_{tt}u - \Delta u = S(x,y),
\end{equation}
and
\begin{equation}
    \label{eq:field-solver 2-D periodic source}
    S(x,y) = 3 e^{-t} \sin(x)\cos(y).
\end{equation}
We apply two-way periodic boundary conditions on the domain $[0,2\pi] \times [0,2\pi]$ and use the initial data
\begin{equation}
    u(x,y,0) = \sin(x)\cos(y), \quad \partial_t u(x,y,0) = -\sin(x)\cos(y).
\end{equation}
The problem \eqref{eq:field-solver 2-D periodic} is associated with the manufactured solution
\begin{equation}
     \label{eq:field-solver 2-D periodic solution}
    u(x,y,t) = e^{-t} \sin(x)\cos(y),
\end{equation}
and defines the source function \eqref{eq:field-solver 2-D periodic source}. The partial derivatives of this solution are calculated to be 
\begin{align}
    \partial_{x} u(x,y,t) &= e^{-t} \cos(x)\cos(y), \label{eq:field-solver 2-D periodic ddx solution} \\
    \partial_{y} u(x,y,t) &= -e^{-t} \sin(x)\sin(y). \label{eq:field-solver 2-D periodic ddy solution}
\end{align}

\begin{figure}[!htb]
    \centering
    \subfloat[][]{
    \includegraphics[width=0.85\textwidth]{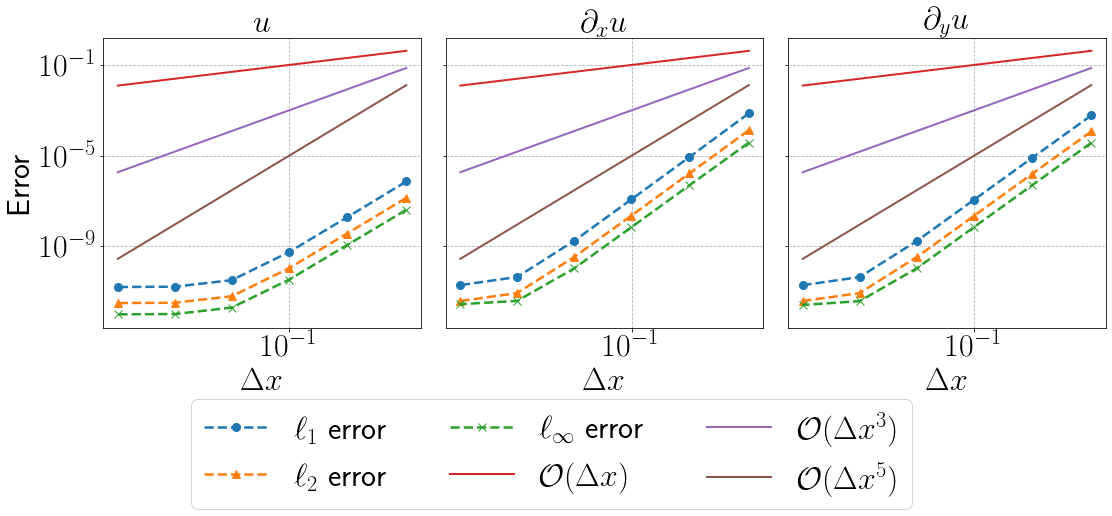}
    \label{fig:2d BDF-1 space refinement periodic}}
    \\
    \subfloat[][]{
    \includegraphics[width=0.85\textwidth]{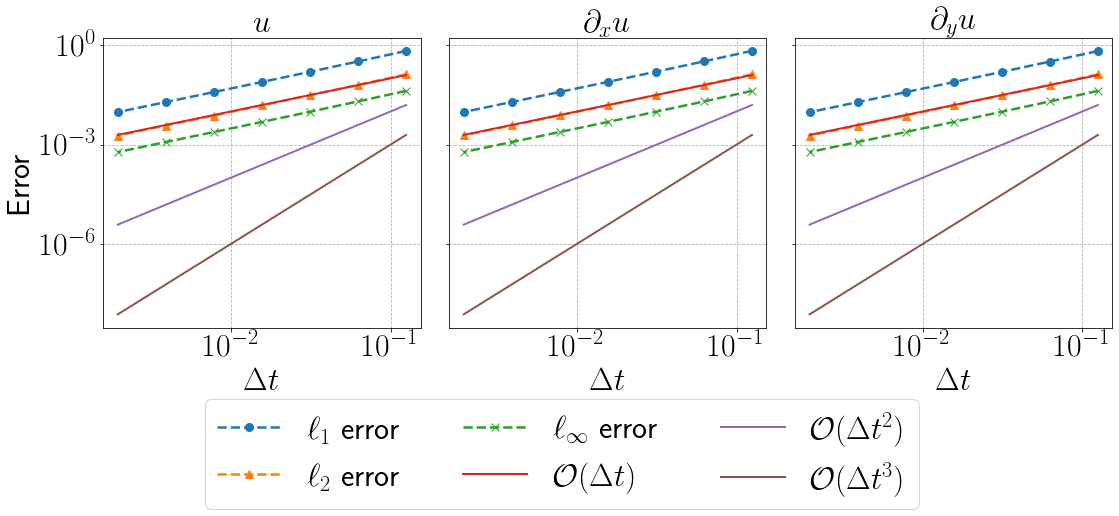}
    \label{fig:2d BDF-1 time refinement periodic}}
    \caption{Space-time refinement of the solution and its spatial derivatives for the two-dimensional periodic example \ref{subsubsec:field-solver periodic results} obtained with the first-order BDF method.
    }
    \label{fig:2d space-time refinement periodic}
\end{figure}

For the space refinement experiment, we varied the spatial mesh in each direction from $16$ points to $512$ points. To keep the temporal error in the methods small during the refinement, we applied the methods for 1 time step using a step size of $\Delta t = 1\times 10^{-4}$. The refinement plots in Figure \ref{fig:2d space-time refinement periodic} indicate fifth-order accuracy in space for all methods. We note that the derivatives in the methods begin to level-off as the error approaches $1\times 10^{-11}$. This is likely due to a different error coefficient in time, which arises from the differentiation process. A smaller time step would be necessary to remove this feature, but this requires some modification of the quadrature. 

In the temporal refinement study, the solution is computed until a final time of $T=1$ using a fixed $256 \times 256$ mesh in space. We successively double the number of time steps from $N_{t} = 8$ until $N_{t} = 512$. We use the analytical solution to initialize the method, since it is available. The results of the temporal refinement study are presented in Figure \ref{fig:2d space-time refinement periodic}, in which all methods, including those for the derivatives, display the expected first-order convergence rate in time.

\subsubsection{Dirichlet Boundary Conditions}
\label{subsubsec:field-solver dirichlet results}

For the Dirichlet problem, we consider the two-dimensional inhomogeneous scalar wave equation
\begin{equation}
    \label{eq:field-solver 2-D dirichlet}
    \partial_{tt}u - \Delta u = S(x,y),
\end{equation}
and
\begin{equation}
    \label{eq:field-solver 2-D dirichlet source}
    S(x,y) = 3 e^{-t} \sin(x)\sin(y).
\end{equation}
We apply homogeneous Dirichlet boundary conditions on the domain $[0,2\pi] \times [0,2\pi]$ and use the initial data
\begin{equation}
    u(x,y,0) = \sin(x)\sin(y), \quad \partial_t u(x,y,0) = -\sin(x)\sin(y).
\end{equation}
The problem \eqref{eq:field-solver 2-D periodic} is associated with the manufactured solution
\begin{equation}
     \label{eq:field-solver 2-D dirichlet solution}
    u(x,y,t) = e^{-t} \sin(x)\sin(y),
\end{equation}
and defines the source function \eqref{eq:field-solver 2-D dirichlet source}. The partial derivatives of this solution are calculated to be 
\begin{align}
    \partial_{x} u(x,y,t) &= e^{-t} \cos(x)\sin(y), \label{eq:field-solver 2-D dirichlet ddx solution} \\
    \partial_{y} u(x,y,t) &= e^{-t} \sin(x)\cos(y). \label{eq:field-solver 2-D dirichelt ddy solution}
\end{align}

\begin{figure}[!htb]
    \centering
    \subfloat[][]{
    \includegraphics[width=0.85\textwidth]{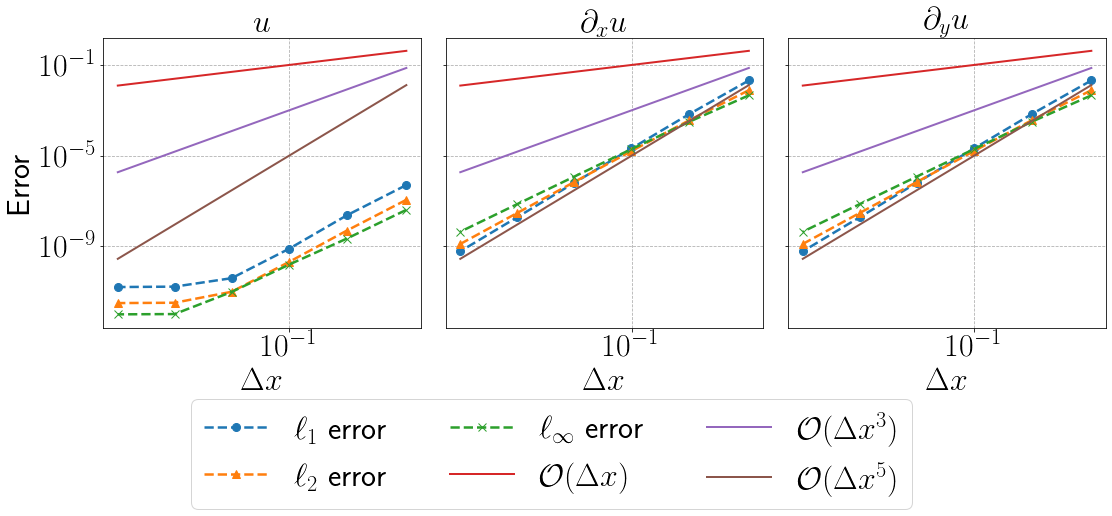}
    \label{fig:2d BDF-1 space refinement dirichlet}}
    \\
    \subfloat[][]{
    \includegraphics[width=0.85\textwidth]{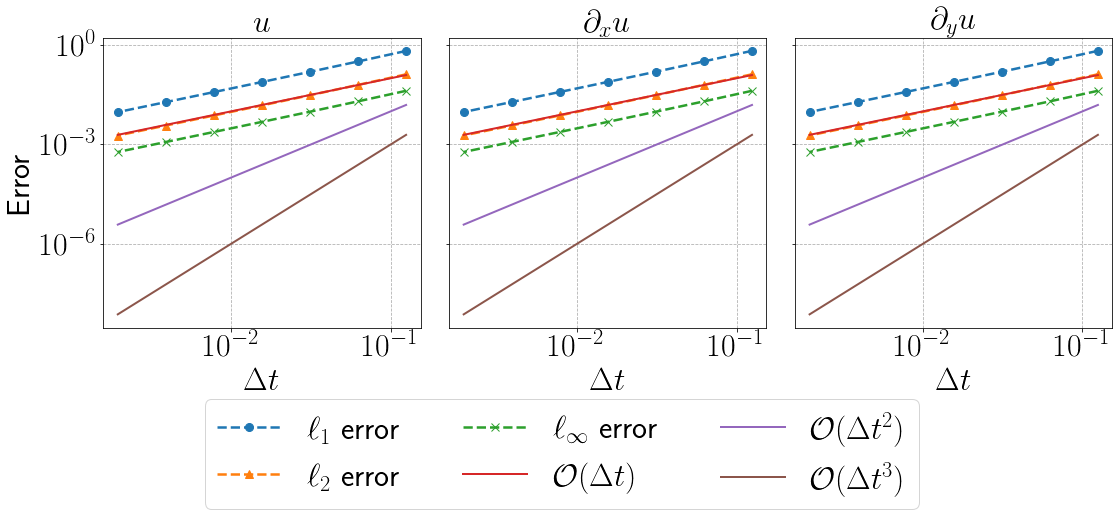}
    \label{fig:2d BDF-1 time refinement dirichlet}}
    \caption{Space-time refinement of the solution and its spatial derivatives for the two-dimensional Dirichlet problem \ref{subsubsec:field-solver dirichlet results} obtained with the first-order BDF method.
    }
    \label{fig:2d space-time refinement dirichlet}
\end{figure}

We performed the spatial refinement study by varying the number of mesh points in each direction from $16$ points to $512$ points. Again, to keep the temporal error in the methods small while space is refined, we applied the methods for only 1 time step with a step size of $\Delta t = 1\times 10^{-4}$. The same remark about small time step sizes mentioned in the space refinement experiment for the periodic case applies here, as well (see section \ref{subsubsec:field-solver periodic results}). The refinement plots in Figure \ref{fig:2d space-time refinement dirichlet} indicate that the methods refine, approximately, to fifth-order in space. In both the mixed and pure BDF approaches, the error in the derivatives behaves differently from what was observed in the periodic example. In particular, we do not observe a flattening of the error when the spacing $\Delta x$ is small.

In the temporal refinement study, the solution is computed until a final time of $T=1$. We use a fixed $256 \times 256$ spatial mesh and the number of time steps in each case is successively doubled from $N_{t} = 8$ until $N_{t} = 512$. Errors can be directly measured with the analytical solution and its derivatives. The results of the temporal refinement study are presented in Figure \ref{fig:2d space-time refinement dirichlet}, in which all methods, including those for the derivatives, display the expected first-order convergence rate. The behavior is essentially identical to the results obtained for the periodic problem presented in Figure \ref{fig:2d space-time refinement periodic}.

\subsection{Plasma Test Problems}
\label{subsec:Plasma test problems}

In this section, we present numerical results that demonstrate the performance of the proposed methods for fields in PIC applications. The benchmark PIC methods used in the comparisons implement standard conservative charge weighting for electrostatic problems and conservative current weighting \cite{VillasenorChargeConservation92} for electromagnetic problems in 2D. The electrostatic problems use the FFT to solve Poisson's equation, while the electromagnetic problems use the staggered FDTD grid introduced by Yee \cite{Yee1966}. First, we test the particle methods and test the formulation with a single particle moving through known fields. We then focus on applying the methods to problems involving fields that respond to the motion of the particles. This includes the well known two-stream instability as well as more challenging simulations of plasma sheaths and particle beams. In particular, the last problem we consider is the Mardahl beam problem \cite{Mardahl97conservation}, which is a popular benchmark problem for relativistic beams. Note that in each of the plasma experiments presented in this paper, we use the physical constants listed in Table \ref{tab:physical constants}. We remark that the implementation used to obtain the results presented in this section is based on a non-dimensionalization whose details can be found in Appendix \ref{app:Non-dim}. We provide the relevant parameters used to setup each of the test problems, so that the results can be more easily reproduced and compared with other methods.

\begin{table}[!htb]
    \centering
    \def\arraystretch{1.2}
    \begin{tabular}{ | c || c | }
        \hline
        \textbf{Parameter}  & \textbf{Value} \\
        \hline
        Ion mass ($m_i$) [kg] & $9.108379025973462\times 10^{-29}$ \\
        %\hline
        Electron mass ($m_e$) [kg] & $9.108379025973462\times 10^{-31}$ \\
        %\hline
        Boltzmann constant ($k_B$) [kg m$^2$ s$^{-2}$ K$^{-1}$] & $1.38064852\times 10^{-23}$ \\
        %\hline
        Permittivity of free space ($\epsilon_{0})$ [kg$^{-1}$ m$^{-3}$ s$^{4}$ A$^{2}$] & $8.854187817\times 10^{-12}$ \\
        %\hline
        Permittivity of free space ($\mu_{0}$) [kg m s$^{-2}$ A$^{-2}$] & $1.25663706\times 10^{-6}$ \\
        %\hline
        Speed of light ($c$) [m/s] & $2.99792458\times 10^{8}$ \\
        \hline
    \end{tabular}
    \caption{Table of the physical constants, given in SI units, used in the numerical experiments.}
    \label{tab:physical constants}
\end{table}

\subsubsection{Motion of a Charged Particle}

We first compare the time integration methods for non-separable Hamiltonians with the well-known Boris method \cite{Boris1970}. This is a natural first step before applying the method to problems with dynamic ``self-fields" that respond to particle motion. Here, we consider a simple model for the motion of a single charged particle that is given by
\begin{align*}
    \frac{d\mathbf{x}}{dt} &= \mathbf{v}, \quad \frac{d\mathbf{v}}{dt} = \frac{q}{m}\left( \mathbf{E} + \mathbf{v} \times \mathbf{B} \right).
\end{align*}

We use electro- and magneto-static fields here and suppose that the magnetic field lies along the $\hat{\mathbf{z}}$ unit vector
\begin{equation*}
    \mathbf{B} = B_{0} \hat{\mathbf{z}}, \quad
    \mathbf{E} = E^{(1)} \hat{\mathbf{x}} + E^{(y)} \hat{\mathbf{y}} + E^{(z)} \hat{\mathbf{z}},
\end{equation*}
where $B_{0}$ is a constant. Again, component-based definitions have been used for the fields $\mathbf{E} = \left( E^{(1)}, E^{(2)}, E^{(3)} \right)$ and $\mathbf{B} = \left( B^{(1)}, B^{(2)}, B^{(3)} \right)$. Consequently, we have that
\begin{equation*}
    \mathbf{v} \times \mathbf{B} = v^{(2)} B_{0} \hat{\mathbf{x}} - v^{(1)} B_{0} \hat{\mathbf{y}},
\end{equation*}
so the full equations of motion are
\begin{empheq}[left=\empheqlbrace]{align*}
    \frac{ d x^{(1)} }{dt} &= v^{(1)}, \quad \frac{ d v^{(1)} }{dt} = \frac{q}{m}\left( E^{(1)} + v^{(2)} B_{0} \right), \\
   \frac{ d x^{(2)} }{dt} &= v^{(2)}, \quad \frac{ d v^{(2)} }{dt} = \frac{q}{m}\left( E^{(2)} - v^{(1)} B_{0} \right), \\
   \frac{ d x^{(3)} }{dt} &= v^{(3)}, \quad \frac{ d v^{(3)} }{dt} = \frac{q}{m} E^{(3)}.
\end{empheq}
We can then use the linear momentum $\mathbf{p} = m \mathbf{v} $ to obtain
\begin{empheq}[left=\empheqlbrace]{align*}
    \frac{ d x^{(1)} }{dt} &= \frac{1}{m} p^{(1)}, \quad \frac{ d p^{(1)} }{dt} = q\left( E^{(1)} + \frac{1}{m} p^{(2)} B_{0} \right), \\
    \frac{ d x^{(2)} }{dt} &= \frac{1}{m} p^{(2)}, \quad \frac{ d p^{(2)} }{dt} = q\left( E^{(2)} - \frac{1}{m} p^{(1)} B_{0} \right), \\
    \frac{ d x^{(3)} }{dt} &= \frac{1}{m} p^{(3)}, \quad \frac{ d p^{(3)} }{dt} = q E^{(3)}.
\end{empheq}
% Next, we show how to convert the electric and magnetic fields to potentials for the generalized momentum formulation.

Using the potentials $\phi$ and $\mathbf{A} \equiv \left( A^{(1)}, A^{(2)}, A^{(3)} \right)$, one can compute the electric and magnetic fields via \eqref{eq:Convert potentials to EB}.
% Using the potentials $\phi$ and $\mathbf{A} \equiv \left( A^{(1)}, A^{(2)}, A^{(3)} \right)$, one can compute the electric and magnetic fields via \eqref{eq:Convert potentials to EB}, which is equivalent to writing
% \begin{align*}
%     E^{(1)} &= - \partial_{x} \phi - \partial_{t} A^{(1)},  \quad  B^{(1)} = \partial_{y} A^{(3)} - \partial_{z} A^{(2)} , \\
%     E^{(2)} &= - \partial_{y} \phi - \partial_{t} A^{(2)}, \quad B^{(2)} = - \partial_{x} A^{(3)} + \partial_{z} A^{(1)}, \\
%     E^{(3)} &= - \partial_{z} \phi - \partial_{t} A^{(3)}, \quad B^{(3)} = \partial_{x} A^{(2)} - \partial_{y} A^{(1)}.
% \end{align*}
The time-independence of the magnetic field for this problem implies that $\partial_{t} \mathbf{A} = 0$, so that 
$$ \mathbf{E} = -\nabla \phi \implies  E^{(1)} = - \partial_{x} \phi, \quad E^{(2)} = - \partial_{y} \phi, \quad E^{(3)} = - \partial_{z} \phi.$$
Therefore, for this problem, we can use
\begin{equation*}
    \phi = -E^{(1)} x -E^{(2)} y -E^{(3)} z. 
\end{equation*}
Moreover, the magnetic field contains only a z-component, which implies that it can be written as
\begin{align*}
    \mathbf{B} = (0, 0, B_{0}) = (0, 0, \partial_{x} A^{(2)} - \partial_{y} A^{(1)} ).
\end{align*}
As the choice of functions for gauges is not unique, it suffices to pick
\begin{equation*}
    A^{(1)} \equiv 0, \quad A^{(2)} = B_0 x, \quad A^{(3)} \equiv 0.
\end{equation*}
In summary, the non-zero values and required derivatives for the potentials are given by
\begin{align*}
    - \partial_{x} \phi &= E^{(1)}, \quad - \partial_{y} \phi = E^{(2)}, \quad - \partial_{z} \phi = E^{(3)}, \quad
    A^{(2)} = B_{0} x, \quad \partial_x A^{(2)} = B_{0},
\end{align*}
% \begin{align*}
%     - \partial_{x} \phi &= E^{(1)}, \quad - \partial_{y} \phi = E^{(2)}, \quad - \partial_{z} \phi = E^{(3)}, \\
%     A^{(1)} &= 0, \quad A^{(2)} = B_{0} x, \quad A^{(3)} = 0, \\
%     \partial_x A^{(1)} &= 0, \quad \partial_y A^{(1)} = 0, \quad \partial_z A^{(1)} = 0, \\
%     \partial_x A^{(2)} &= B_{0}, \quad \partial_y A^{(2)} = 0, \quad \partial_z A^{(2)} = 0, \\
%     \partial_x A^{(3)} &= 0, \quad \partial_y A^{(3)} = 0, \quad \partial_z A^{(3)} = 0,
% \end{align*}
which results in the simplified equations of motion for the Hamiltonian system
\begin{empheq}[left=\empheqlbrace]{align*}
    \frac{ d x^{(1)} }{dt} &= \frac{1}{m} P^{(1)}, \quad \frac{ d P^{(1)} }{dt} = q E^{(1)} + \frac{q }{m} \Bigg[ B_{0} \left( P^{(2)} - q B_{0} x^{(1)} \right) \Bigg], \\
    \frac{ d x^{(2)} }{dt} &= \frac{1}{m} \left( P^{(2)} - q B_{0} x^{(1)} \right), \quad \frac{ d P^{(2)} }{dt} = q E^{(2)}, \\
    \frac{ d x^{(3)} }{dt} &= \frac{1}{m} P^{(3)}, \quad \frac{ d P^{(3)} }{dt} = q E^{(3)}.
\end{empheq}

\begin{figure}[t]
    \centering
    \includegraphics[scale=0.25]{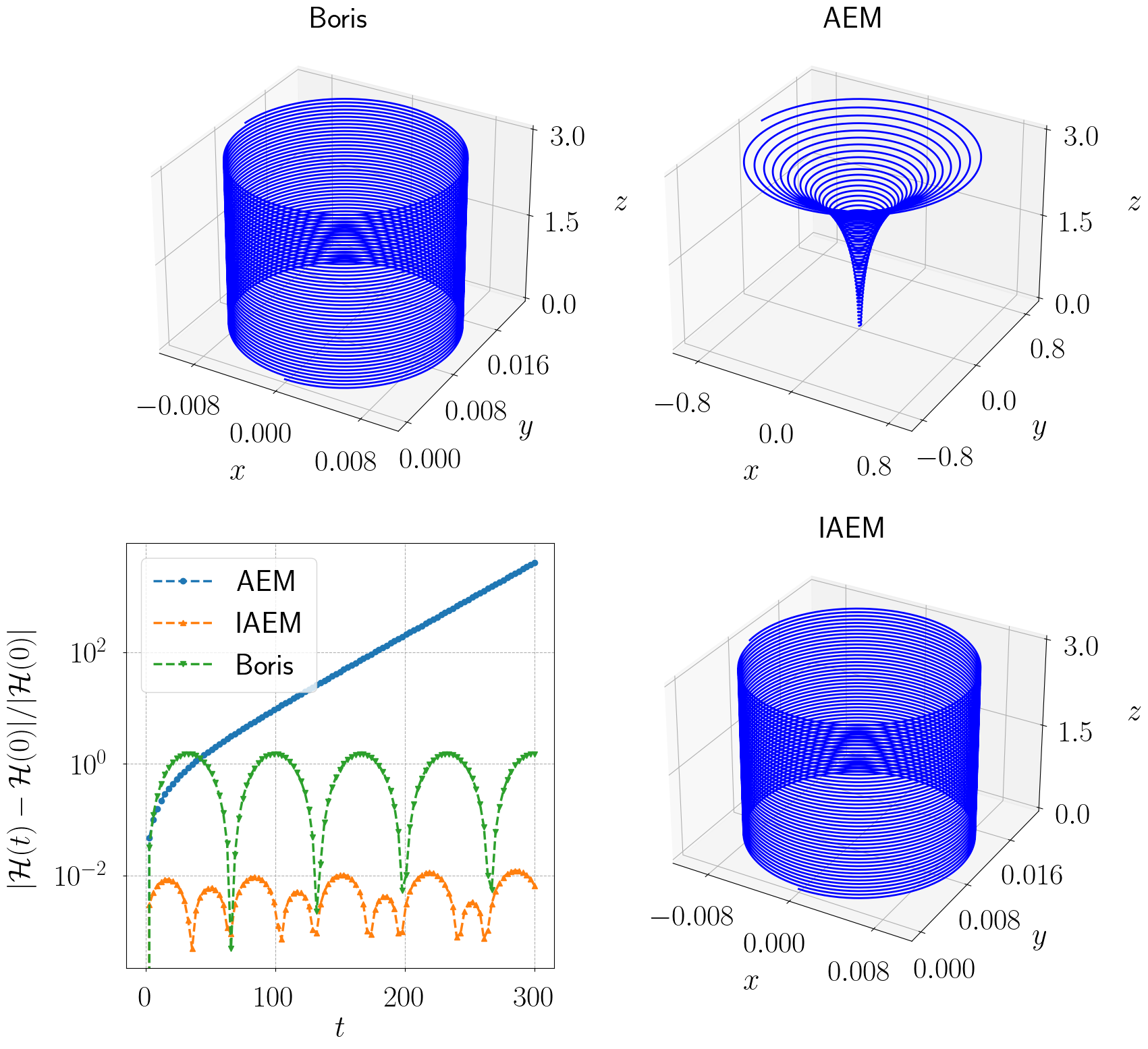}
    \caption{Trajectories for the single particle test obtained using the Boris method \cite{Boris1970}, the AEM \cite{Gibbon2017Hamiltonian}, and the IAEM. The particle rotates about a static magnetic field which points in the $z$-direction. Also shown is the time history of the Hamiltonian generated by each of the methods which is measured relative to the initial data. In particular, the AEM shows a growth in the overall energy causing the gyroradius to increase. This behavior is not observed in the improved method.}
    \label{fig:single particle cyclotron}
\end{figure}

The setup for the test consists of a single particle with mass $m = 1.0$ and charge $q = -1.0$ whose initial position is at the origin of the domain $\mathbf{x}(0) = \left( 0, 0, 0\right)$. Initially, the particle is given non-zero momenta in the $x$ and $z$ directions so as to generate so called ``cyclotron" motion. We choose the initial momenta to be $\mathbf{p}(0) = \mathbf{P}(0) = \left( 1.0 \times 10^{-2}, 0, 1.0 \times 10^{-2} \right)$. The strength of the magnetic field in the $z$ direction is selected to be $B_{0} = 1.0$, and we ignore the contributions from the electric field, so that $\mathbf{E} = \left( 0 ,0, 0 \right)$. Each method is run to a final time of $T = 300.0$, using a total of $1\times 10^{4}$ time steps, so that $\Delta t = 0.03$. The position of the particle is tracked through time and plotted as a curve in three-dimensions. In Figure \ref{fig:single particle cyclotron}, we compare the particle trajectories and the relative error in the Hamiltonian obtained with each of the methods. We note that the gyroradius for the AEM increases over time because the method is not volume-preserving. Over time, this causes the total energy to increase, as substantiated by the error plots for the Hamiltonian. In contrast, we see that the simple correction used in the IAEM reduces this behavior; however, the correction does not completely eliminate this behavior in the case of longer simulations, as the truncation errors accumulate over time.

\begin{figure}[t]
    \centering
    \includegraphics[scale=0.4]{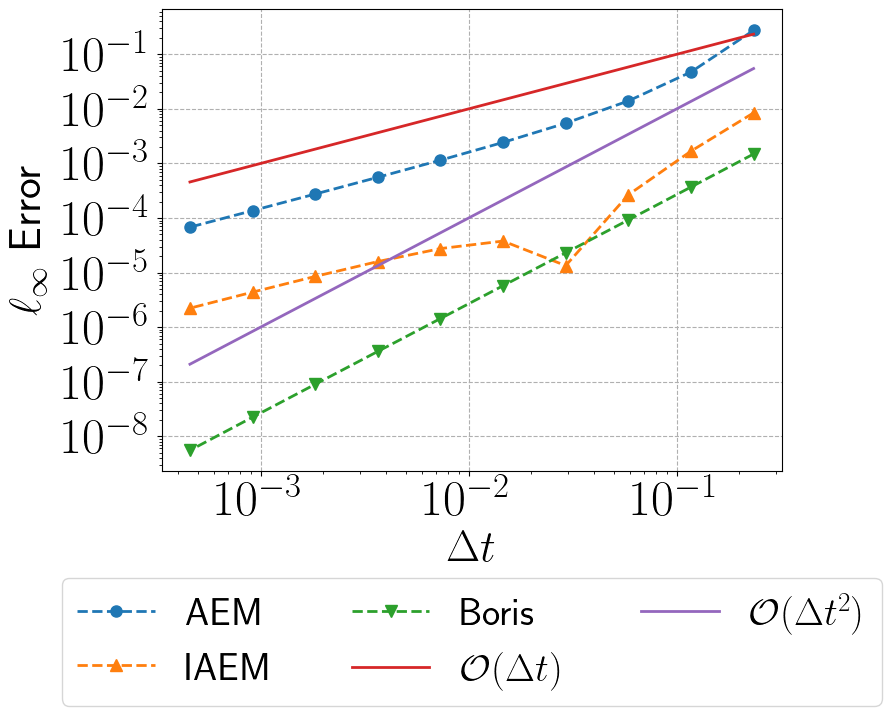}
    \caption{Refinement study for the trajectory of a single particle obtained using the Boris method \cite{Boris1970}, AEM \cite{Gibbon2017Hamiltonian}, and the IAEM that uses a Taylor correction. Errors are measured in the $\ell_{\infty}$-norm against a reference solution obtained using the Boris method. Even though the AEM with the Taylor correction remains globally first-order accurate in time, its improvement over the AEM is quite apparent (roughly an order of magnitude).}
    \label{fig:single particle refinement}
\end{figure}

Next, we perform a refinement study of the methods to examine their error properties using the same experimental parameters from the cyclotron test. We reduce the final time to $T = 30.0$ and measure the errors with the $\ell_{\infty}$ norm using a reference solution computed with $10^{6}$ time steps, so that $\Delta t = 3.0 \times 10^{-5}$. The test successively doubles the number of steps, starting with 100 steps and uses, at most, $1.28 \times 10^{4}$ steps. The results of the refinement study are shown in Figure \ref{fig:single particle refinement}. Despite the fact that both the base and improved versions of the AEM refine to first-order accuracy, we see that the Taylor correction decreases the error in the base method by roughly an order of magnitude. For coarser time step sizes, the improved method has errors that are (in some sense) comparable to the Boris method, which is second-order accurate. Of course, the second-order method will outperform both versions of the AEM as the time step decreases.

\subsubsection{The Cold Two-stream Instability}

We consider the motion of ``cold" streams of electrons restricted to a one-dimensional periodic domain by means of a sufficiently strong (uniform) magnetic field in the two remaining directions. Ions are taken to be uniformly distributed in space and sufficiently heavy compared to the electrons so that their motion can be neglected in the simulation. The ions, which remain stationary, act as a neutralizing background against the dynamic electrons. The electron velocities are represented as a sum of two Dirac delta distributions that are symmetric in velocity space:
\begin{equation*}
    f(v) = \delta (v - v_b) + \delta (v + v_b).
\end{equation*}
The stream velocity $v_b > 0$ is set according to a drift velocity whose value ultimately controls the interaction of the streams. A slight perturbation in the electron velocities is then introduced to force a charge imbalance, which generates an electric field that attempts to restore the neutrality of the system. This causes the streams to interact or ``roll-up," corresponding to regions of trapped particles.

In order to describe the models used in the simulation, let us denote the components of the position and momentum vectors for particle $i$ as $\mathbf{x}_i \equiv \left( x_{i}^{(1)}, x_{i}^{(2)}, x_{i}^{(3)} \right)$ and $\mathbf{P}_i \equiv \left( P_{i}^{(1)}, P_{i}^{(2)}, P_{i}^{(3)} \right)$, respectively. Then, the equations for the motion of particle $i$ assume the form
\begin{align*}
    \frac{d x_{i}^{(1)}}{d t} &= \frac{1}{m_i} P_{i}^{(1)}, \quad \frac{d P_{i}^{(1)}}{d t} = - q_i \partial_{x} \phi.
\end{align*}
The motion in this plane requires knowledge of $\phi$, $\partial_{x} \phi$, which can be obtained by solving a two-way wave equation for the scalar potential:
\begin{equation}
    \label{eq:TSI wave model potential}
    \frac{1}{c^2} \partial_{tt} \phi - \partial_{xx} \phi = \frac{1}{\epsilon_0} \rho.
\end{equation}
As this is an electrostatic problem, the gauge condition can be safely ignored. In the limit where the characteristic thermal velocity $V$ of the particles become well-separated from the speed of light $c$, so that $\kappa = c/V \gg 1$, one instead solves the Poisson equation
\begin{equation}
    \label{eq:TSI Poisson model potential}
    - \partial_{xx} \phi = \frac{1}{\epsilon_0} \rho.
\end{equation}
Using asymptotic analysis, it can be shown that the approximation error made by using the Poisson model for the scalar potential is $\mathcal{O}(1/\kappa)$ \cite{Wolf-Thesis}.

We establish the efficacy of the proposed algorithms for time stepping particles and evolving fields by comparing with well-known methods. The setup for this test problem consists of a non-dimensional spatial mesh defined on the interval $[-10\pi/3, 10\pi/3]$, which is discretized using 128 total grid points and is supplied with periodic boundary conditions. The non-dimensional final time for the simulation is taken to be \mbox{$T_{f} = 100$} with 4,000 time steps being used to evolve the system. The plasma is represented with a total of 30,000 macroparticles, consisting of 10,000 ions and 20,000 electrons. As mentioned earlier, the positions of the ions and electrons are taken to be uniformly spaced along the grid. Ions remain stationary in the problem, so we set their velocity to zero. The construction of the streams begins by first splitting the electrons into two equally sized groups, whose respective (non-dimensional) drift velocities are set to be $\pm 1.$ To generate an instability we add a perturbation to the electron velocities of the form
\begin{equation*}
    \delta v(x) = \epsilon \sin \left( \frac{2\pi k (x - a)}{L_x} \right).
\end{equation*}
Here, $\epsilon = 5.0\times 10^{-4}$ controls the strength of the perturbation, $k = 1$ is the wave number for the perturbation, $x$ is the position of the particle (electron), $a$ is the left-most grid point, and $L_x$ is the length of the domain. In a more physically realistic simulation, the perturbation would be induced by some external force, which would also result in a perturbation of the position data for the particles. Such a perturbation of the position data requires a self-consistent field solve to properly initialize the potentials. In our simulation, we assume that no spatial perturbation is present, so that the fields are identically zero at the initial time step. The plasma parameters used in the non-dimensionalization for this test problem are displayed in Table \ref{tab:two-stream plasma parameters}. Note that in this configuration, the normalized speed of light is $\kappa = 50$, and the corresponding normalized permittivity is $\sigma_{1} = 1$. We find that this configuration adequately resolves the plasma Debye length ($\approx 6$ cells/$\lambda_{D}$), angular plasma period ($\approx 40$ steps/$\omega_{pe}^{-1}$), and the particle CFL $< 1$, which are typically used to ensure maintain stability in explicit PIC methods.

\begin{table}[!h]
    \centering
    \def\arraystretch{1.2}
    \begin{tabular}{ | c || c | }
        \hline
        \textbf{Parameter}  & \textbf{Value} \\
        \hline
        Average number density ($\bar{n}$) [m$^{-3}$] & $7.856060\times 10^{1}$ \\
        %\hline
        Average temperature ($\bar{T}$) [K] & $2.371698\times 10^{6}$ \\
        %\hline
        Debye length ($\lambda_{D})$ [m] & $1.199170\times 10^{4}$ \\
        %\hline
        Inverse angular plasma frequency ($\omega_{pe}^{-1}$) [s/rad] & $2.000000\times 10^{-3}$ \\
        %\hline
        Thermal velocity ($v_{th} = \lambda_{D} \omega_{pe}$) [m/s] & $5.995849\times 10^{6}$ \\
        \hline
    \end{tabular}
    \caption{Table of the plasma parameters used in the two-stream instability example.}
    \label{tab:two-stream plasma parameters}
\end{table}

% For this problem, we note that the particular IC leads to an equivalence between the two methods.

\begin{figure}[!htb]
    \centering
    \subfloat[][Leapfrog]{
    \includegraphics[width=0.38\textwidth]{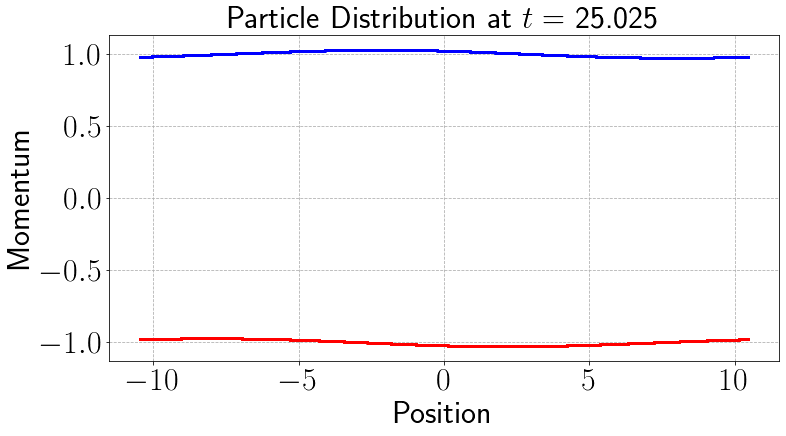}}
    \subfloat[][AEM]{
    \includegraphics[width=0.38\textwidth]{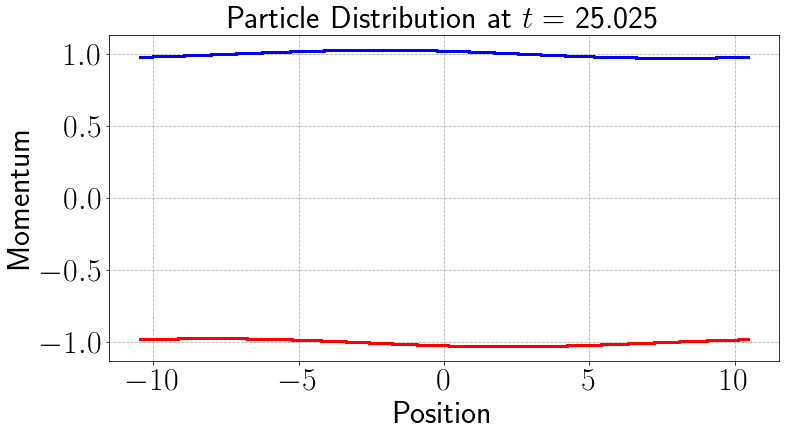}} \\
    \subfloat[][Leapfrog]{
    \includegraphics[width=0.38\textwidth]{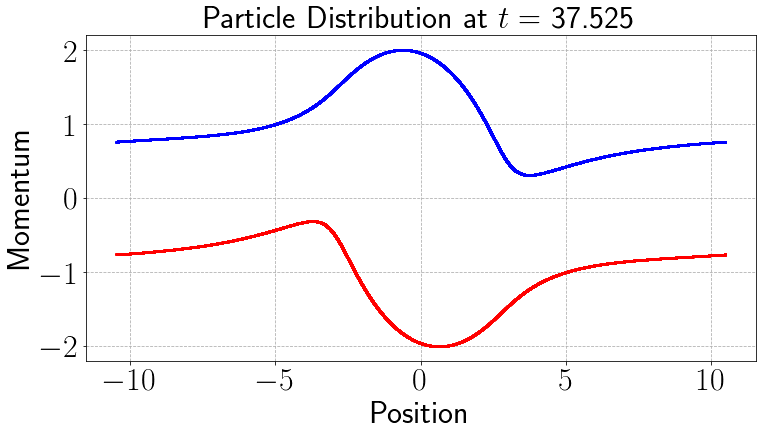}}
    \subfloat[][AEM]{
    \includegraphics[width=0.38\textwidth]{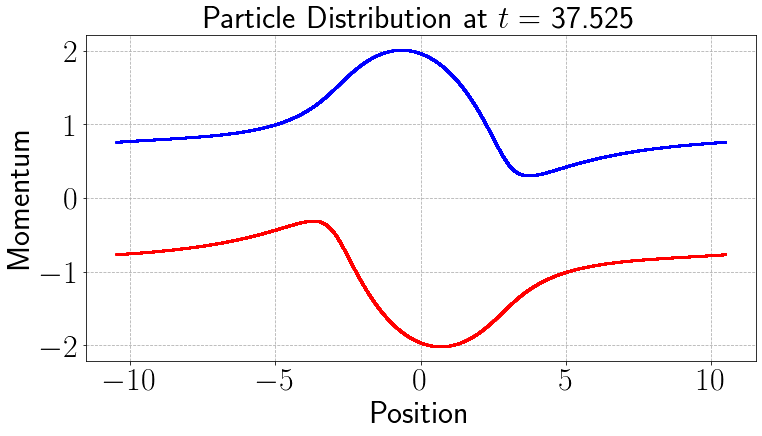}} \\
    \subfloat[][Leapfrog]{
    \includegraphics[width=0.38\textwidth]{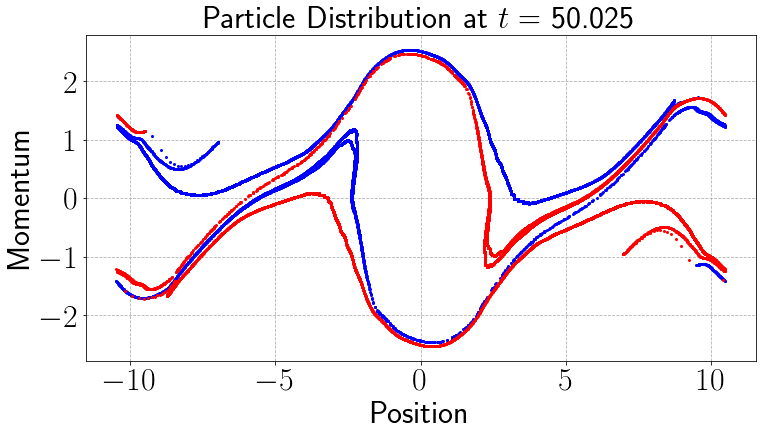}}
    \subfloat[][AEM]{
    \includegraphics[width=0.38\textwidth]{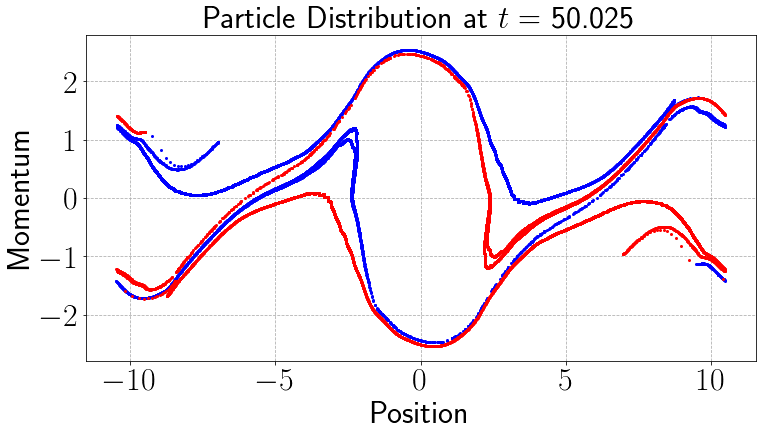}} \\
    \subfloat[][Leapfrog]{
    \includegraphics[width=0.38\textwidth]{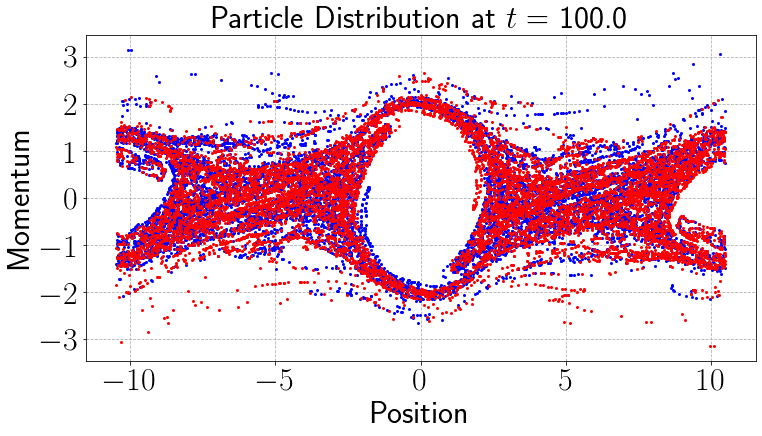}}
    \subfloat[][AEM]{
    \includegraphics[width=0.38\textwidth]{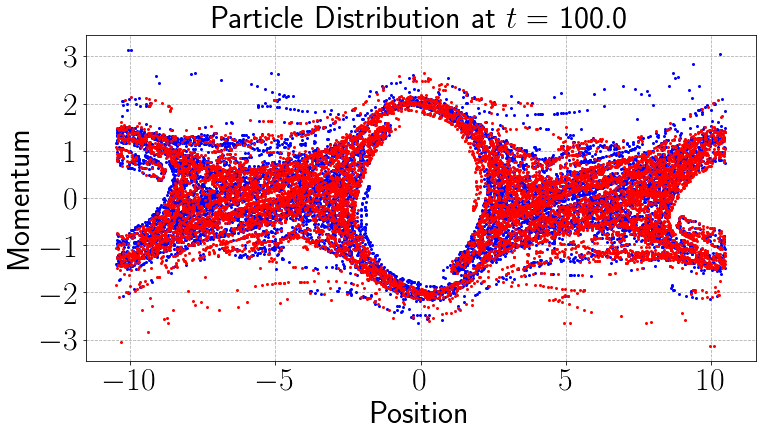}}
    \caption{We plot the electrons in phase space obtained with the Poisson model for the two-stream instability problem at different times given in units of $\omega_{pe}^{-1}$. Results obtained using leapfrog time integration are shown in the left column, while the right column applies the AEM. The IAEM, which applies the Taylor correction, is not considered here because the contributions from the magnetic fields are ignored. The FFT is used to compute the scalar potential (as well as its gradient). Identical results are observed with both approaches.}
    \label{fig:Poisson TSI LF and AEM + FFT}
\end{figure}

\begin{figure}[!htb]
    \centering
    \subfloat[][Leapfrog]{
    \includegraphics[width=0.35\textwidth]{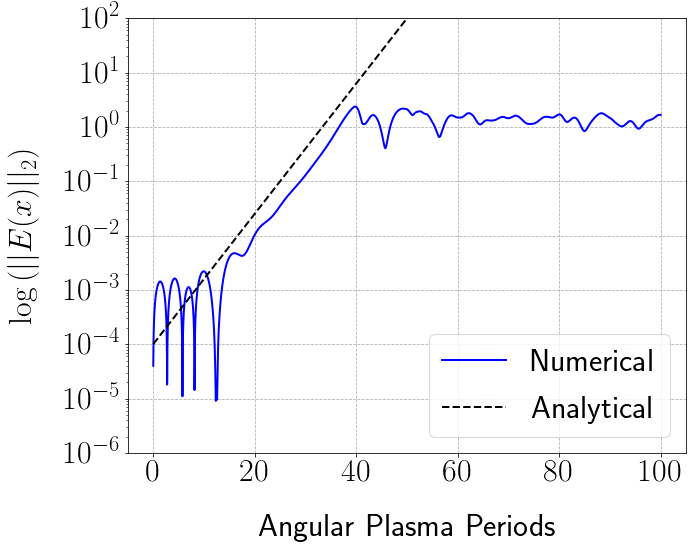}}
    \subfloat[][AEM]{
    \includegraphics[width=0.35\textwidth]{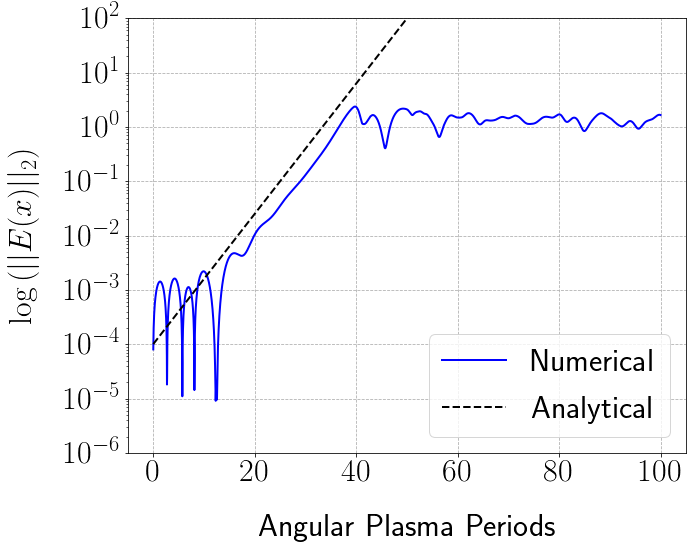}} \\
    \caption{We compare the growth rate in the $\ell_{2}$-norm of the electric field for the Poisson model using both methods against an analytical growth rate obtained from linear response theory. Using this experimental configuration, the analytical growth rate is determined to be $\text{Im}(\omega) \approx 0.2760$. We can see that the AEM reproduces the growth rate of the leapfrog method.}
    \label{fig:E-field decay Poisson TSI LF and AEM + FFT}
\end{figure}

We first assessed the behavior of the particle integrator by considering the Poisson model \eqref{eq:TSI Poisson model potential} for the scalar potential. The Taylor correction version of the AEM was not considered in this problem because the contributions from the magnetic field are ignored. Since the combination of leapfrog time integration with an FFT field solver is such a commonly used approach to this problem, it allowed us to identify key differences attributed solely to the choice of time integration method used for particles. We note that the particular initial condition for this problem leads to a special case in which the AEM is equivalent to leapfrog integration. Since the problem starts out as charge neutral, there is no electric field at time $t=0$. This means that there is no modification to the particle velocities in the step that generates the staggering required by the leapfrog method. Of course, this is no longer true for problems which have an initial charge imbalance because the electric field would be non-zero. Plots that compare the evolution of the electron beams, obtained with both methods and an FFT field solver, are presented in Figure \ref{fig:Poisson TSI LF and AEM + FFT}. As expected, we see that the AEM produces structures that are identical to those which are generated with the leapfrog scheme. Using basic linear response theory (see e.g., \cite{liboff}) one obtains the dispersion relation for the cold problem as
\begin{equation*}
    \omega^4 - 2\omega^2 \left( \omega_{pe} + k^2 v_{b}^{2} \right) + k^{2} v_{b}^{2} \left( k^2 v_{b}^{2} - 2\omega_{pe} \right) = 0.
\end{equation*}
While the dispersion relation for the warm problem could also be considered \cite{ichimaru-v1,BittencourtBook}, its evaluation is slightly more complicated than the cold problem. We remark that cold problems cannot be not be adequately represented in mesh-based discretizations, which is a key advantage offered by a particle-based approach. Additionally, cold problems eliminate artifacts introduced by sampling methods during the initialization phase. Taking $v_{b} = 1$, $\omega_{pe} = 1$, and $k = 2\pi/L_{x}$ in the dispersion relation yields the growth rate $\text{Im}(\omega) \approx 0.2760$. In Figure \ref{fig:E-field decay Poisson TSI LF and AEM + FFT}, we compare the growth rate of the electric field in from both methods with the analytical growth rate. Again, we see identical results among both methods, which reproduce the correct growth rate. 

\begin{figure}[!htb]
    \centering
    \subfloat[][Leapfrog]{
    \includegraphics[width=0.38\textwidth]{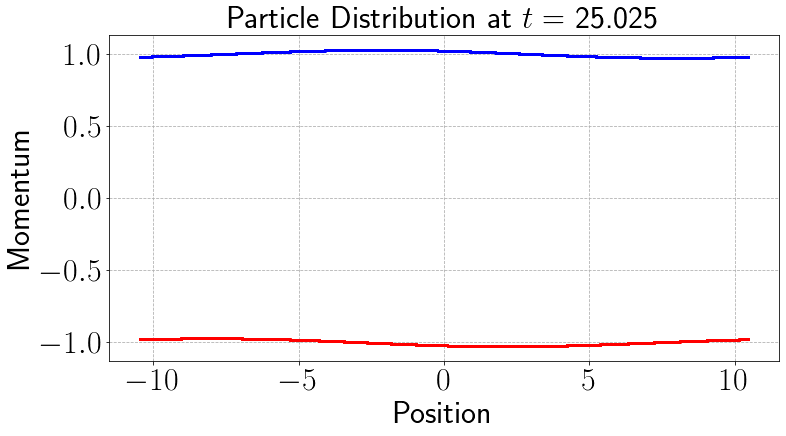}}
    \subfloat[][AEM]{
    \includegraphics[width=0.38\textwidth]{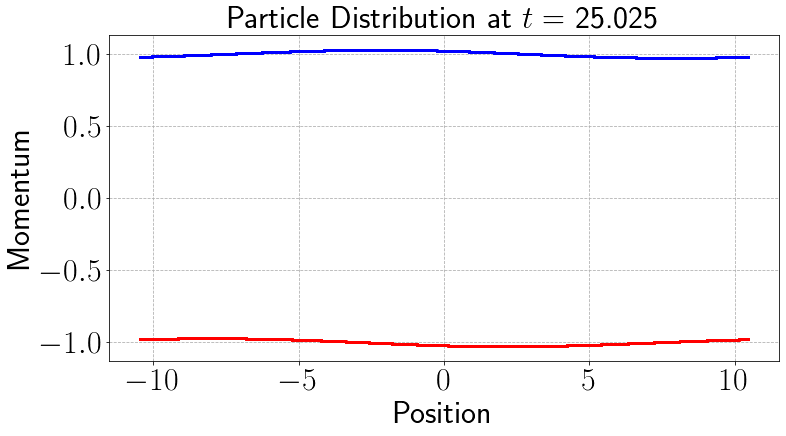}} \\
    \subfloat[][Leapfrog]{
    \includegraphics[width=0.38\textwidth]{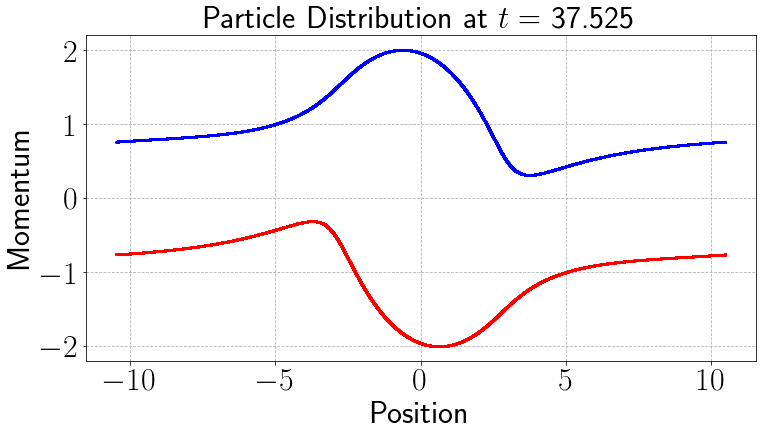}}
    \subfloat[][AEM]{
    \includegraphics[width=0.38\textwidth]{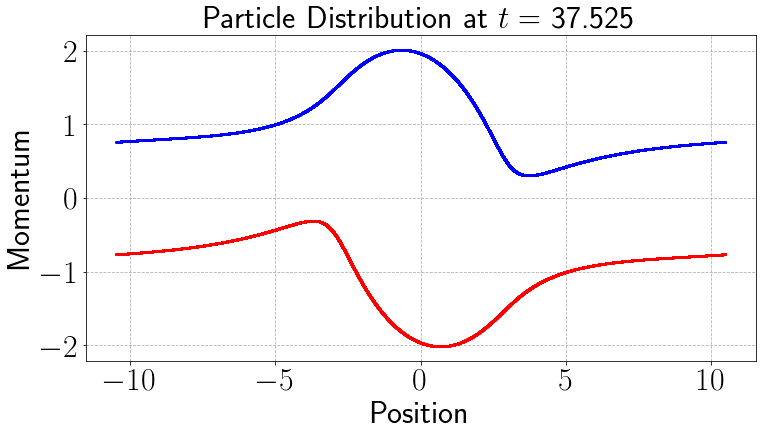}} \\
    \subfloat[][Leapfrog]{
    \includegraphics[width=0.38\textwidth]{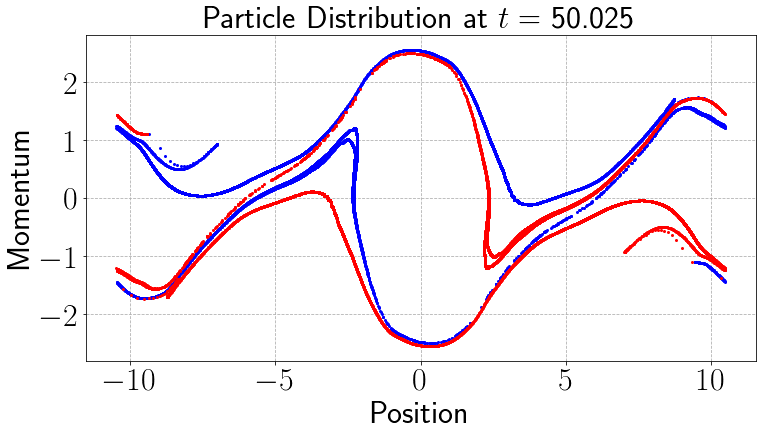}}
    \subfloat[][AEM]{
    \includegraphics[width=0.38\textwidth]{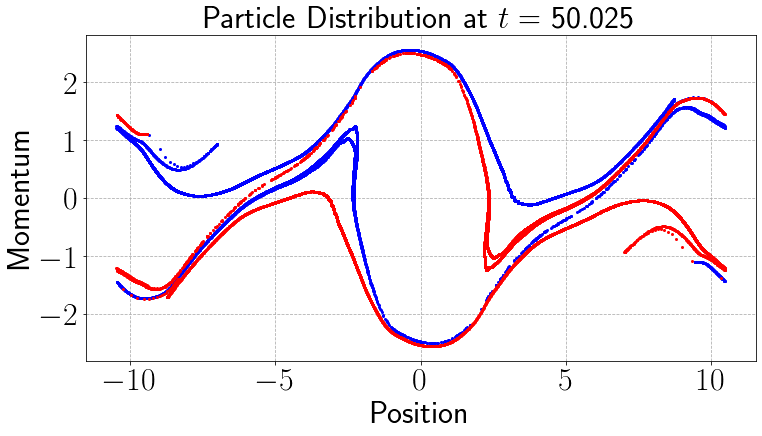}} \\
    \subfloat[][Leapfrog]{
    \includegraphics[width=0.38\textwidth]{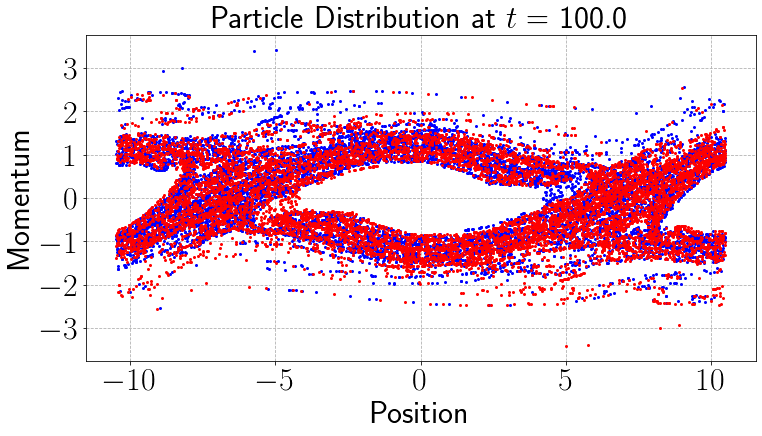}}
    \subfloat[][AEM]{
    \includegraphics[width=0.38\textwidth]{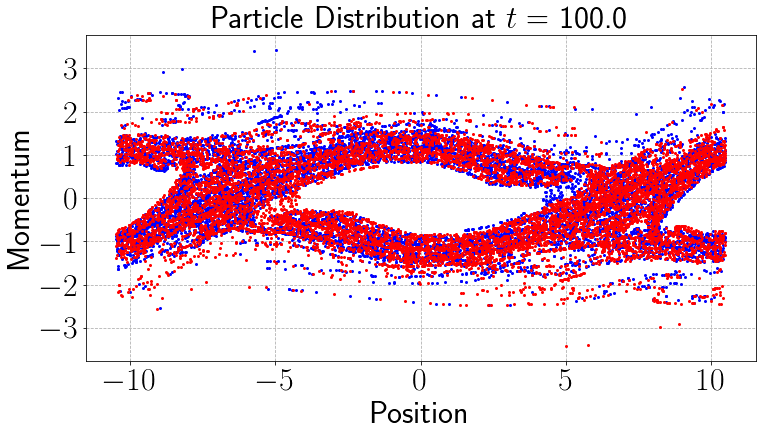}}
    \caption{We plot the electrons in phase space obtained using the wave model for the two-stream example at different times given in units of $\omega_{pe}^{-1}$. Results obtained using leapfrog time integration are shown in the left column, while the right column applies the AEM. The first-order (diffusive) BDF scheme (BDF-1) is used to compute the scalar potentials and their derivatives in both approaches. Unlike the results obtained with the Poisson model, in which the FFT was used as the field solver (see Figure \ref{fig:Poisson TSI LF and AEM + FFT}), we observe differences in the structure of the trapping regions. Such regions in the wave model appear to be more compressed than those in the elliptic model.}
    \label{fig:Wave TSI LF and AEM + BDF-1}
\end{figure}

\begin{figure}[!htb]
    \centering
    \subfloat[][Leapfrog]{
    \includegraphics[width=0.35\textwidth]{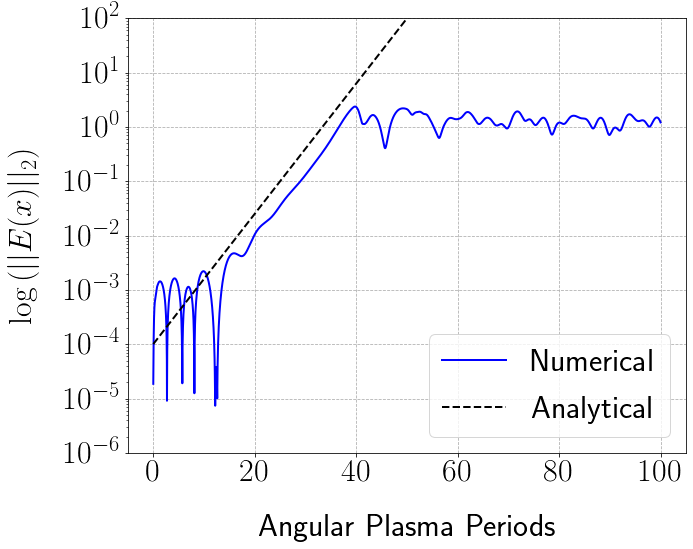}}
    \subfloat[][AEM]{
    \includegraphics[width=0.35\textwidth]{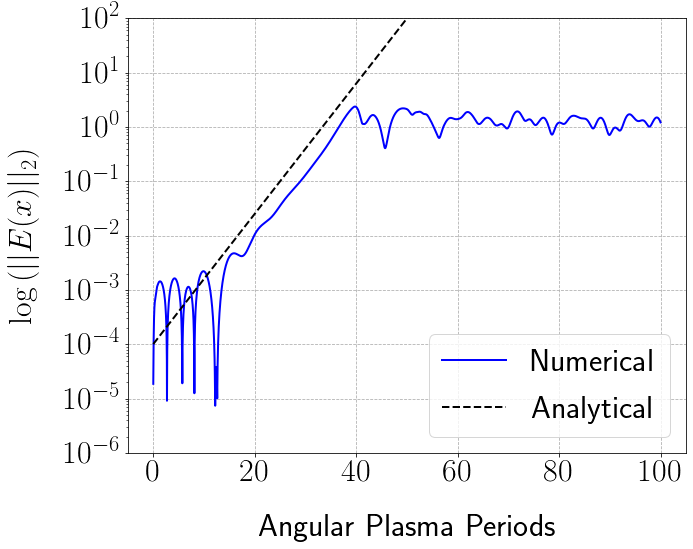}} \\
    \caption{The growth rate in the $\ell_{2}$-norm of the electric field obtained with both methods, which uses the BDF-1 field solver to evolve the wave equation for the scalar potential and compute its derivatives. This data is compared against an analytical growth rate obtained from linear response theory for the Poisson model. Using this experimental configuration, the analytical growth rate is determined to be $\text{Im}(\omega) \approx 0.2760$. We, again, see that the AEM reproduces the growth rate of the leapfrog method.}
    \label{fig:E-field decay wave TSI LF and AEM + BDF-1}
\end{figure}

The same experiment was repeated using a two-way wave model \eqref{eq:TSI wave model potential} in the place of the Poisson model \eqref{eq:TSI Poisson model potential} for the scalar potential. For strongly electrostatic problems ($\kappa \gg 1$), the wave model should produce results which are similar to those of the Poisson model shown in Figure \ref{fig:Poisson TSI LF and AEM + FFT}. This feature allows us to benchmark the performance of the wave solver and proposed methods for derivatives by comparing against the elliptic model. The results for the two methods are displayed in Figure 
\ref{fig:Wave TSI LF and AEM + BDF-1}. We see that the early behavior is quite similar to the results in Figure \ref{fig:Poisson TSI LF and AEM + FFT} obtained with the elliptic model. At later times, however, the trapping regions are more ``compressed" than those generated with the Poisson model. This is a likely consequence of the finite speed of propagation in the wave model, where the potential responds more slowly to an imbalance in charge. We can also check the growth rate in the electric field, as we did with the Poisson model. In this configuration, Poisson's equation is a good approximation to the wave model for the potential, so we can use the growth rate presented above to assess the validity of the method. The growth rates for the wave model are displayed in Figure \ref{fig:E-field decay wave TSI LF and AEM + BDF-1}, which show good agreement with theory.

\subsubsection{Numerical Heating}
\label{subsubsec:numerical heating}

We now discuss the numerical heating study, which is used to characterize the effect of resolving the Debye length $\lambda_{D}$. These numerical properties turn out to be connected to the symplecticity of the method. Explicit PIC methods are not symplectic because the fields are not self-consistent with the particles that represent the plasma. Consequently, the grid should be sufficiently fine so that a given particle can ``see" the correct potential that is otherwise screened by particles of opposite charge. In other words, with explicit PIC methods, one needs to resolve the charge separation in the plasma, whose characteristic length scale is set according to the Debye length to prevent aliasing errors. A general rule of thumb for explicit PIC simulations is that the grid spacing $\Delta x$ should satisfy $4 \Delta x < \lambda_{D}$ to prevent substantial numerical heating. Otherwise, the temperature of the plasma increases until a ``new" Debye length is obtained that is adequately resolved on the given mesh. In practice, however, heating generally behaves in an uncontrollable manner, growing without bound, leading to highly unphysical behavior.

% Implicit PIC methods, which break this restriction (see e.g., \cite{Chacon2011es-implicit-pic}), allow for a substantially coarser mesh to be used for a given calculation and will be investigated in our future work.

The setup for this problem is slightly different from the two-stream example discussed earlier. Here, we provide, as input, a Debye length $\lambda_{D}$ and a thermal velocity $v_{th}$, which can be used to calculate the average number density $\bar{n}$ and macroscopic temperature $\bar{T}$ for the plasma. The remaining parameters can be derived from these values and are shown in Table \ref{tab:heating plasma parameters}. The normalized speed of light for both the electrostatic and electromagnetic problems is $\kappa = 50$, and the normalized permittivity is $\sigma_{1} = 1$. For the electromagnetic problem, the normalized permeability obtained with these experimental parameters is $\sigma_{2} = 4.0 \times 10^{-4}$. Here we consider both electrostatic (1D-1V/1D-1P) and electromagnetic (2D-2V/2D-2P) configurations that consist of ions and electrons in a periodic domain. The spatial domain for the electrostatic case is $[-25 \lambda_{D}, 25 \lambda_{D}]$, while the electromagnetic case uses $[-25 \lambda_{D},25 \lambda_{D}]^{2}$. In both cases, the spatial domain is refined by successively doubling the number of mesh points from 16 to 256 in each dimension. The simulations use $1 \times 10^{6}$ time steps with a final time of $T_f = 1 \times 10^{3}$ angular plasma periods. In the electrostatic simulation, we use $5 \times 10^{3}$ macroparticles for each species, and increase this to $2.50632 \times 10^5$ for the electromagnetic simulation. As before, we assume that the ions remain stationary since they are heavier than the electrons. Electrons are given uniform positions in space and their velocities are obtained by sampling from a Maxwellian distribution using the parameters in Table \ref{tab:heating plasma parameters}. We make the problem current neutral by splitting the electrons into two equally sized groups whose velocities differ only in sign. A drift velocity is not used in these tests. To ensure consistency across the runs, we also seed the random number generator prior to sampling.

\begin{table}[!h]
    \centering
    \def\arraystretch{1.2}
    \begin{tabular}{ | c || c | }
        \hline
        \textbf{Parameter}  & \textbf{Value} \\
        \hline
        Average number density ($\bar{n}$) [m$^{-3}$] & $1.129708\times 10^{14}$ \\
        %\hline
        Average temperature ($\bar{T}$) [K] & $2.371698\times 10^{6}$ \\
        %\hline
        Debye length ($\lambda_{D})$ [m] & $1.0\times 10^{-2}$ \\
        %\hline
        Inverse angular plasma frequency ($\omega_{pe}^{-1}$) [s/rad] & $1.667820\times 10^{-9}$ \\
        %\hline
        Thermal velocity ($v_{th} = \lambda_D  \omega_{pe}$) [m/s] & $5.995849 \times 10^{6}$ \\
        \hline
    \end{tabular}
    \caption{Table of the plasma parameters used in the numerical heating examples.}
    \label{tab:heating plasma parameters}
\end{table}

We monitor heating during the simulations by computing the variance in the components of the electron velocities, since this is connected to the temperature of a Maxwellian distribution. In the one-dimensional case, the variance data at a given time step is converted to a temperature (in units of Kelvin) using the relation
\begin{equation*}
    \bar{T} = \frac{m_e V^{2}}{k_B} \text{Var}(v^{(1)}),
\end{equation*}
where we have used ``Var" to denote variance and $V$ is the normalization used for velocity. Similarly, for the two-dimensional case, we compute the average of the variance for each component of the velocity, which is similarly converted to a temperature (in units of Kelvin) using
\begin{equation*}
    \bar{T} = \frac{m_e V^{2} \left( \text{Var}(v^{(1)}) + \text{Var}(v^{(2)}) \right) }{2k_B}.
\end{equation*}
We use the superscripts in the above metrics to refer to the individual velocity components across all of the particles. The factor of two is used to average the variance among these components. When assessing the temperatures produced by different methods, we rescale the temperatures so they have the proper units of Kelvin. This allows us compare the different methods in a more realistic setting in which we might be interested in comparing the raw temperatures predicted by different methods.

The models used in the electrostatic tests are identical to the ones presented for the two-stream instability example, so we shall skip these details for brevity. In the case of the electromagnetic experiment, the particle equations in the non-relativistic Hamiltonian formulation are
\begin{empheq}[left=\empheqlbrace]{align*}
    \frac{d x_{i}^{(1)}}{d t} &= \frac{1}{m_i} \left( P_{i}^{(1)} - q_i A^{(1)} \right), \\
    \frac{d x_{i}^{(2)}}{d t} &= \frac{1}{m_i} \left( P_{i}^{(2)} - q_i A^{(2)} \right), \\
    \frac{d P_{i}^{(1)}}{d t} &= - q_i \partial_{x} \phi + \frac{q_i }{m_i} \bigg[ \left( \partial_{x} A^{(1)} \right) \left( P_{i}^{(1)} - q_i A^{(1)} \right) + \left( \partial_{x} A^{(2)} \right) \left( P_{i}^{(2)} - q_i A^{(2)} \right) \bigg], \\
    \frac{d P_{i}^{(2)}}{d t} &= - q_i \partial_{y} \phi + \frac{q_i }{m_i} \bigg[ \left( \partial_{y} A^{(1)} \right) \left( P_{i}^{(1)} - q_i A^{(1)} \right) + \left( \partial_{y} A^{(2)} \right) \left( P_{i}^{(2)} - q_i A^{(2)} \right) \bigg].
\end{empheq}
The contributions from the fields are obtained by solving a system of wave equations for the potentials, which take the form
\begin{empheq}[left=\empheqlbrace]{align*} 
    &\frac{1}{c^2} \partial_{tt} \phi - \partial_{xx} \phi - \partial_{yy} \phi = \frac{1}{\epsilon_0}\rho, \\
    &\frac{1}{c^2} \partial_{tt} A^{(1)} - \partial_{xx} A^{(1)} - \partial_{yy} A^{(1)} = \mu_0 J^{(1)}, \\
    &\frac{1}{c^2} \partial_{tt} A^{(2)} - \partial_{xx} A^{(2)} - \partial_{yy} A^{(2)} = \mu_0 J^{(2)}.
\end{empheq}

\begin{figure}[!htb]
    \centering
    % trim={<left> <lower> <right> <upper>}
    \includegraphics[width=0.35\textwidth, trim={22cm 0 8cm 0},clip]{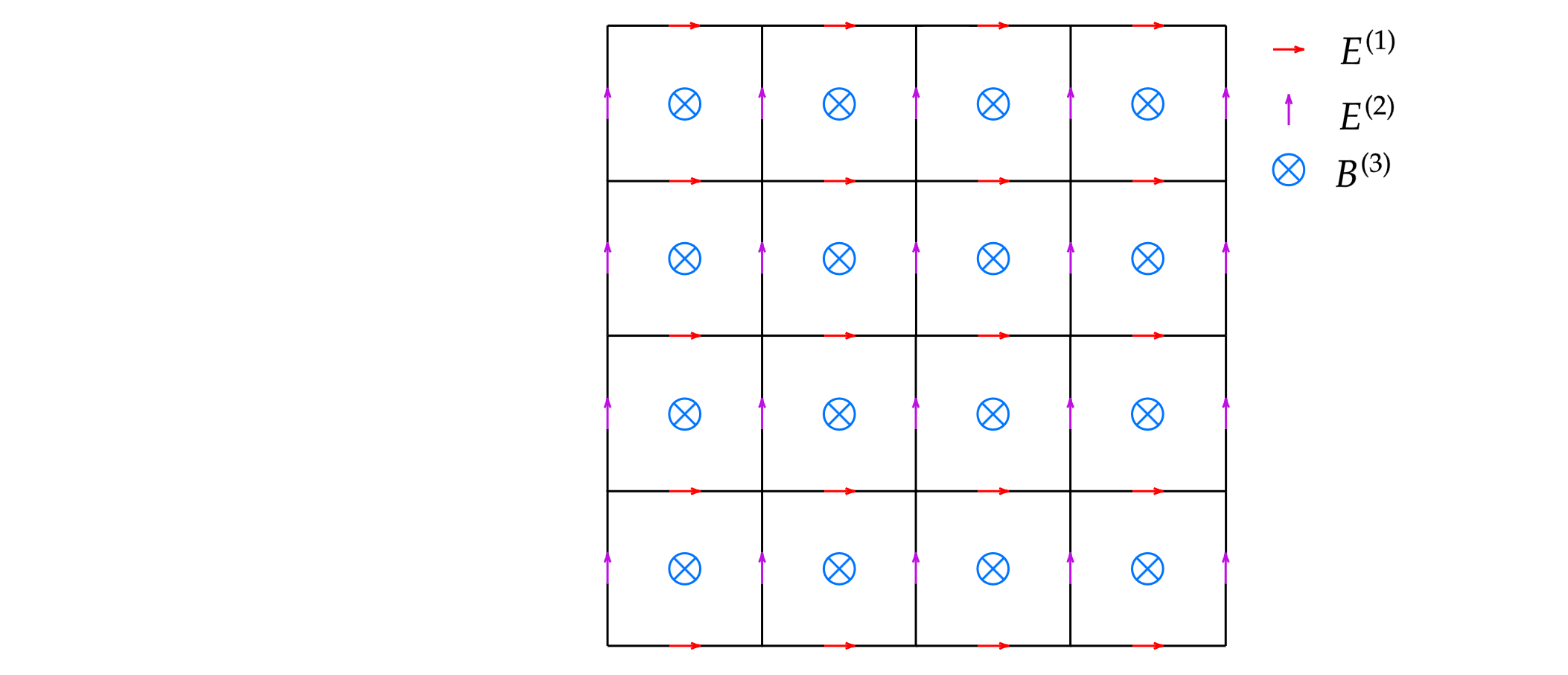}
    \caption{The staggered FDTD grid in TE mode.}
    \label{fig:Yee-Grid-Bz}
\end{figure}

To establish the heating properties of the proposed methods in an electromagnetic setting, an identical experiment is performed using a standard FDTD-PIC approach in which the equations of motion for the particles are expressed in terms of $\mathbf{E}$ and $\mathbf{B}$. For this example, these equations take the form
\begin{empheq}[left=\empheqlbrace]{align*}
    \frac{d x_{i}^{(1)}}{d t} &= v_{i}^{(1)}, \quad \frac{d v_{i}^{(1)}}{d t} = \frac{q_i}{m_i} \Bigg(E^{(1)} +  v_{i}^{(2)} B^{(3)} \Bigg), \\
    \frac{d  x_{i}^{(2)}}{d t} &= v_{i}^{(2)}, \quad \frac{d v_{i}^{(2)}}{d t} = \frac{q_i}{m_i} \Bigg(E^{(2)} -  v_{i}^{(1)} B^{(3)} \Bigg),
\end{empheq}
and are evolved in a leapfrog format through the Boris method \cite{Boris1970}. Since we have restricted the system to two spatial dimensions, the curl equations decouple into the so-called transverse electric (TE) and transverse magnetic (TM) modes. We retain the curl equations
\begin{empheq}[left=\empheqlbrace]{align*}
    & \partial_{x} E^{(2)} - \partial_{y} E^{(1)} = -\partial_{t} B^{(3)}, \\
    & -\partial_{z} B^{(2)} = \mu_0 J^{(1)} + \frac{1}{c^2} \partial_{t} E^{(1)}, \\
    & -\partial_{x} B^{(3)} = \mu_0 J^{(2)} + \frac{1}{c^2} \partial_{t} E^{(2)},
\end{empheq}
which are discretized using the staggered FDTD mesh \cite{Yee1966} based on the TE mode (see Figure \ref{fig:Yee-Grid-Bz}).

\begin{figure}[!htb]
    \centering
    \subfloat[][Leapfrog with FFT]{
    % trim: L, B, R, T
    \includegraphics[clip, trim={0cm, 0cm, 11.5cm, 0cm}, width=0.357\textwidth]{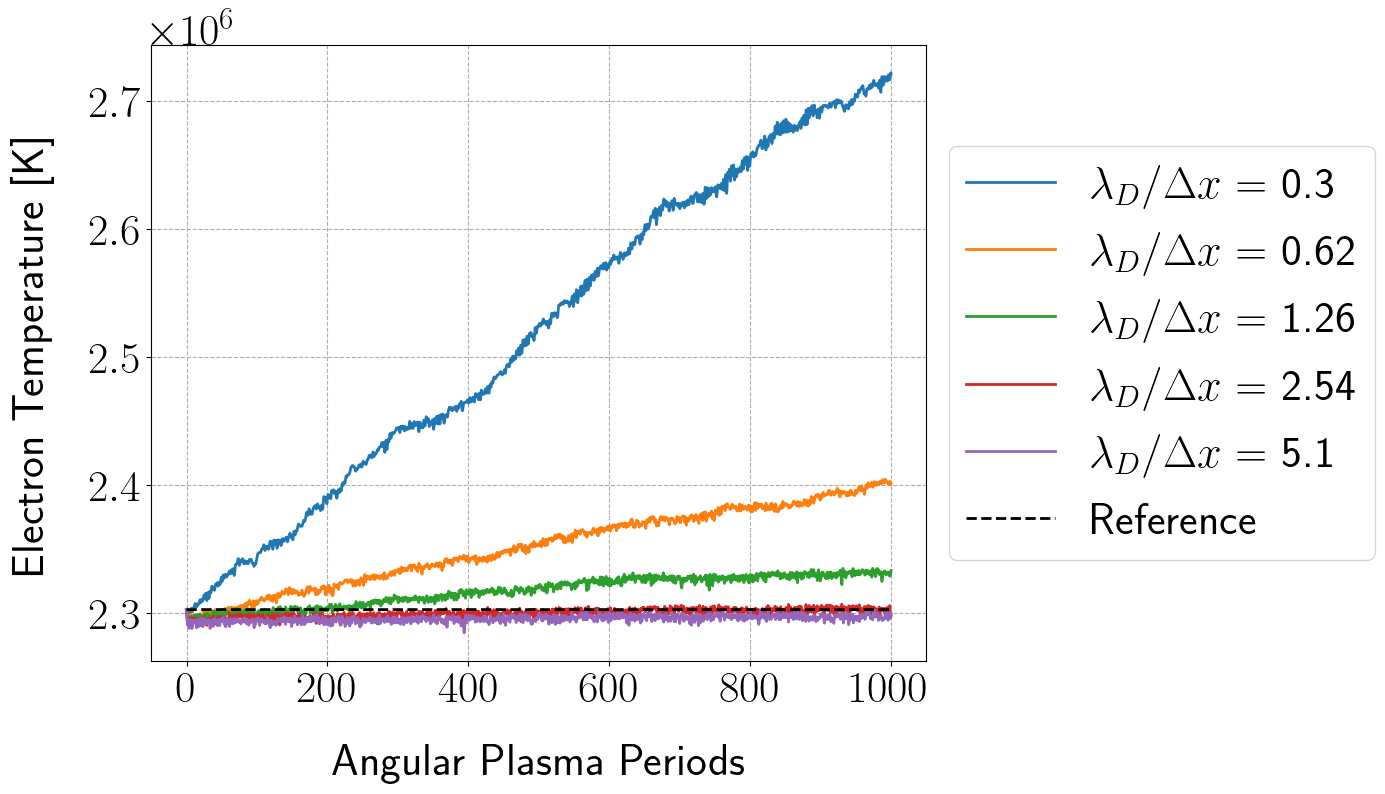}}
    \subfloat[][AEM with FFT]{
    % trim: L, B, R, T
    \includegraphics[clip, trim={1.8cm, 0cm, 0cm, 0cm}, width=0.5\textwidth]{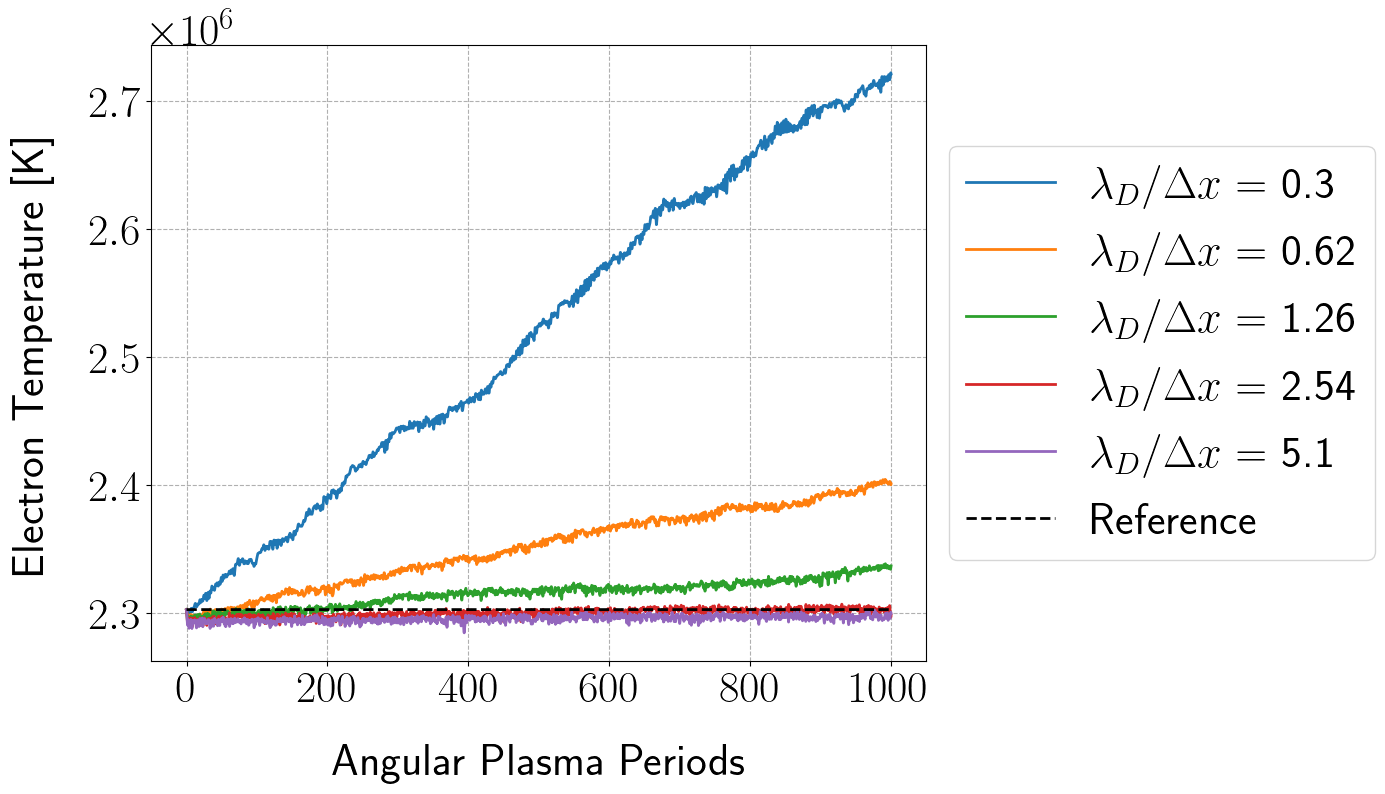}} 
    \caption{Results from the electrostatic heating tests with a Poisson model for the potential. We plot the average electron temperature as a function of the number of angular plasma periods. On the left, we show the results obtained with the standard leapfrog time integration scheme, while the plot on the right uses the AEM \cite{Gibbon2017Hamiltonian}. Since the magnetic field is ignored in the model, the AEM and its improved version are identical. Additionally, the since the problem is charge neutral, the results for the two methods will be identical. Fields and their derivatives are constructed using a FFT field solver. The results suggest that heating can be prevented if we use $\approx 2.54$ grid cells per plasma Debye length. This is quite close to the usual rule of thumb which recommends $\approx 4$ grid cells per Debye length.}
    \label{fig:1D-1V ES heating test Poisson}
\end{figure}

\begin{figure}[!htb]
    \centering
    \subfloat[][Leapfrog with BDF-1]{
    % trim: L, B, R, T
    \includegraphics[clip, trim={0cm, 0cm, 11.5cm, 0cm}, width=0.357\textwidth]{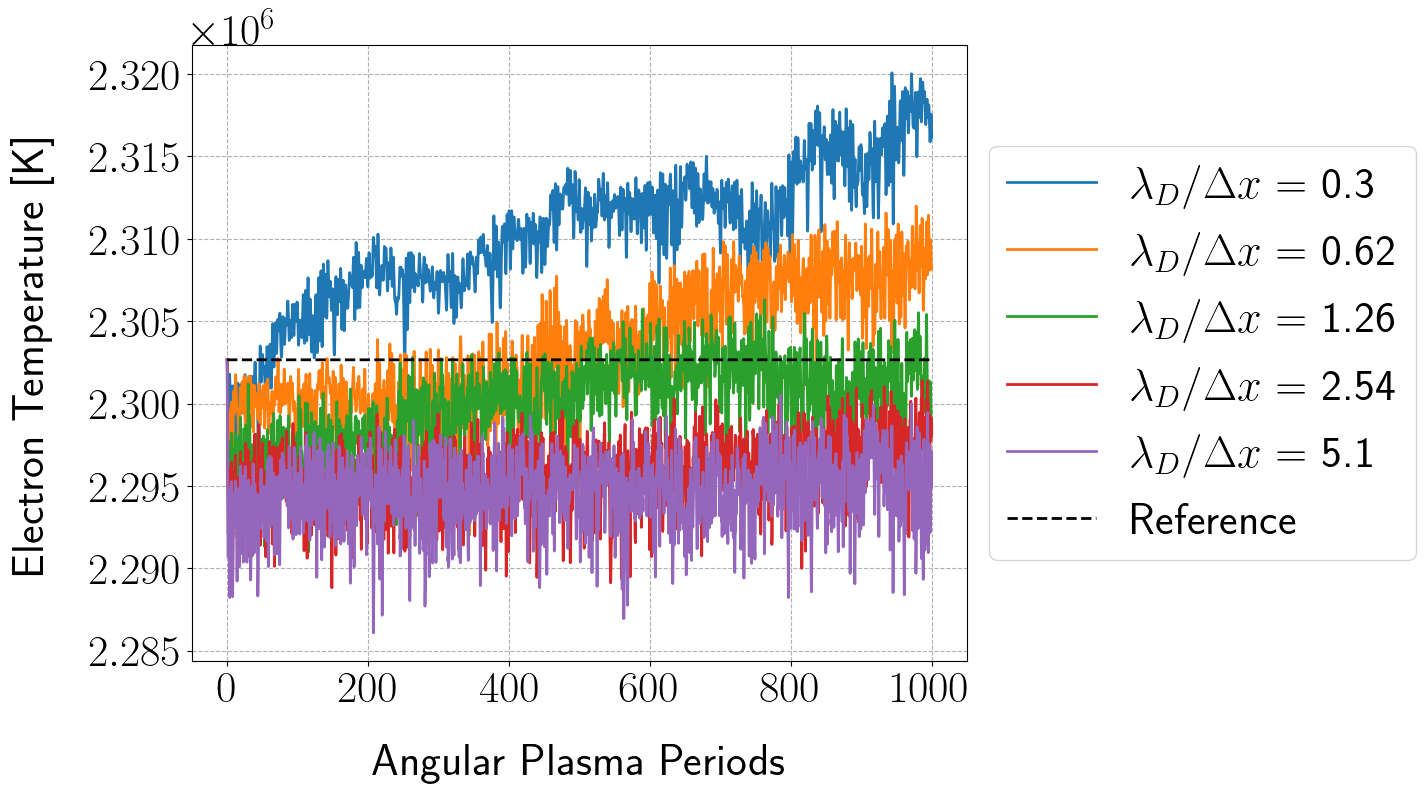}}
    \subfloat[][AEM with BDF-1]{
    % trim: L, B, R, T
    \includegraphics[clip, trim={1.8cm, 0cm, 0cm, 0cm}, width=0.5\textwidth]{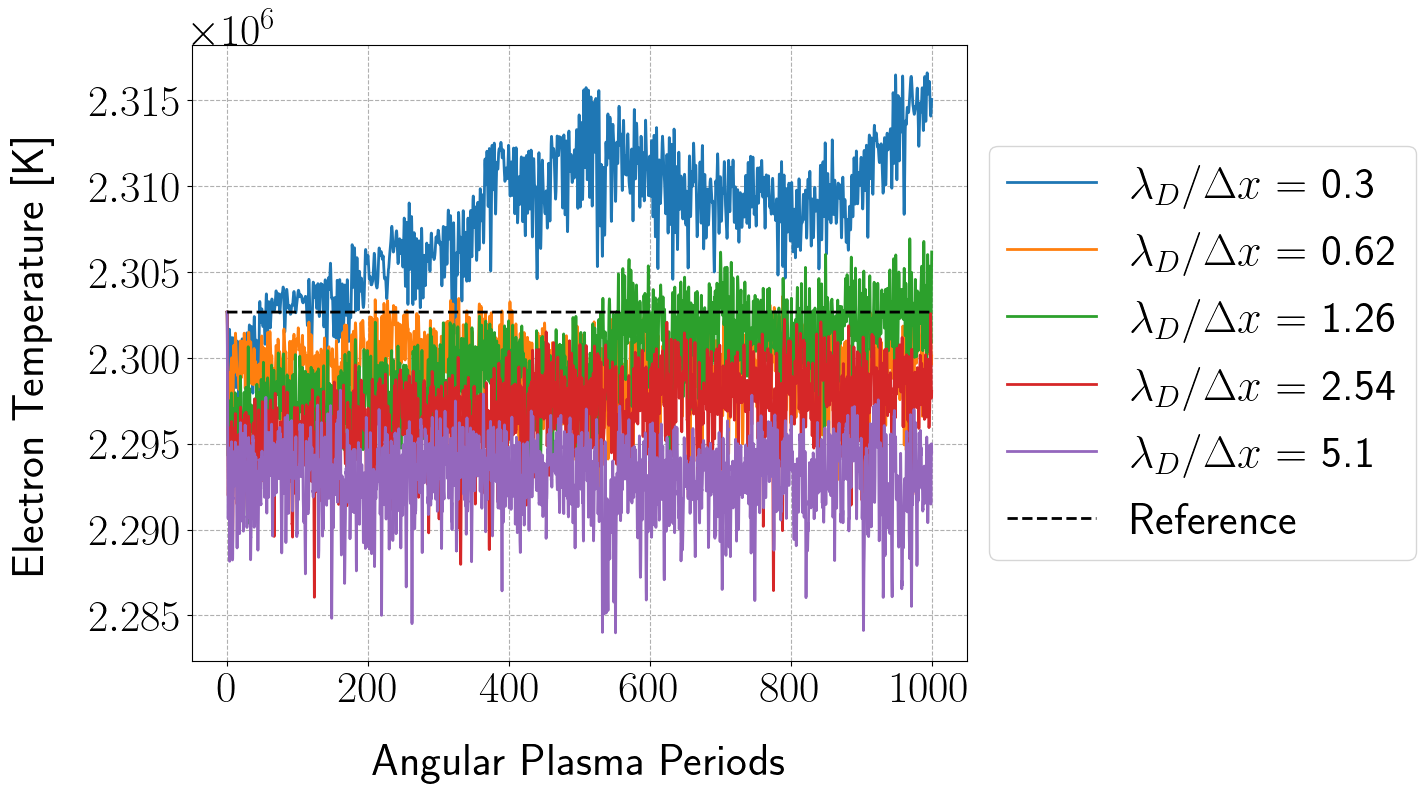}} 
    \caption{Results from the electrostatic heating tests with a wave model for the potential. We plot the average electron temperature as a function of the number of angular plasma periods. On the left, we show the results obtained with the standard leapfrog time integration scheme, while the plot on the right uses the base AEM \cite{Gibbon2017Hamiltonian}. Since the magnetic field is ignored in the model, the AEM and its improved version are identical. Additionally, the since the problem is charge neutral, the results for the two methods will be identical. Fields and their derivatives are constructed using the proposed BDF-1 field solver. In contrast to the results obtained with the Poisson model (see Figure \ref{fig:1D-1V ES heating test Poisson}) we observe less severe fluctuations in temperature due to the finite speed of propagation in the wave model. Furthermore, the temperature fluctuations are not severe even in grid configurations which do not adequately resolve the Debye length.}
    \label{fig:1D-1V ES heating test wave}
\end{figure}

\begin{figure}[!htb]
    \centering
    \subfloat[][FDTD-PIC]{
    % trim: L, B, R, T
    \includegraphics[clip, trim={0cm, 0cm, 11.5cm, 0cm}, scale=0.28
]{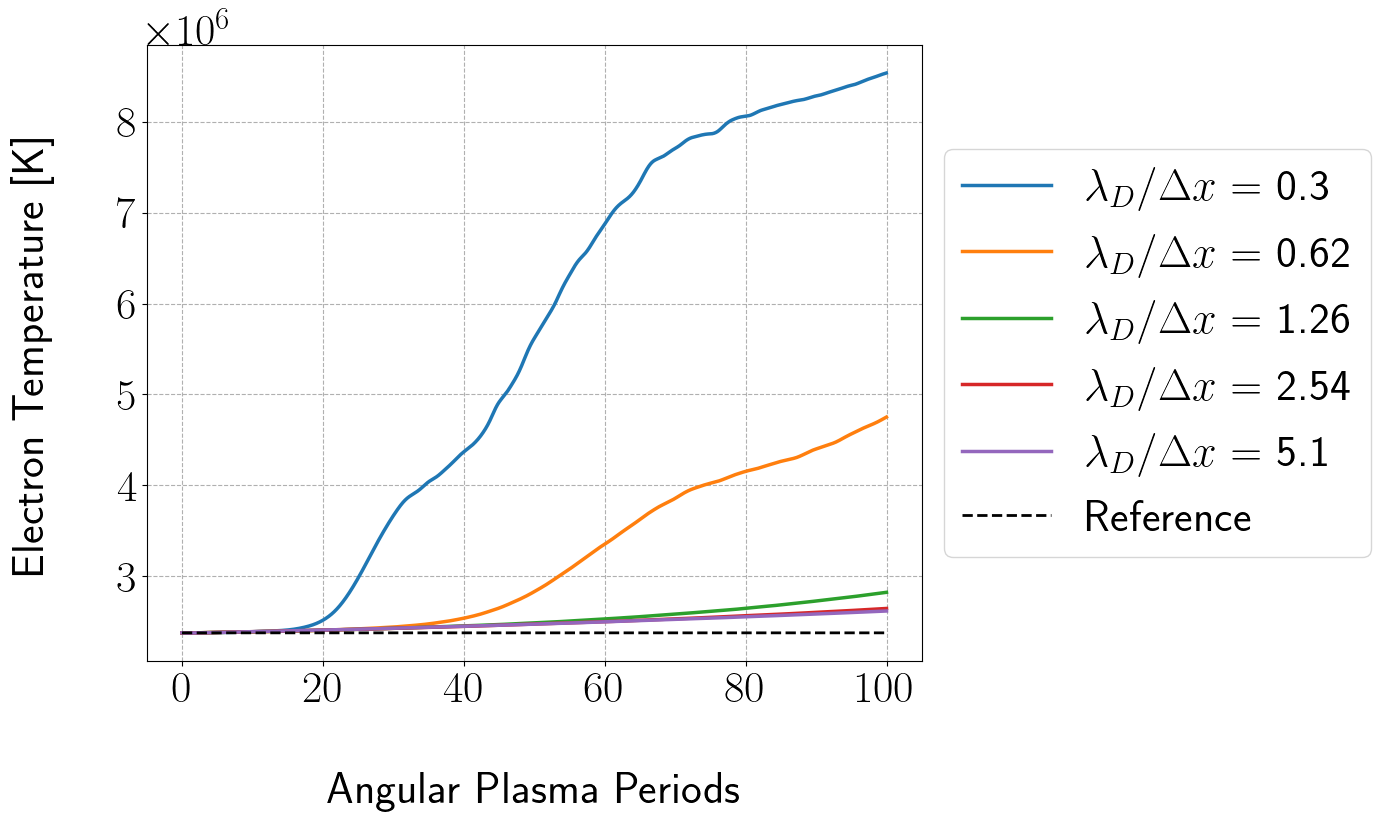}}
    \subfloat[][IAEM with BDF-1]{
    % trim: L, B, R, T
    \includegraphics[clip, trim={1.8cm, 0cm, 0cm, 0cm}, scale=0.2725]{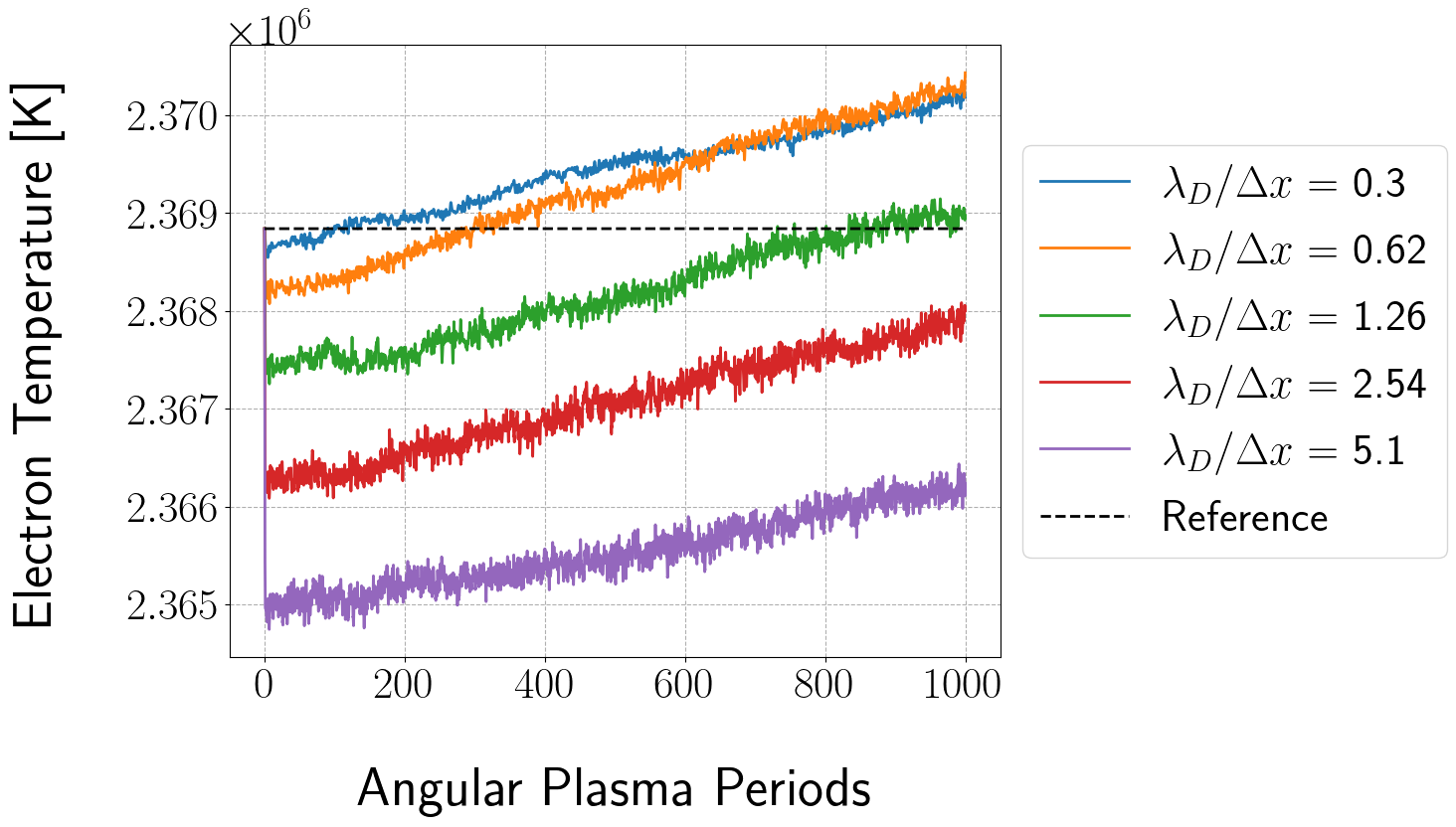}} 
    \caption{We present results from the electromagnetic heating experiments that plot the average electron temperature as a function of the number of angular plasma periods. The plot on the left was obtained with the FDTD-PIC method, while the plot on the right uses the IAEM and the proposed BDF-1 field solver. Since the IAEM is not symplectic, it will, over time, generate additional sources of energy causing the simulation to heat even if the plasma Debye length is sufficiently resolved; however, the results indicate that fluctuations in the temperature are not substantial, even in cases where the Debye length would normally be considered underresolved by the mesh. In contrast, the FDTD-PIC method displays more significant fluctuations over a smaller time window, even when the Debye length is resolved by the mesh. Note the differences in the magnitude of the electron temperature between the plots.}
    \label{fig:2D-2V EM heating test}
\end{figure}

\begin{figure}[!htb]
    \centering
    \subfloat[][Mass conservation]{
    % trim: L, B, R, T
    \includegraphics[clip, trim={0cm, 0cm, 11.5cm, 0cm}, scale=0.2725
]{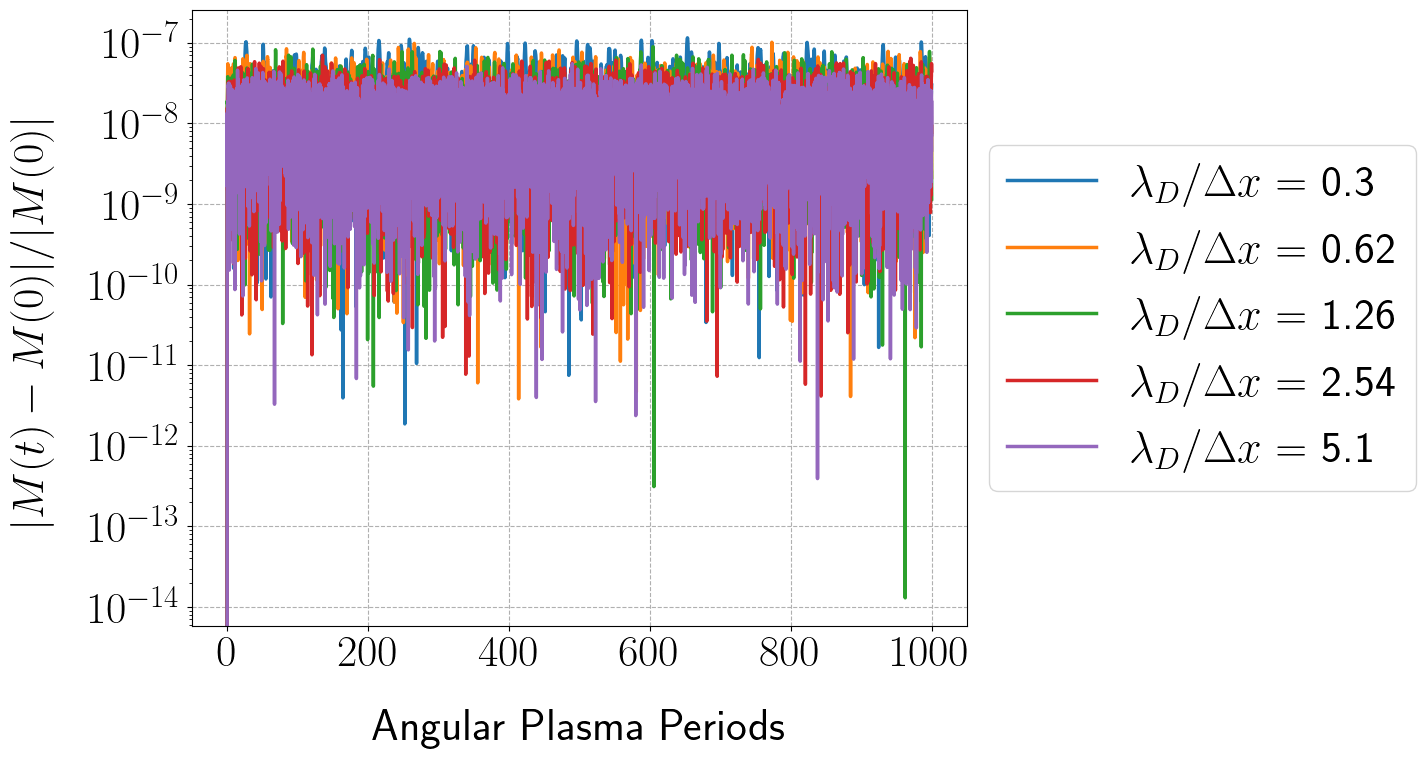}}
    \subfloat[][Lorenz gauge error ($\ell_{2}$)]{
    % trim: L, B, R, T
    \includegraphics[clip, trim={0cm, 0cm, 0cm, 0cm}, scale=0.2725]{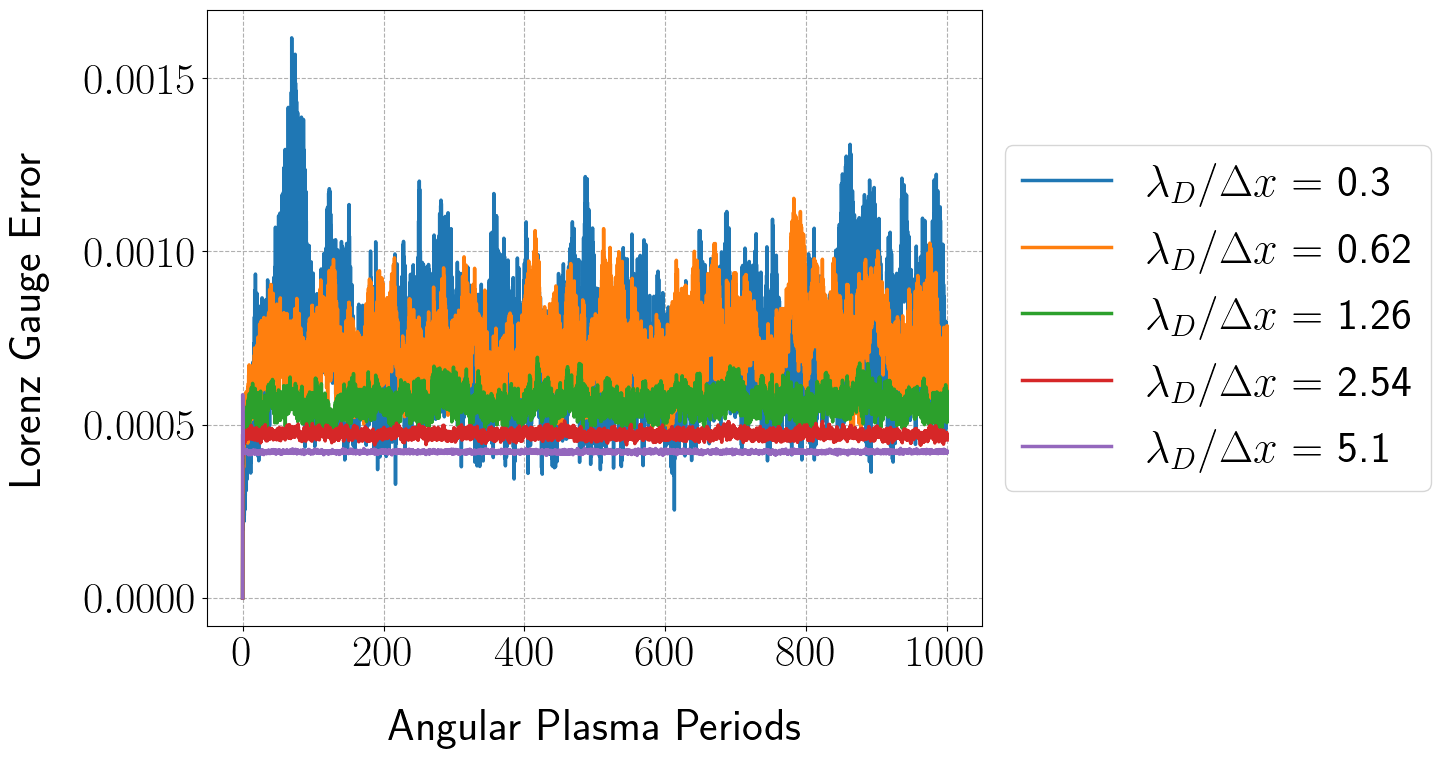}}
    \caption{We plot the change in the total mass (left) and the error in the Lorenz gauge (right) for the BDF-1 field solver with the IAEM in the 2D-2V heating experiment. The change in the total mass is measured relative to its value at the initial time. Since the gauge condition is satisfied by the initial data for the problem, its value is initially zero, so we, instead, measure errors in the absolute sense using the $\ell_{2}$-norm. The relative error in the Hamiltonian $\mathcal{H}$ (total energy) is not presented for brevity, as this information can be inferred from the temperature plot in Figure \ref{fig:2D-2V EM heating test}. The proposed method demonstrates reasonable mass conservation, and the gauge condition appears to be controlled over many plasma periods despite the absence of a cleaning method.}
    \label{fig:2D-2V EM heating test conserved}
\end{figure}

The results of our heating experiments can be found in Figures \ref{fig:1D-1V ES heating test Poisson} - \ref{fig:2D-2V EM heating test conserved}. Figures \ref{fig:1D-1V ES heating test Poisson} and \ref{fig:1D-1V ES heating test wave} present the results for the electrostatic problem, in which, we consider both Poisson and wave equation models, respectively, for the scalar potential. Figure \ref{fig:2D-2V EM heating test} shows the results for the electromagnetic heating experiment that compares the FDTD-PIC method and the IAEM. For the electrostatic experiments, we observe significant differences in the heating behavior due to the choice of models used for the scalar potential. In particular, comparing Figures \ref{fig:1D-1V ES heating test Poisson} and \ref{fig:1D-1V ES heating test wave}, we can clearly see the rate of heating is far more significant when the Poisson model is used instead of the wave equation. This is partly due to the finite speed of propagation offered by the wave model, which causes the potential to respond more slowly to variations in the charge density $\rho$. We find this to be true even in cases where the Debye length would normally be considered underresolved by practitioners, resulting in far less severe fluctuations in temperature. Similar behaviors are also observed in the more complicated electromagnetic experiment whose data is presented in Figure \ref{fig:2D-2V EM heating test} in which the proposed method demonstrates mesh-independent heating properties, with notably smaller temperature fluctuations across the grid configurations, over many plasma periods. For example, over 1000 plasma periods, the relative increase in temperature across all meshes is $ < 0.1\%$. In contrast, the benchmark FDTD-PIC approach displays significant fluctuations in the temperature, even in cases where the grid spacing resolves the Debye length. These results indicate that the new method permits the use of a much coarser grid than current explicit particle methods. Moreover, in simulations on bounded domains, particles may not be present in the domain long enough to see any noticeable effects of heating, as they could be removed through an absorption mechanism or a particle collector. We also include some plots of conserved quantities for the electromagnetic case, namely, the total mass and the residual in the Lorenz gauge condition in Figure \ref{fig:2D-2V EM heating test conserved} obtained with the proposed method. We observe reasonable mass conservation and control of the gauge error over many plasma periods despite the fact that we are not enforcing the gauge condition.

% 1D-1V ES heating test (Poisson)

% 1D-1V ES heating test (Wave)

% 2D-2V EM heating test (temperatures)

% 2D-2V EM heating test (conserved quantities)

%%%%%%%%%%%%%%

\subsubsection{Plasma Sheath}
\label{subsubsec:sheath problem}

Plasma sheaths are a fundamental concept in the physics of plasmas and can be simulated using PIC methods. The study of sheaths was pioneered by Langmuir \cite{Langmuir_1932}, who described an ionized gas contained within a glass apparatus in a rather captivating manner:
\begin{quote}
[W]hen a current of a few milliamperes from a hot cathode is flowing in a glass tube containing mercury vapor saturated at room temperature, the voltage being above about 20 volts, the tube is largely filled with the characteristic green-blue glow of the mercury discharge, but the glow does not quite reach the walls. A dark space separates the glow from the walls, as if the glow were being repelled by the glass.
\end{quote}
The formation of sheaths is not an uncommon event, with two examples being the insertion of a conducting probe \cite{BittencourtBook} and a basic matrix sheath with uniform ion charge density, which occurs in a DC discharge. Such discharges can be created using a pulsed negative electrode voltage during plasma immersion ion implantation \cite{LichtenbergLiebermanPrinciples1994}.

In our computational model of a sheath, a macroscopically neutral plasma is deposited in a two-dimensional box with perfectly electrically conducting (PEC) walls, which have zero tangential components in their electric fields. As the problem is charge neutral, the electron drift velocity causes some of the electrons to move towards the wall.  When an electron comes into contact with a PEC wall, it is effectively neutralized by its associated image charge (of opposite sign), which results in a cancellation of the electric field on the surface. Hence, it is removed from the simulation. Consequently, as the lighter and hotter electrons are eliminated from the domain, the charge imbalance forms a potential well that draws the electrons back in towards the heavier and cooler (stationary) ions. When the electrons rush back to the center of the box, they repel each other, and the process begins anew, forming a ``breathing" pattern over time. The loss of the hotter electrons to the wall results in the formation of a potential well, which, in turn, forms a sheath near the domain boundaries. In the domain of the sheath, quasi-neutrality no longer holds on the scale of the initial Debye length. In other words, the Debye length varies substantially between the quasi-neutral interior and the sheath region \cite{Allen_2009}. The results of the numerical heating experiment have clear computational implications to the study of sheaths, as one needs to ensure that the mesh appropriately resolves the smallest Debye length set by the high density regions. Therefore, methods that are less susceptible to (artificial) numerical heating would provide a clear advantage over those that suffer from heating effects because they permit the use of a coarser mesh.  Of course, one needs to have adequate resolution of the sheath, but methods with high-order accuracy in space have the potential to resolve the sheath using fewer mesh points.

The procedure used to setup the simulation is nearly identical to the one used for the numerical heating experiment in section \ref{subsubsec:numerical heating}. Slight modifications to the plasma parameters are made to emphasize sheath formation (See Table \ref{tab:sheath plasma parameters}). The resulting normalized speed of light for this problem is $\kappa = 7.700159 \times 10^{2}$. Additionally, the normalized permittivity and permeability for this problem are $\sigma_{1} = 1$ and $\sigma_{2} = 1.686555 \times 10^{-6}$, respectively. A neutral plasma consisting of ions and electrons is deposited within a two-dimensional box whose dimensions are $\left[-8\lambda_{D}, 8 \lambda_{D}\right]^2$. The equations for the fields and particles in this experiment are identical to those used in the 2D-2V (2D-2P) electromagnetic heating experiment discussed in section \ref{subsubsec:numerical heating}. The proposed method was compared with the popular FDTD-PIC method, which (again) adopts the TE mode convention; however, this selection for the staggering with the FDTD mesh is motivated by the PEC boundary conditions for the fields, which, respectively, require $E^{(1)}$ to be zero along the $x$-axis and $E^{(2)}$ to be zero along the $y$-axis. Furthermore, since $B^{(3)}$ is cell-centered, its values along the boundary do not need to be specified. This leads to the staggered grid depicted in Figure \ref{fig:Yee-Grid-Bz}. In our implementation of the FDTD-PIC method, we extend the $E^{(1)}$ field a half-cell in the $x$ direction and the $E^{(2)}$ field a half-cell in the $y$ direction. This ensures that $E^{(1)}$ and $E^{(2)}$ are positioned on the boundary and that $B^{(3)}$ remains on the interior of the domain. We note that this is not a requirement for the numerical heating problem, which uses periodic boundaries and therefore has no such half-cell extensions.

\begin{table}[!htb]
    \centering
    \def\arraystretch{1.2}
    \begin{tabular}{ | c || c | }
        \hline
        \textbf{Parameter}  & \textbf{Value} \\
        \hline
        Average number density ($\bar{n}$) [m$^{-3}$] & $2.5 \times 10^{12}$ \\
        %\hline
        Average temperature ($\bar{T}$) [K] & $1.0 \times 10^{4}$ \\
        %\hline
        Debye length ($\lambda_{D})$ [m] & $4.364992\times 10^{-3}$ \\
        %\hline
        Inverse angular plasma frequency ($\omega_{pe}^{-1}$) [s/rad] & $1.121147\times 10^{-8}$ \\
        %\hline
        Thermal velocity ($v_{th} = \lambda_D  \omega_{pe}$) [m/s] & $3.893328 \times 10^{5}$ \\
        \hline
    \end{tabular}
    \caption{Table of the plasma parameters used in the sheath problem.}
    \label{tab:sheath plasma parameters}
\end{table}

In the first test, we perform a series of refinements to understand how the solution is impacted as a function of spatial resolution and the number of macroparticles. More specifically, we fix the total number of macroparticles and adjust the number of cells per dimension, taking $N_x = N_y = N$. During the refinement, $N$ is successively doubled from $16$ to $64$. We use $2.50632\times10^5$ macroparticles for each species, for all meshes, which results in slightly more than $60$ macroparticles of each species per mesh cell, in the case of a $64 \times 64$ mesh. We also fix $\Delta t = \text{CFL} \Delta x / c$, where $\text{CFL}=1/ \sqrt{2}$ for both the FDTD and BDF-1 field solvers. As noted, the initial positions for ions and electrons are sampled from a uniform distribution over the domain $\left[-8\lambda_{D}, 8 \lambda_{D}\right]^{2}$ with a fixed random seed across all runs. The heavier ions are treated as stationary, so their velocities are set to zero in this test. Electrons velocities are obtained by sampling from a Maxwellian distribution using the parameters shown in Table \ref{tab:sheath plasma parameters}. The problem is made current neutral by splitting the electrons into two equally sized groups whose velocities differ only in sign. The simulation is run for 60 angular plasma periods. At each time step, we record the particle count, electron temperature, as well as the potentials and the corresponding fields on the mesh.

\begin{figure}[!htb]
    \centering
    \includegraphics[clip, scale=0.35
]{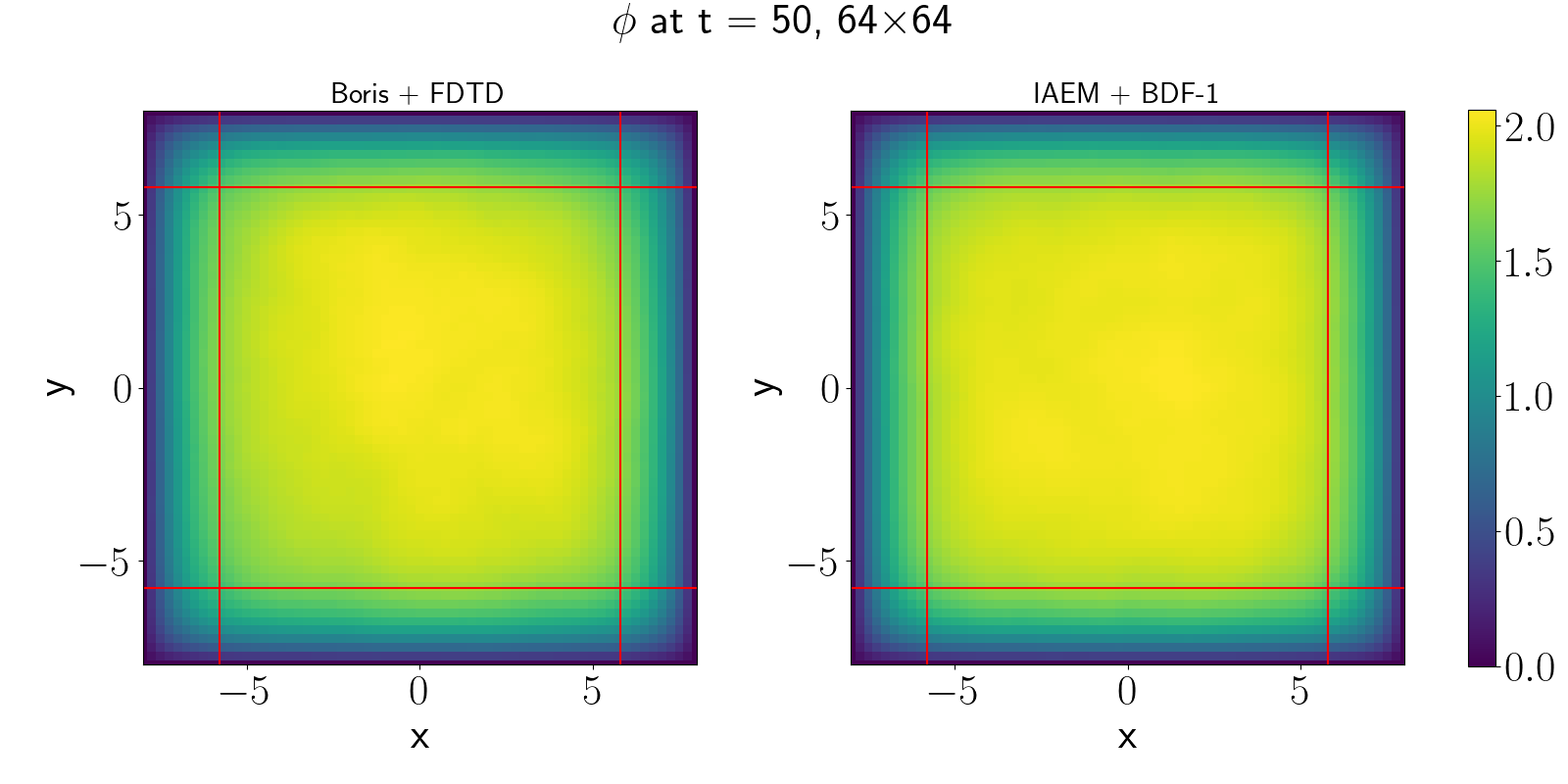}
    \caption{A comparison of the scalar potential obtained with the benchmark approach (Boris + FDTD) and the proposed scheme (IAEM + BDF-1). The formula for calculating the expected sheath width $s = \lambda_{D}\sqrt{2\phi_{0}/T_{e}}$, given in \cite{LichtenbergLiebermanPrinciples1994}, is also included in the plot (marked with red lines). Here, $\phi_{0}$ is the magnitude of the potential at the center of the domain, where the potential is flat. The sheaths produced by both methods agree with the width predicted from theory.}
    \label{fig:2D-2V Sheath Length}
\end{figure}

\begin{figure}[!htb]
    \centering

    \subfloat[IAEM + BDF-1]{
    % trim: L, B, R, T
    \includegraphics[clip, trim={0cm, 0cm, 0cm, 1.1cm}, scale=0.5
]{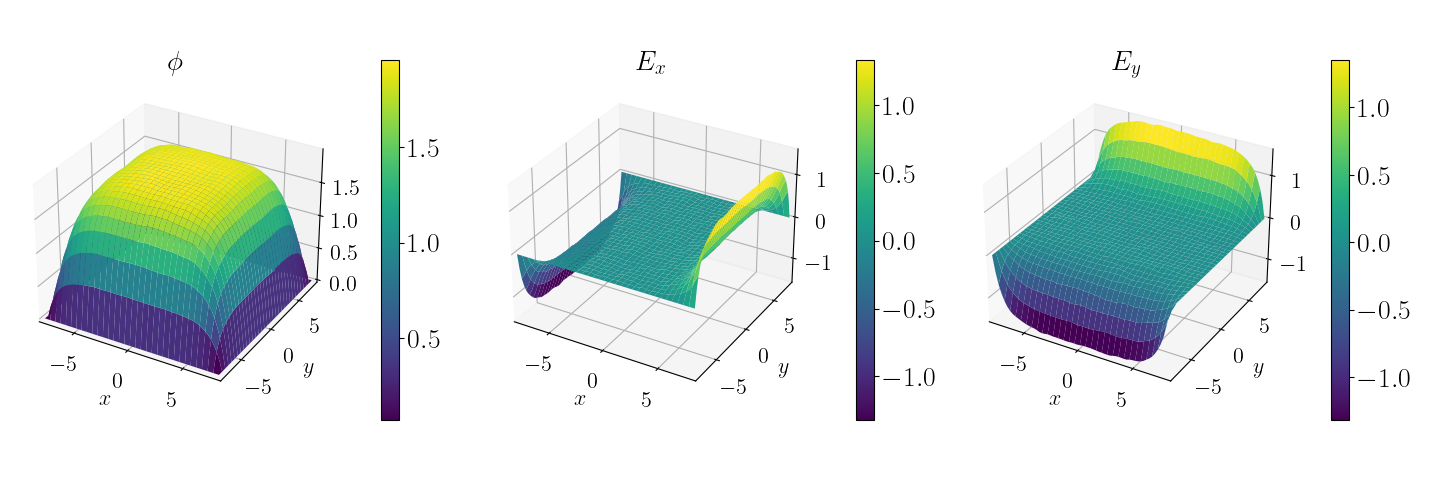}}
    \\
    \vspace{10pt}
    \subfloat[Boris + FDTD]{
    % trim: L, B, R, T
    \includegraphics[clip, trim={0cm, 0cm, 0cm, 1.1cm}, scale=0.5]{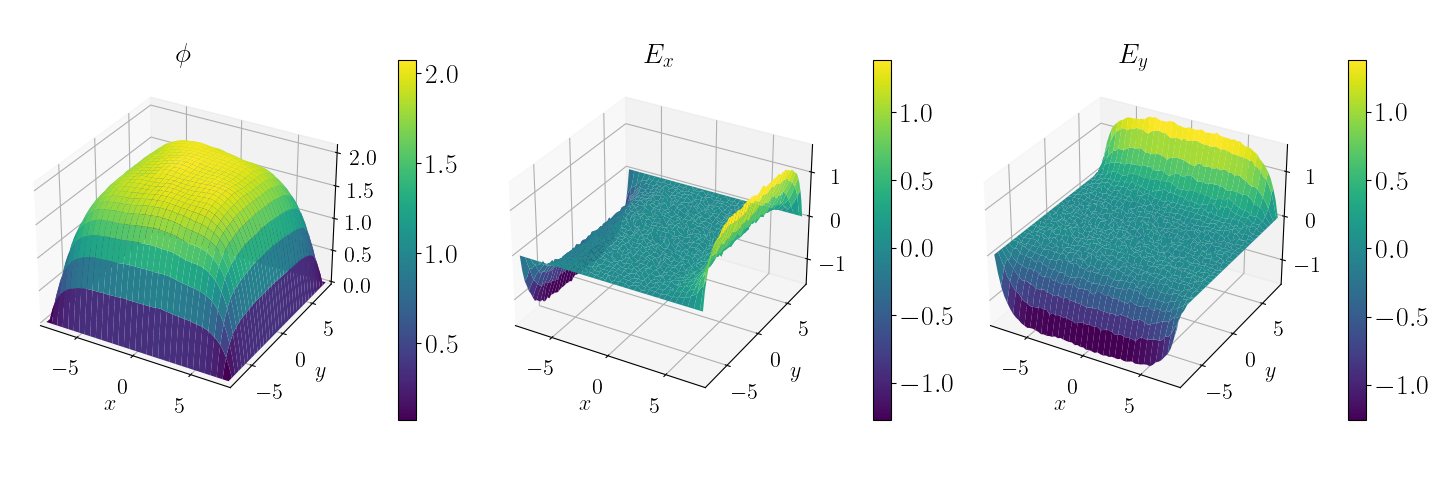}}
    \caption{Time-averaged scalar potential along with the time-averaged $x$ and $y$ components of the \mbox{electric} fields for the sheath problem. The breathing mode is well established by $t = 40$, and the fields are averaged over the next $20$ angular plasma periods. The potential and fields produced by the \mbox{benchmark} method (Boris + FDTD) are much rougher than those obtained with the proposed method (IAEM + BDF-1).}
    \label{fig:2D-2V Sheath Fields}
\end{figure}

Figure \ref{fig:2D-2V Sheath Length} is a plot of the scalar potential for both methods at a time of 50 angular plasma periods using a $64 \times 64$ mesh. On the left, we plot the solution obtained with the \mbox{Boris + FDTD} method, while the plot on the right was obtained using the \mbox{IAEM + BDF-1} approach. At 50 angular plasma periods, the transients arising from the sheath formation have mostly settled, and the fields become flat in the middle of the domain. What remains is a steady breathing mode. We address the breathing mode in two ways. One way is to choose a time snapshot were the field is flat on the interior, as seen in figure \ref{fig:2D-2V Sheath Length}. The second is to time-average the fields, which is done in Figure \ref{fig:2D-2V Sheath Fields} that follows. To get a sense of the sheath size, we can appeal to the well-known analytical theory of the matrix sheath \cite{LichtenbergLiebermanPrinciples1994}. Marked in red is the analytical solution to the 1D matrix sheath calculated from the plasma parameters from the simulation. We see that the predicted locations of the sheath obtained with both methods are in reasonable agreement with the analytical solution.

% 2D-2V Sheath Results - Time averaged fields with fixed total particle count
% TO-DO: The font size needs to be increased and the font style needs to be modified to match other figures
%
%

In Figure \ref{fig:2D-2V Sheath Fields}, we plot the time-averaged potentials and fields obtained with both methods using a $64 \times 64$ mesh. Solutions are time-averaged across the time interval ranging from $t = 40$ to $t = 60$ angular plasma periods. We observe more noise in the fields obtained with the \mbox{Boris + FDTD} approach when compared to the fields computed with the \mbox{IAEM + BDF-1} method. Additionally, the \mbox{Boris + FDTD} method produces a clear asymmetry in the potential that is not present in the potential obtained with the IAEM + BDF-1 method. This suggests that the breathing mode is less adequately maintained with the \mbox{Boris + FDTD} method. If we look at the non-averaged solutions at a later time, this is indeed true. For this problem, one expects a symmetric solution, which indicates that the new method is an improvement to the benchmark approach.

\begin{figure}[!htb]
    \centering
    % trim: L, B, R, T
    \includegraphics[scale=0.275
]{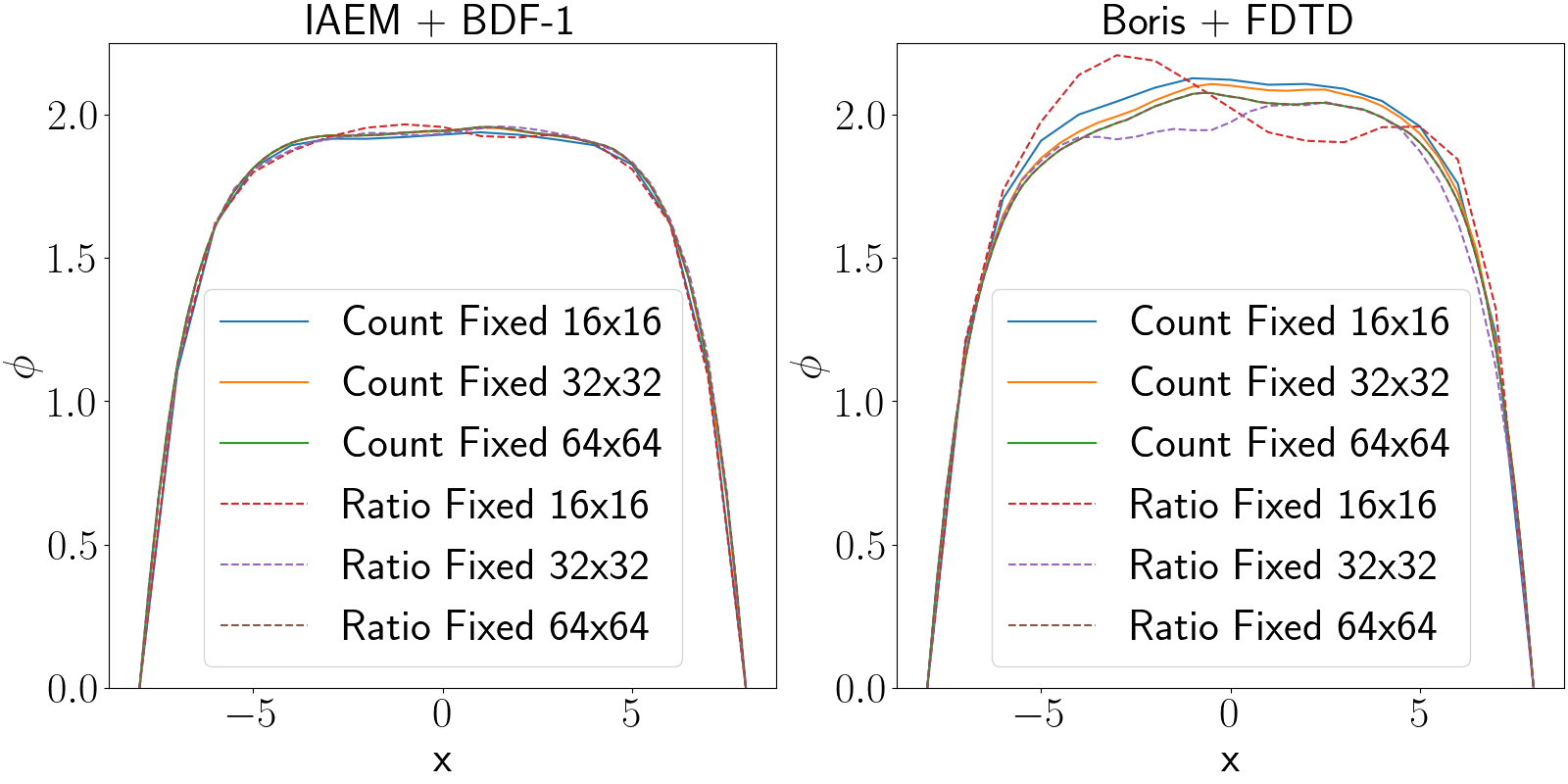}
    \caption{Cross-sections of the time-averaged potentials computed with both methods. Results obtained using a fixed total number of macroparticles per species (``Count Fixed") and a fixed number of particles per cell (``Ratio Fixed") are presented for comparison. Each simulation begins the time averaging procedure after 40 angular plasma periods, at which point both methods are well into the ``breathing mode." The cross-sections are taken about $y=0$ and are similar to those obtained by \cite{LichtenbergLiebermanPrinciples1994}. We note that the time-averaged potentials obtained with the Boris + FDTD approach are substantially ``rougher" near the center.}
    \label{fig:2D-2V Sheath Potential Slices fixed count}
\end{figure}

In Figure \ref{fig:2D-2V Sheath Potential Slices fixed count}, we plot the time averaged center line potential about $y=0$.  As in Figure \ref{fig:2D-2V Sheath Fields}, the solution is time-averaged from $t = 40$ to $t = 60$ angular plasma periods.  The figure on the left is the IAEM + BDF-1 method and the figure on the right is the Boris + FDTD method.  The solid lines in both figures use a total of $2.50632\times10^5$ macroparticles, per species, for all mesh resolutions, while the dashed lines represent results with the number of macroparticles per cell set to $61$.  We see that the solution for the IAEM + BDF-1 approach is more symmetric and is more or less identical across different mesh resolutions. While we observe convergence in the Boris+FDTD (in the sense of macroparticle count and mesh resolution), the solution is clearly noisier and not as robust.  Issues around convergence are further discussed when we consider the time trace of the particle count and the temperature as the sheath forms and the problem settles into a breathing mode dynamic.

% 2D-2V Sheath Results - 1D Sheath time averaged slices (fixed number of simulation particles)

\begin{figure}[!htb]
    \centering
    \includegraphics[scale=0.28]{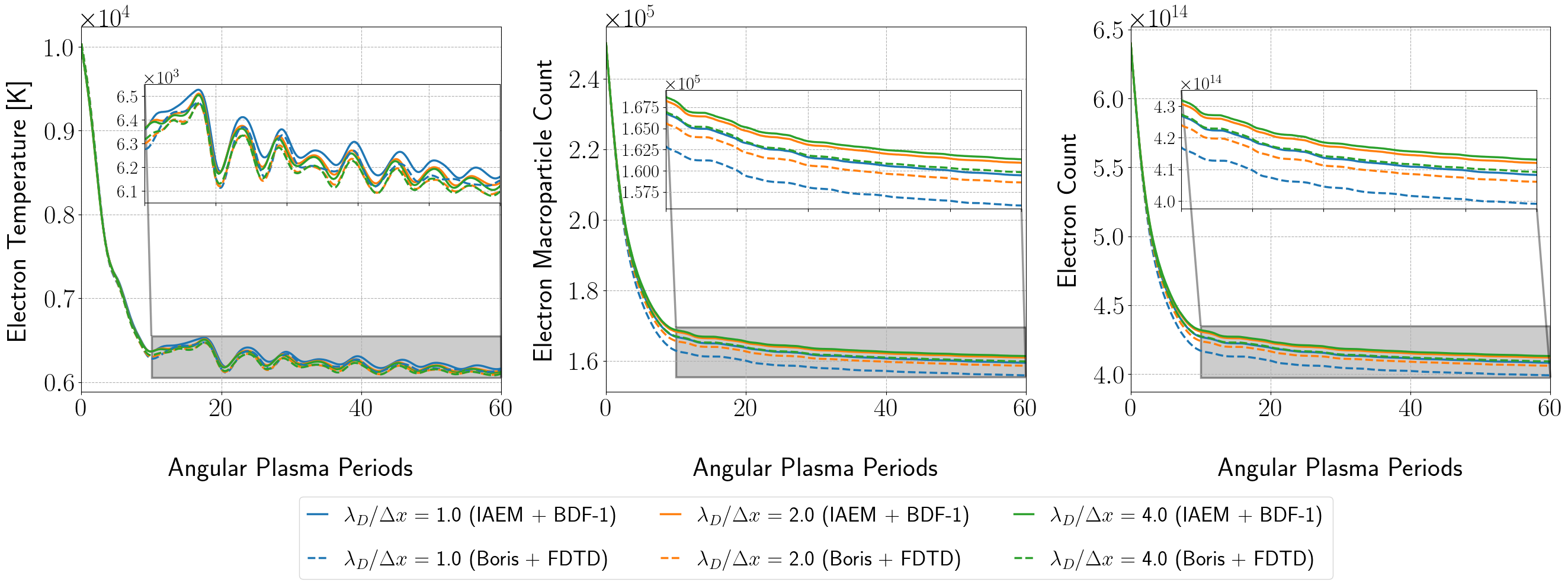}
    \caption{Electron temperature and particle counts collected for the sheath experiment using different grid resolutions and a fixed total number of macroparticles. As the mesh is refined, the number of macroparticles per cell decreases. However, the total number of physical particles is scaled so that the runs start in an identical manner. The data obtained with the new method (IAEM + BDF-1) is plotted using solid lines, while the results for the benchmark method (Boris + FDTD) are plotted on dashed lines. In the plot of the electron counts we focus on the region in which the potential begins to settle, highlighting the differences in the electron count. We see that the retention of the faster electrons with the new method results in a larger electron temperature when compared with the benchmark scheme. The observed electron temperatures in the proposed method are consistent with the results of the electromagnetic heating experiment in Figure \ref{fig:2D-2V EM heating test}. We note that the new method experiences less significant fluctuations in the temperature and physical particle counts than the benchmark scheme, despite the  variations in the number of macroparticles per cell.}
    \label{fig:sheath temp and count fixed total}
\end{figure}

\begin{figure}[!htb]
    \centering
    \includegraphics[scale=0.28]{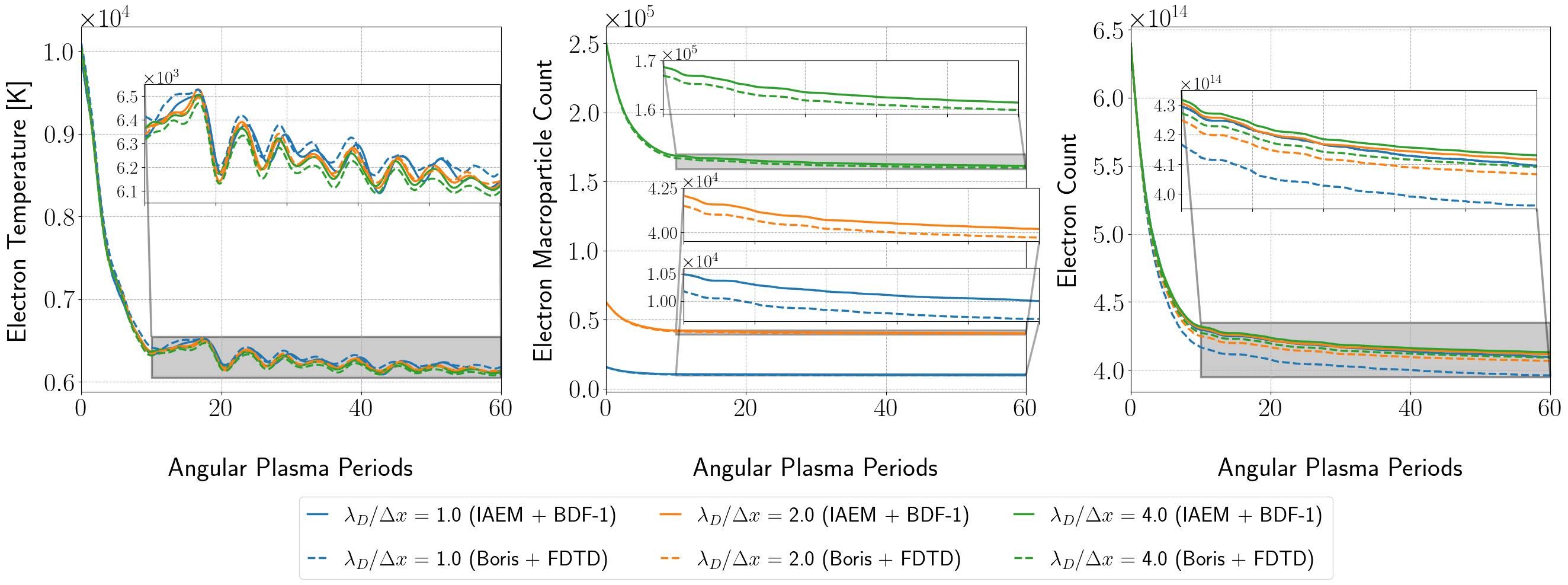}
    \caption{Electron temperature and count data collected for the sheath experiment with different grid resolutions. In contrast to the data presented in Figure \ref{fig:sheath temp and count fixed total}, the number of macroparticles is increased as the mesh is refined so that the number of macroparticles per mesh cell is identical across the runs. The data obtained with the new method \mbox{(IAEM + BDF-1)} are plotted using solid lines, while the results for the benchmark method (Boris + FDTD) are plotted on dashed lines. The new method produces more qualitatively consistent results across the mesh resolutions than the benchmark scheme. In particular, these results suggest that the new method self-refines at a faster rate to the data obtained with the finest mesh than the benchmark scheme.}
    \label{fig:sheath temp and count fixed ppc}
\end{figure}

Next, in Figure \ref{fig:sheath temp and count fixed total}, we show the electron temperature (left), the macroparticle count (middle) and electron count (right) for the range of spatial resolutions described earlier. In this study, the total macroparticle count for each species is initially $2.50632\times10^5$ for all runs. We see that the temperature and particle counts for the \mbox{Boris + FDTD} and \mbox{IAEM + BDF-1} methods converge as the cell resolution increases. In particular, the \mbox{Boris + FDTD} method converges more slowly than the IAEM+BDF-1 method. We note that there are two mechanisms associated with the faster convergence of the \mbox{IAEM + BDF-1} method. First, the IAEM+BDF-1 approach is high-order in space, so the sheath will be more resolved when compared to the \mbox{Boris + FDTD} approach on a similar computational mesh. Second, the \mbox{Boris + FDTD} approach contains a higher level of noise than the BDF scheme because the latter is dissipative, while the former is dispersive. Additional noise in the \mbox{Boris + FDTD} approach is due to the current weighting scheme \cite{VillasenorChargeConservation92}, which is used to enforce charge conservation. We might also expect that issues such as numerical heating could be impacting the \mbox{Boris + FDTD} results when the Debye length is not adequately resolved. In contrast, the \mbox{IAEM + BDF-1} method displays more robust behavior even when the Debye length is ``under-resolved.'' We repeat this experiment but we fix the number of macroparticles per cell to be 61. Similar phenomena is observed in Figure \ref{fig:sheath temp and count fixed ppc}, namely, we see faster convergence in the number of electrons for the \mbox{IAEM + BDF-1} method. Further, the results for the \mbox{IAEM + BDF-1} approach on the $16\times 16$ mesh and $61$ macroparticles per cell are similar to those obtained with a $16\time 16$ mesh and $2.50632\times10^5$ total macroparticles. This is also true for the $32\times 32$ mesh in the case of the \mbox{IAEM + BDF-1} method.  In contrast, the results for the \mbox{Boris + FDTD} approach, while qualitatively similar, are not nearly as close together. These results, together with the data presented in Figures \ref{fig:2D-2V Sheath Fields} and \ref{fig:2D-2V Sheath Potential Slices fixed count}, seem to suggest that the \mbox{IAEM + BDF-1} method can be used with far fewer simulation particles than the traditional \mbox{Boris + FDTD} method. This feature will be the study of future work.

% 2D-2V Sheath Results - Comparison of electron temperature and particle count history for both methods 
% This plot uses a fixed number of simulation particles

% 2D-2V Sheath Results - Comparison of MOLT and FDTD particle count history with fixed particle count per cell

\begin{figure}[!htb]
    \centering
    \includegraphics[clip, trim={0cm, 0cm, 0cm, 0cm}, scale=0.275]{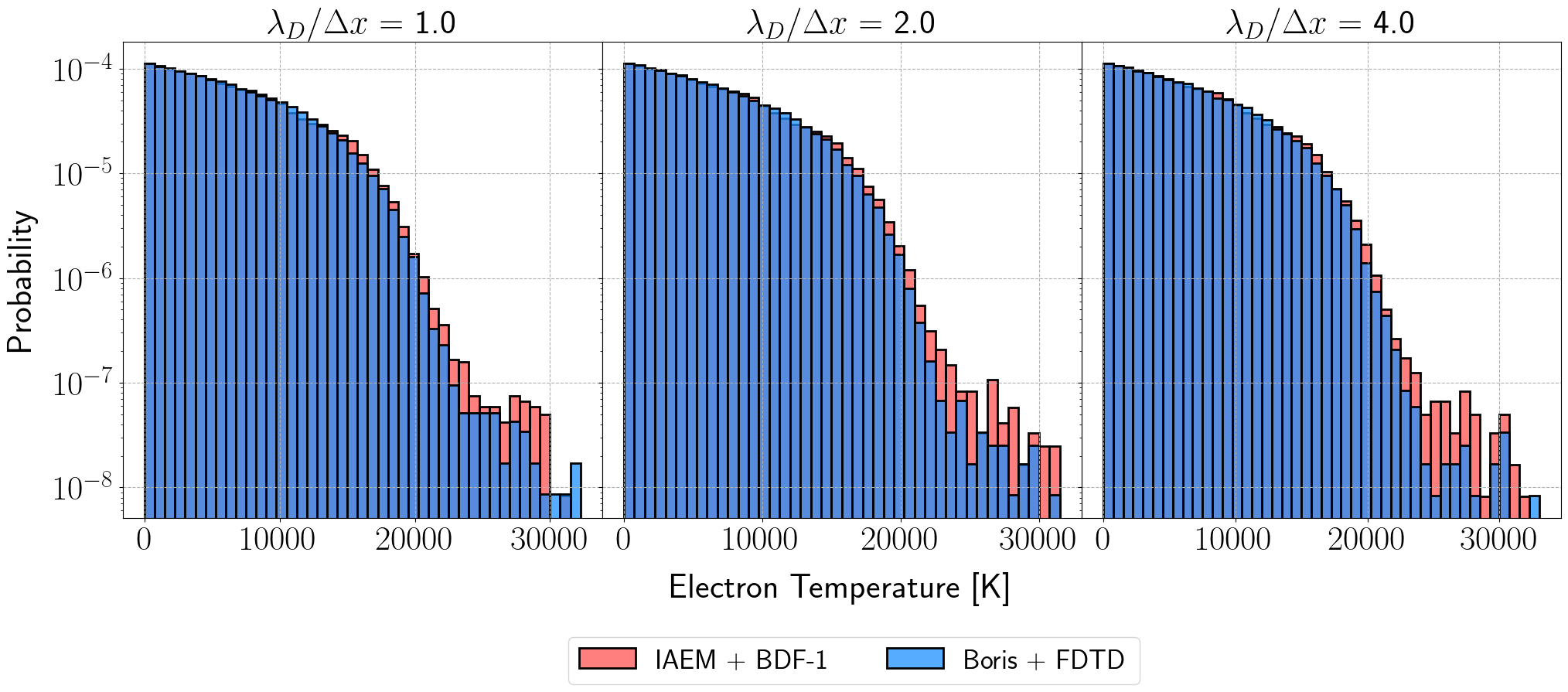}
    \caption{Stacked histograms of the particle temperature distributions, at the final time, obtained with both methods, for several different grid resolutions, with the total number of macroparticles held fixed. While the bulk properties are similar among the approaches, the new method retains more of the warmer electrons than the benchmark scheme, contributing to a larger overall temperature.}
    \label{fig:sheath energy distributions}
\end{figure}

In Figure \ref{fig:sheath energy distributions}, we plot the electron temperatures as probability densities with a fixed total number of macroparticles set to $2.50632\times10^5$. From left to right in the figure, the plots correspond to data obtained using $16\times 16$, $32\times 32$, and $64\times 64$ spatial meshes. The distribution function for IAEM + BDF-1 is in red and the distribution function for Boris + FDTD is in blue. In each case, the warmer tails of the IAEM + BDF-1 densities contain more simulation particles than the Boris + FDTD method. This is likely due to the high-order spatial resolution of the IAEM + BDF-1, as it will be able to resolve the sheath with fewer points.  It could also have to do with the reduced noise and improved symmetry observed in the solution to the fields from the IAEM + BDF-1 method, as noise or asymmetry in the breathing mode could easily push a hot particle in the tail out of the domain.  

% 2D-2V Sheath Results - Energy distributions (fixed total number of simulation particles)

\begin{figure}[!htb]
    \centering
    \includegraphics[clip, trim={0cm, 0cm, 0cm, 0cm}, scale=0.275]{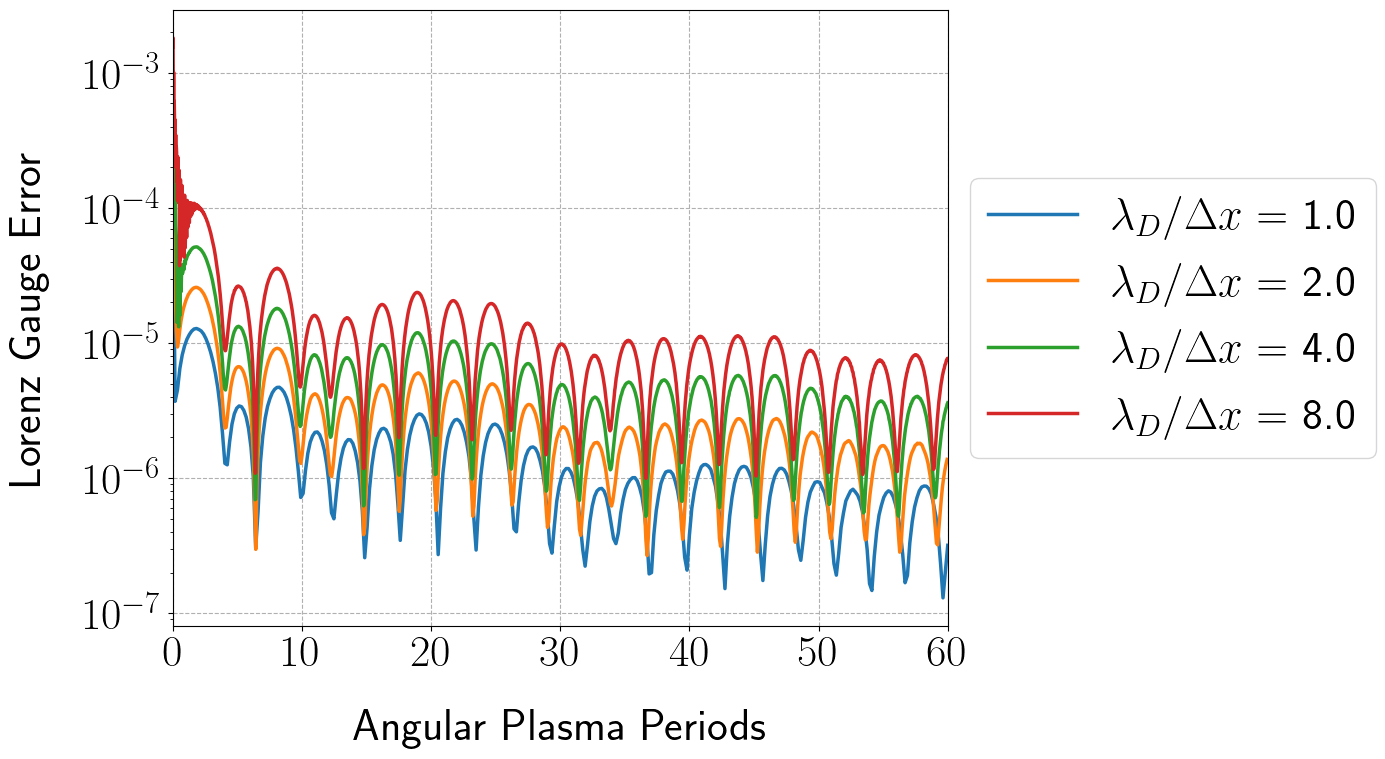}
    \caption{The $\ell_{2}$-norm of the residual in the Lorenz gauge for the new method using a fixed total number of particles. Although the magnitude of the error increases as we refine the mesh, it is reasonable given that no method is used to enforce the gauge condition. In light of the time-consistency theorem, this implies that the total charge is also reasonably conserved.}
    \label{fig:sheath gauge error IAEM + BDF-1}
\end{figure}

Lastly, we check for any significant violations of the Lorenz gauge condition for the IAEM + BDF-1 method. Figure \ref{fig:sheath gauge error IAEM + BDF-1} plots the $\ell_{2}$-norm of the gauge error as a function of time. We change the spatial resolution but keep the number of macroparticles fixed at $2.50632\times10^5$. In these experiments, the method maintains a bounded gauge error for all time.  However, as the number of particles per cell decreases, we observe an increase in size of the Lorenz gauge error. We note that there is a sign to the gauge error, depending on whether it is ions or electrons. In a system that is truly equal, in the sense of number of particles, this error cancels. We think the increase in the error with increased mesh resolution is simply a result of less local charge cancellation. This will also be explored as part of our future work.

% 2D-2V Sheath Results - Gauge error for IAEM + BDF-1 (fixed total number of simulation particles)

%\ref{fig:2D-2V Sheath Potential Slices fixed count} and  The plot in Figure \ref{fig:2D-2V Sheath Potential Slices fixed count} shows one-dimensional cross-sections of the time-averaged sheaths taken about the center line $y=0$, while Figure \ref{fig:2D-2V Sheath Fields} shows the analogous two-dimensional time-averaged potential along with the corresponding fields.

%%%%%%%%%%%%%%

\subsubsection{Non-relativisitic Expanding Particle Beam}
\label{subsubsec:non-relativistic expanding beam problem}

We now consider an application of the proposed methods to expanding particle beams \cite{OconnorPIC-Benchmark}. This example is well-known for its sensitivity to issues concerning charge conservation, so it is typically considered when evaluating methods used to enforce charge conservation. While this particular example is normally solved in cylindrical coordinates, the simulations presented in this work use a two-dimensional rectangular grid that retains the fields $E^{(1)}$ and $E^{(2)}$, as well as $B^{(3)}$. An injection zone is placed on one of the faces of the box and injects a steady beam of particles into the domain. The beam expands as particles move along the box due to the electric field and eventually settles into a steady-state. Similar to the sheath problem, particles are absorbed or ``collected" once they reach the edge of the domain and are removed from the simulation. Along the boundary of the domain, the electric and magnetic fields are prescribed PEC boundary conditions, which, in two spatial dimensions, is equivalent to enforcing homogeneous Dirichlet boundary conditions on the potentials $\phi$, $A^{(1)}$, and $A^{(2)}$. Since the problem is PEC, there can be no (tangential) currents or charge on the boundary.

\begin{figure}[!htb]
    \centering
    \includegraphics[width=0.4\textwidth]{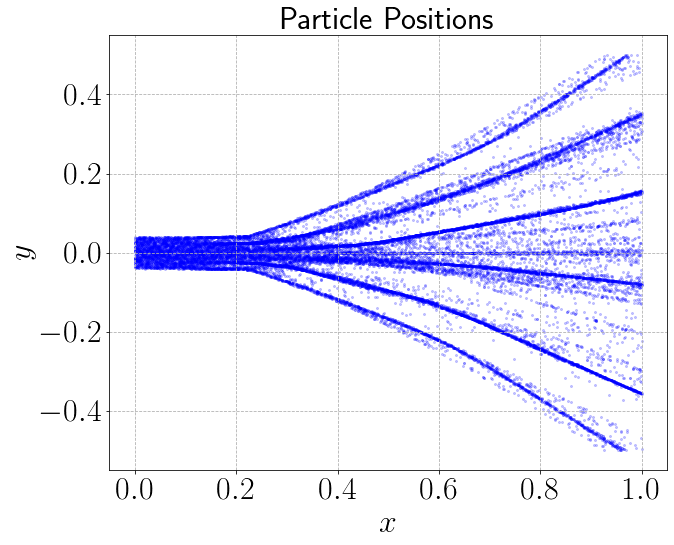}
    \caption{Striation patterns in a non-relativistic expanding beam simulation using the FDTD-PIC method with bilinear current mappings (area weighting). Irrotational errors in the electric field introduced by the mapping cause the particles to focus in regions of the domain.}
    \label{fig:7.8025e14_naive}
\end{figure}

% Should we keep this section here? I think this is fine. Plus, it shows what happens when charge conservation issues occur
As discussed earlier, the FDTD method is known to preserve the involutions for Maxwell's equations in the absence of moving charge \cite{TafloveHagnessFDTDbook}; however, this is not applicable to the examples considered in this work. In order to update the fields in the FDTD approach, we need to map the current density components $J^{(1)}$ and $J^{(2)}$ to mesh points that are collocated with $E^{(1)}$ and $E^{(2)}$, respectively, according to the mesh staggering. As mentioned earlier, it is well known that the use of bilinear maps for depositing current to the mesh results in catastrophic errors due to violations of charge conservation. The resulting fields cause the charged particles to ``focus" in certain regions, leading to the appearance of striation patterns. An example of this phenomenon is presented in Figure \ref{fig:7.8025e14_naive}, which shows the formation of non-physical striations after twenty particle crossings. However, indicators for such patterns can appear as early as two particle crossings. The map for the FDTD method used in the comparison is based on the technique discussed in the paper by Villasenor and Buneman \cite{VillasenorChargeConservation92}. Particles are advanced using the Boris method.

%%% Base method (Boris + FDTD) with bi-linear maps

To setup the simulation, we first create a box specified by the region $[0,1] \times [-1/2,1/2]$ that is normalized by the length scale $L$, which corresponds to the physical distance along the $x$-axis of the box. We assume that the beam consists only of electrons, which are prescribed some injection velocity $v_{\text{injection}}$ through their $x$-components.  An estimate of the crossing time for a particle can be obtained from the injection velocity and the length of the domain, which sets the time scale $T$ for the simulation. The duration of the simulation is given in terms of particle crossings, which are then used to set the time step $\Delta t$. In each time step, particles are initialized in an injection region specified by the interval $[- L_{\text{ghost}},0) \times [-R_{b}, R_{b}]$, where $R_{b}$ is the radius of the beam, and the width of the injection zone $L_{\text{ghost}}$ is chosen such that $$ L_{\text{ghost}} = v_{\text{injection}} \Delta t. $$ This ensures that all particles placed in the injection zone will be in the domain after one time step. The positions of particles in the injection region are set according to samples taken from a uniform distribution, and the number of particles injected for a given time step is set by the injection rate. In each time step, the injection procedure is applied before the particle position update, so that, at the end of the time step, the injection zone is empty. To prevent the introduction of an impulse response in the fields due to the initial injection of particles, a linear ramp function is applied to the macroparticle weights over one particle crossing. The methods were applied using both wide and narrow beam configurations, whose parameters can be found in Tables \ref{tab:Wide beam parameters} and \ref{tab:Narrow beam parameters}, respectively. The normalized speed of light in the simulations is $\kappa = 5.995849$. Using the wide beam configuration listed in Table \ref{tab:Wide beam parameters}, we obtain $\sigma_{1} = 1.006865 \times 10^{-1}$ and $\sigma_{2} = 2.762661 \times 10^{-1}$ for the normalized permittivity and permeability, respectively. For the narrow configuration provided in Table \ref{tab:Narrow beam parameters}, these parameters change to $\sigma_{1} = 5.061053 \times 10^{-1}$ and $\sigma_{2} = 5.496139 \times 10^{-2}$, respectively.

\begin{table}[!htb]
    \centering
    \begin{subtable}[t]{0.45\textwidth}
        \centering
        \def\arraystretch{1.2}
        \begin{tabular}{ | c || c | }
            \hline
            \textbf{Parameter}  & \textbf{Value} \\
            \hline
            Beam radius ($R_{b})$ [m] & $8.0 \times 10^{-3}$ \\
            %\hline
            Average number density ($\bar{n}$) [m$^{-3}$] & $7.8025\times 10^{14}$ \\
            %\hline
            Physical domain length ($L$) [m] & $1.0\times 10^{-1}$ \\
            %\hline
            Injection velocity ($v_{\text{injection}}$) [m/s] & $5.0\times 10^{7}$ \\
            %\hline
            Injection rate ($r_{\text{injection}}$) [s$^{-1}$] & $1 \times 10^{2}$ \\
            %\hline
            Crossing time ($T$) [s] & $2.0 \times 10^{-9}$ \\
            \hline
        \end{tabular}
        \caption{Wide beam configuration}
        \label{tab:Wide beam parameters}
    \end{subtable}
    \hspace{32pt}
    \begin{subtable}[t]{0.45\textwidth}
        \centering
        \def\arraystretch{1.2}
        \begin{tabular}{ | c || c | }
            \hline
            \textbf{Parameter}  & \textbf{Value} \\
            \hline
            Beam radius ($R_{b})$ [m] & $8.0 \times 10^{-3}$ \\
            %\hline
            Average number density ($\bar{n}$) [m$^{-3}$] & $1.5522581\times 10^{14}$ \\
            %\hline
            Physical domain length ($L$) [m] & $1.0\times 10^{-1}$ \\
            %\hline
            Injection velocity ($v_{\text{injection}}$) [m/s] & $5.0\times 10^{7}$ \\
            %\hline
            Injection rate ($r_{\text{injection}}$) [s$^{-1}$] & $1 \times 10^{2}$ \\
            %\hline
            Crossing time ($T$) [s] & $2.0 \times 10^{-9}$ \\
            \hline
        \end{tabular}
        \caption{Narrow beam configuration}
        \label{tab:Narrow beam parameters}
    \end{subtable}
    \hfill
    \caption{Parameters used in the setup for the non-relativistic expanding particle beam problems.}
\end{table}

\begin{figure}[!htb]
    \centering
    \subfloat[][]{
    \includegraphics[width=0.37\textwidth]{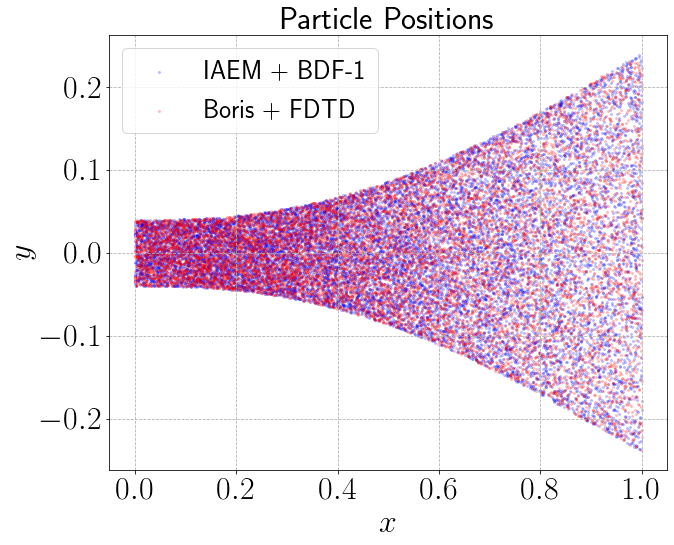}
    \label{fig:EBP + IAEM + BDF-1 wide particles}} 
    \subfloat[][]{
    \includegraphics[width=0.38\textwidth]{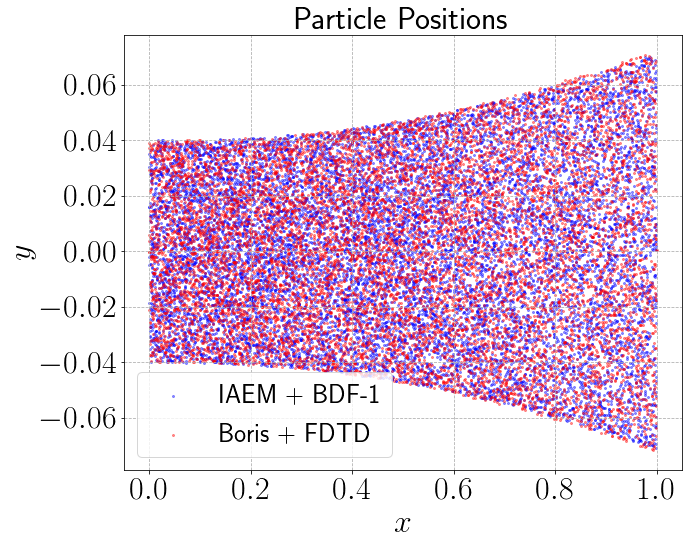}
    \label{fig:EBP + AEM + BDF-1 gauge slim particles}} \\
    \caption{We compare the proposed PIC method against the standard FDTD-PIC method for the non-relativistic expanding beam configurations specified in Tables \ref{tab:Wide beam parameters} (shown on the left) and \ref{tab:Narrow beam parameters} (shown on the right). In each case, the particle positions generated by the two methods after 3000 crossings are plotted together to track the shape of the beam. The beams in the proposed methods remain intact after many particle crossings without the use of a cleaning method. Moreover, the beams generated by the proposed methods show excellent agreement with the beam profiles from the benchmark FDTD-PIC method.}
    \label{fig:EBP comparison of methods}
\end{figure}

\begin{figure}[!htb]
    \centering
    \includegraphics[width=\textwidth]{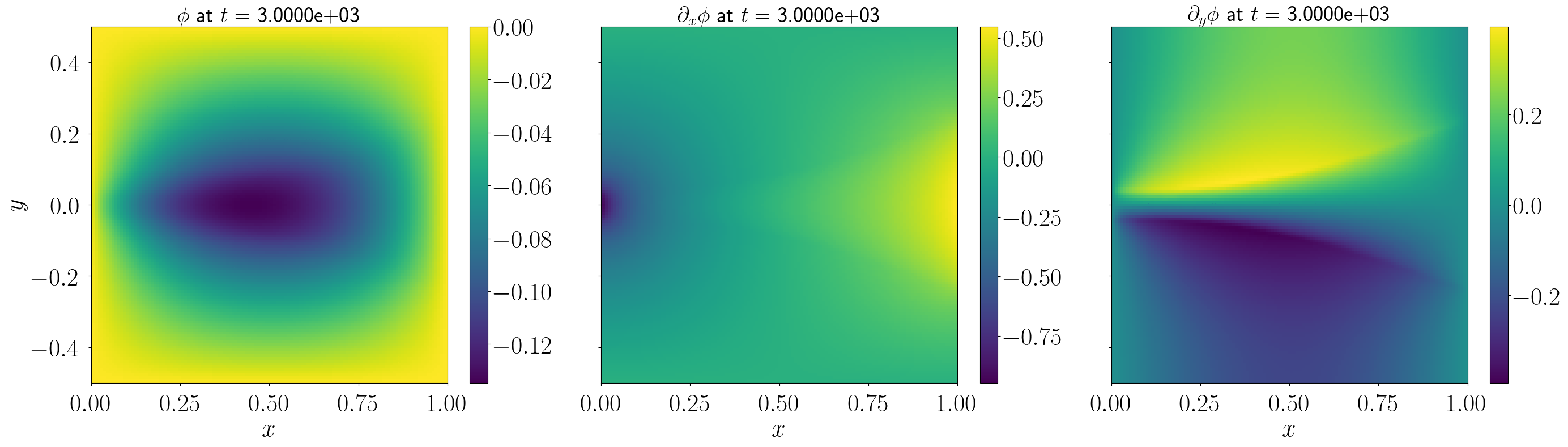}
    \caption{The scalar potential $\phi$ and its spatial derivatives for the non-relativistic expanding beam problem after 3000 particle crossings. The fields shown above correspond to parameters listed in Table \ref{tab:Wide beam parameters}. No methods are used to enforce the gauge condition in this experiment. We can see that the proposed methods generate smooth fields for subsequent use in the particle update.}
    \label{fig:EBP + IAEM + BDF-1 fields at end}
\end{figure}

The proposed method was compared against the Boris + FDTD method using the problem configurations specified in Tables \ref{tab:Wide beam parameters} and \ref{tab:Narrow beam parameters}. Each simulation was evolved to a final time corresponding to 3000 particle crossings with a $128 \times 128$ mesh. A total of $4 \times 10^6$ time steps were used, which gave a CFL $\approx 0.576$ for the fields. In Figure \ref{fig:EBP comparison of methods}, we plot the particles in the beams generated using the IAEM + BDF-1 solver and the Boris + FDTD method. We observe excellent agreement with the benchmark FDTD PIC method despite the first-order time accuracy of the new method. Additionally, we find that the steady-state structure of the beam is well-preserved with the proposed method despite the fact that charge conservation is not strictly enforced. We find that the potentials and their spatial derivatives, which are computed using the BDF-1 wave solver are quite smooth and do not show signs of excessive dissipation even after 3000 particle crossings. Plots of the scalar potential and its gradient obtained with the proposed methods are displayed in Figure \ref{fig:EBP + IAEM + BDF-1 fields at end}. We show this data for the wide beam configuration provided in Table \ref{tab:Wide beam parameters}, and note that the results are quite similar for the narrow beam configuration provided in Table \ref{tab:Narrow beam parameters}. While the goal of our work is to build higher-order field solvers for plasma applications, these results are interesting from the perspective of practicality, as they demonstrate that it is possible to obtain a solution of reasonable quality in a fairly inexpensive manner.

%%% Compare the plots of the beams

%%% Plots of the fields for the wide configuration
%%% This shows what the fields look like (derivatives)

%%%%%%%%%%%%%%

\subsubsection{The Mardahl Beam Problem}
\label{subsubsec:Mardahl beam problem}

We conclude the numerical experiments with the Mardahl beam problem, which is a benchmark relativistic beam problem proposed by Mardahl and Verboncoeur \cite{Mardahl97conservation}. In this problem, electrons are injected into a PEC cavity with relativistic velocities ($v_{\text{injection}} = 0.967 c$), and the number density is relatively small, so the beam moves across the domain mostly unperturbed. Once the electrons reach the boundary, they are removed from the simulation. A complete list of parameters for our experimental setup, which were derived from \cite{Mardahl97conservation}, is provided in Table \ref{tab:Mardahl beam parameters}. The normalized speed of light for this problem is $\kappa = 1.034126$, and the corresponding normalized permittivity and permeabilities are $\sigma_{1} = 1.226639 \times 10^{3}$ and $\sigma_{2} = 7.623181 \times 10^{-4}$ As with its non-relativistic counterpart, this problem is also sensitive to violations of charge conservation \cite{wolf2016particle}. It also serves as as useful demonstration of the formulation presented in this work in the relativistic setting, which is the state space for applications that will be considered in future work.

\begin{table}[!htb]
    \centering
    \def\arraystretch{1.2}
    \begin{tabular}{ | c || c | }
        \hline
        \textbf{Parameter}  & \textbf{Value} \\
        \hline
        Beam radius ($R_{b})$ [m] & $5.0 \times 10^{-1}$ \\
        %\hline
        Average number density ($\bar{n}$) [m$^{-3}$] & $2.15299207054\times 10^{10}$ \\
        %\hline
        Physical domain length ($L$) [m] & $1.0\times 10^{0}$ \\
        %\hline
        Injection velocity ($v_{\text{injection}}$) [m/s] & $2.89899306886\times 10^{8}$ \\
        %\hline
        Injection rate ($r_{\text{injection}}$) [s$^{-1}$] & $1 \times 10^{2}$ \\
        %\hline
        Crossing time ($T$) [s] & $3.44947358 \times 10^{-9}$ \\
        \hline
    \end{tabular}
    \caption{Table of the parameters used in the setup for the Mardahl beam problem.}
    \label{tab:Mardahl beam parameters}
\end{table}

The setup for this test case is nearly identical to the non-relativistic expanding beam problems considered in the previous section, so we shall limit the discussion here for brevity. In the original presentation \cite{Mardahl97conservation}, the edge of the particle beam coincides with the boundary of the physical domain. Instead, we extend the normalized domain to $[0,1] \times [-1,1]$, so that the edge of the beam can be clearly seen. Furthermore, the original presentation showed simulation results up to $100$ crossings of the beam. The final time of the simulation was set to $3000$ particle crossings and used $4 \times 10^{6}$ time steps, so the fields and particles have a CFL $\approx 0.01$. We remark that this number is quite small for the BDF field solver, which by the stability result shown in section \ref{subsubsec:BDF stability analysis}, permits a much larger time step. As discussed earlier, the principle concern of this work is the development of a compatible formulation that can leverage the implicit wave solvers developed in previous work, e.g., \cite{causley2014higher,causley2017wave-propagation}. The exploration and integration of these solvers with the methods of this paper is an open area of research. As with the non-relativistic test case, the results of the proposed method show excellent agreement with the benchmark FDTD-PIC method, despite the fact that no method is used to explicitly enforce the gauge condition.

%%% Compare the plots of the beams
\begin{figure}[!htb]
    \centering
    \includegraphics[width=0.45\textwidth]{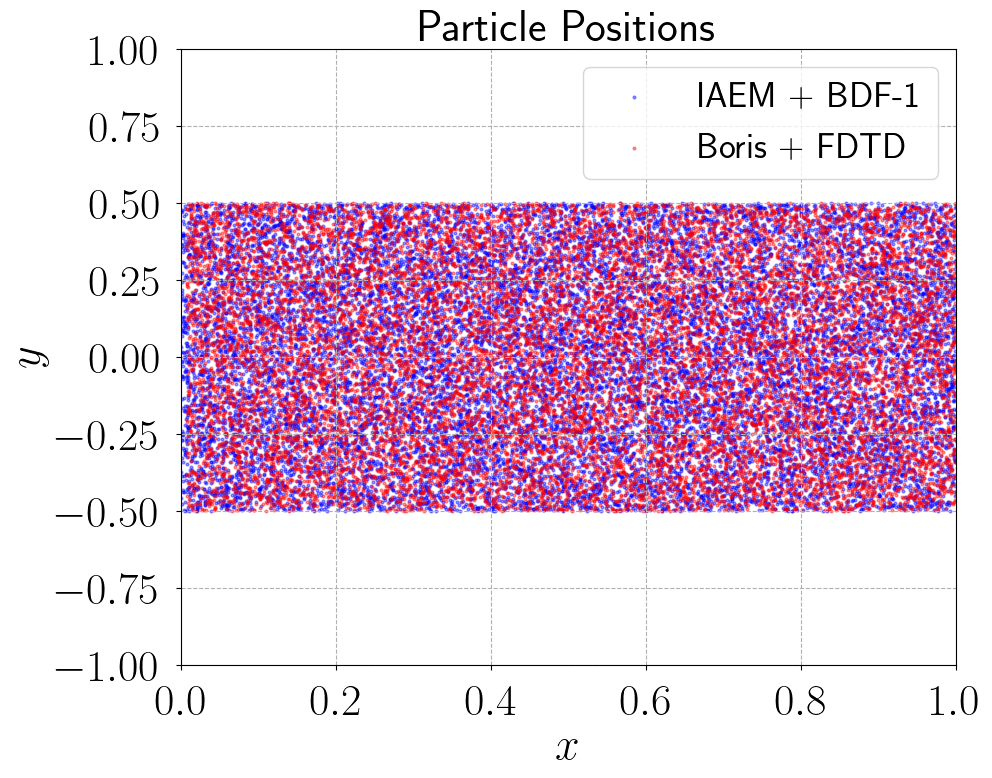}
    \caption{We compare the proposed PIC method against the standard FDTD-PIC method for the Mardahl beam problem whose configuration is specified in Table \ref{tab:Mardahl beam parameters}. The particle positions generated by the two methods are plotted together after 3000 crossings to track the shape of the beam. The beam simulated with the proposed method remains intact after many particle crossings without the use of a cleaning method. Unlike the original paper \cite{Mardahl97conservation}, we extended the physical domain beyond the edge of the beam to investigate its structure. Similar to the non-relativistic problem, the beam structure obtained with the proposed method shows excellent agreement with the profile from the benchmark FDTD-PIC method.}
    \label{fig:mardahl comparison of methods}
\end{figure}

% Summary of the section
%
% Summary of the section contents
%
\subsection{Summary}
\label{subsec:5 Summary}

In this section, we presented a collection of numerical results for the BDF wave solver that demonstrate its application in a variety of wave propagation problems. This includes applications to plasmas, where the wave solver is used to update fields in a new PIC method that is based on a Hamiltonian formulation. First, we analyzed the methods for evolving the potentials as well as the novel techniques for computing derivatives on the mesh. We considered several types of boundary conditions of relevance to the plasma examples presented in this work. In each case, the proposed methods for derivatives displayed space-time convergence rates that are identical to those of the wave solver used for the potentials. After establishing the convergence properties of the wave solver and the methods for derivatives, we then considered several applications involving plasmas using the new PIC method. The accuracy of the new PIC method was confirmed through a comparison of results obtained with standard PIC approaches. In particular, the new methods displayed superior numerical heating properties over the benchmark Boris + FDTD method used for electromagnetic problems. Additionally, the new method showed notable improvements in the sheath experiment, in terms of stability and preservation of symmetries. We found this to be true even in cases where the grid resolution is comparable to the plasma Debye length, which means that the new method permits coarser grids to be used in simulations. Additionally, the methods showed excellent agreement with the benchmark \mbox{Boris + FDTD} method for the relativistic and non-relativistic particle beam tests without resorting to the use of a method to explicitly enforce charge conservation. While the current results are generally encouraging, aspects concerning the efficiency of the new method will be explored in our future work.

% Give an overall summary of the paper's contributions
\section{Conclusions and Future Directions}
\label{sec:Conclusion and future work}

In this paper, we developed a new PIC method for solving the VM system. Using the Lorenz gauge condition, we expressed Maxwell's equations as a collection of wave equations for potentials. The potentials were evolved on a co-located grid using a wave solver derived from a three-point backwards difference approximation of the time derivatives that is globally first-order accurate in time. The equations of motion for the particles were also expressed in terms of these potentials as well as their spatial derivatives through the use of generalized momentum. We proved that the semi-discrete Lorenz gauge formulation considered in this work induced a corresponding semi-discrete equivalence between the Lorenz gauge condition and the continuity equation. The potential formulation naturally ensures that the involution \mbox{$\nabla \cdot {\bf B}=0$} is satisfied. 

Inspired by integral equation methods, we developed new approaches for the evaluation of spatial derivatives appearing in the particle updates. The proposed methodologies enjoy rates of convergence identical to the wave solver used to evolve the potentials and naturally inherit the stability and geometric flexibility offered by the BDF wave solver. The equations of motion for the particles in this formulation comprise a non-separable Hamiltonian system, which was solved using a semi-implicit Euler update that applies a (linear) Taylor correction. The new PIC method, which combines the BDF wave solver (and related methods for derivatives) with the new particle advance, was applied to a range of classic test problems including a plasma sheath and particle beams. We found that the new method offers notable improvements over conventional explicit PIC methods in critical test problems such as numerical heating and sheath applications. In particular, it is worth noting that the new algorithm displays mesh-independent heating properties. The numerical results presented in this paper suggest that PIC simulations can be performed using coarser grids than what are typically permitted by existing explicit algorithms. For example, in the electromagnetic heating experiment, when $\lambda_D/\Delta x \approx 2.5$, the electron temperature for the new method increased $< 0.1\%$ over 1000 angular plasma periods. Additionally, the results of the new method showed minimal sensitivity to the number of particles used per cell, which indicates that fewer simulation particles are required. The exploration of these qualities and their effect on the overall efficiency of the new method are also of interest to us.

% Future work
There are many opportunities for improving the methods developed in this work. A natural extension of the methods presented in this paper is the development of techniques to enforce the Lorenz gauge condition. Other extensions include the generalization to high-order in time, which requires new methods for the particles as well as higher-order discretizations for the fields. The globally second-order BDF solver presented in \cite{causley2017wave-propagation} is a natural first step in this direction, although we plan to investigate higher-order wave solvers. Fully-implicit discretizations are also being considered. Though not explored in this work, the proposed methodologies for the derivatives naturally retain the geometric capabilities of the method used for the fields, which will allow us to study plasmas in devices with complex geometric features. We remark that the generalized momentum formulation for the particles eliminates the need to compute the time derivatives of the vector potential. This also provides a promising opportunity to develop asymptotic-preserving PIC methods \cite{cheng2017asymptotic}. The evaluation of the boundary integrals in these schemes could be performed using new fast summation methods developed for GPUs \cite{Vaughn_Wilson_GPU_BLTC_2020,WilsonVaughn_GPU_Dual_Tree_2021}.

% Old stuff
% We have introduced a new particle method that is based on working with Maxwell's equations in vector potential form under the Lorenz gauge coupled with  casting the particle advance in the form of the canonical momentum.  The method is designed for the filed data to be co-located on the grid and avoids taking time derivatives of the vector potential. The method ensure the  $\nabla \cdot {\bf B}=0$ and under a current conserving map, that satisfies the semi-discreet continuity equation, will enforce the gauge condition.  The method as proposed was applied to a range of classic test problems, including the two stream instability, numerical heating, the Bennett pinch and the beam in a box.   The new method notably preforms better than traditional particle-in-cell in critical tests such as the  numerical heating and the beam in a box tests and at least as well as PIC in all other cases.  In particular it is worth noting that, at the cost of an explicit method, the new algorithm heats less than one tenth of a percent over 1000 angular plasma periods for when the mesh spacing is three Debye lengths per cell, and has mesh independent heating.
% Give thanks here
\section{Acknowledgements}

Some of the simulations presented in this work were supported by computational resources made available through the Institute for Cyber-Enabled Research at Michigan State University. The authors would like to thank both the Air Force Office of Scientific Research, the National Science Foundation, and the Department of Energy for their support though grants FA9550-19-1-0281,  FA9550-17-1-0394,  DMS-1912183, and DE-SC0023164. We would also like to express our gratitude to Professor John P. Verboncoeur of Michigan State University, Dr.~John W.~Luginsland at the Air Force Office of Scientific Research, and Dr.~Eric M.~Wolf for their valuable feedback and suggestions. Lastly, we wish to dedicate this work to the memory of Professor William N.G. Hitchon, a dear colleague and mentor, who assisted us with revisions and offered many suggestions that significantly improved the quality of the present work.

\appendix

%%% Non-dimensionalization procedure
\section{Non-dimensionalization}
\label{app:Non-dim}

In this section, we discuss the scalings used to non-dimensionalize the components of the models explored in this work. Our choice in exploring the normalized form of these models is simply to reduce the number of floating point operations with small or large numbers. We first non-dimensionalize the field equations under the Lorenz gauge, then focus on the equations of motion for the particles. 

The setup for the non-dimensionalization used in this paper considers the following substitutions:
\begin{empheq}[left=\empheqlbrace]{align*}
    &\mathbf{x} \rightarrow L \tilde{\mathbf{x}}, \quad t \rightarrow T \tilde{t}, \\
    &\mathbf{v} \rightarrow V \tilde{\mathbf{v}} \equiv \frac{L}{T} \tilde{\mathbf{v}}, \quad \mathbf{P} \rightarrow P \tilde{\mathbf{P}} \equiv \frac{ML}{T} \tilde{\mathbf{P}}, \\
    &\phi \rightarrow \phi_{0} \tilde{\phi}, \quad \mathbf{A} \rightarrow A_{0} \tilde{\mathbf{A}}, \\
    &\rho \rightarrow Q \bar{n} \tilde{\rho}, \quad \mathbf{J} \rightarrow Q \bar{n}V \tilde{\mathbf{J}} \equiv  \frac{Q\bar{n} L}{T} \tilde{\mathbf{J}}.
\end{empheq}
Here, we use $\bar{n}$ to denote a reference number density $[\text{m}^{-3}]$, $Q$ is the scale for charge in $[\text{C}]$, and we also introduce $M$, which represents the scale for mass $[\text{kg}]$. The values for $Q$ and $M$ are set according to the electrons, so that $Q = \lvert q_e \rvert$ and $M = m_{e}$. We choose the scales for the potentials $\phi_0$ and $A_0$ to be
\begin{equation}
    \phi_{0} = \frac{M L^2}{Q T^2}, \quad A_{0} = \frac{M L}{Q T}. \label{eq:scales for the potentials}
\end{equation}
A natural choice of the scales for $L$ and $T$ are the Debye length and angular plasma period, which are defined, respectively, by
\begin{equation*}
    L = \lambda_{D} = \sqrt{\frac{\epsilon_0 k_{B} \bar{T} }{ \bar{n} q_{e}^{2}}} \quad [\text{m}], \quad T = \omega_{pe}^{-1} =  \sqrt{\frac{m_e \epsilon_0}{\bar{n} q_{e}^2}} \quad [\text{s/rad}],
\end{equation*}
where $k_{B}$ is the Boltzmann constant, $m_e$ is the electron mass, $q_e$ is the electron charge, and $\bar{T}$ is an average macroscopic temperature for the plasma. We choose to select these scales for all test problems considered in section \ref{subsec:Plasma test problems} with the exception of the beam problems, in which the length scale $L$ corresponds to the longest side of the simulation domain and $T$ is the crossing time for a particle that is injected into the domain. In most cases, the user will need to provide a macroscopic temperature $\bar{T}$ [K] in addition to the reference number density $\bar{n}$ to compute $\lambda_{D}$ and $\omega_{pe}^{-1}$. Note that the plasma period can be obtained from the angular plasma period $T$ after multiplying the latter by $2\pi$. Having introduced the definitions for the normalized variables, we proceed to rescale the models, beginning with the field equations for the potentials before addressing equations of motion for the particles.

\subsection{Maxwell's Equations in the Lorenz Gauge}
\label{subsec:Non-dim fields lorenz gauge}

We non-dimensionalize the field equations by substituting scales introduced at the beginning of the section into the equations \eqref{eq:scalar potential eqn lorenz} - \eqref{eq:Lorenz gauge condition}, which gives
\begin{empheq}[left=\empheqlbrace]{align*} 
&\frac{1}{c^2} \frac{\phi_{0}}{T^2} \frac{\partial^2 \tilde{\phi}}{\partial \tilde{t}^2} - \frac{\phi_0}{L^2} \tilde{\Delta} \tilde{\phi}=  \frac{Q \bar{n}}{\epsilon_0}  \tilde{\rho}, \\
&\frac{1}{c^2} \frac{A_0}{T^2} \frac{\partial^2 \mathbf{ \tilde{A}}}{\partial \tilde{t}^2} -\frac{A_0}{L^2} \tilde{\Delta} \mathbf{ \tilde{A}} =  \frac{ \mu_0 Q \bar{n} L }{T} \mathbf{\tilde{\mathbf{J}}}, \\
&\frac{1}{c^2} \frac{\phi_0}{T} \frac{\partial \tilde{\phi}}{\partial \tilde{t}} + \frac{A_0}{L}\tilde{\nabla} \cdot \mathbf{ \tilde{A}} =0.
\end{empheq}
The first equation can be rearranged to obtain
\begin{equation*}
    \frac{L^2}{c^2 T^2} \frac{\partial^2 \tilde{\phi}}{\partial \tilde{t}^2} - \tilde{\Delta} \tilde{\phi} =  \frac{L^2 Q \bar{n}}{\epsilon_0 \phi_0} \tilde{\rho}.
\end{equation*}
Similarly, with the second equation we obtain
\begin{equation*}
    \frac{L^2}{c^2 T^2} \frac{\partial^2 \mathbf{ \tilde{A}}}{\partial \tilde{t}^2} - \tilde{\Delta} \mathbf{ \tilde{A}} = \frac{Q \bar{n} V L^2}{c^2 \epsilon_0 A_0} \mathbf{ \tilde{\mathbf{J}}},
\end{equation*}
where we have used $V = L T^{-1}$ as well as the fact that $c^{2} = \left( \mu_0 \epsilon_0 \right)^{-1}.$ Finally, the gauge condition becomes
\begin{equation*}
  \frac{\phi_0 V}{ c^2 A_0} \frac{\partial \tilde{\phi}}{\partial \tilde{t}} + \tilde{\nabla} \cdot \mathbf{ \tilde{A}} = 0.    
\end{equation*}
Introducing the normalized speed of light $\kappa = c/V$, and selecting $\phi_0$ and $A_0$ according to \eqref{eq:scales for the potentials}, we find that the above equations simplify to (dropping the tildes)
\begin{empheq}[left=\empheqlbrace]{align}
    &\frac{1}{\kappa^2} \partial_{tt} \phi - \Delta \phi = \frac{1}{\sigma_{1} }\rho, \label{eq:non-dim psi lorenz} \\
    &\frac{1}{\kappa^2} \partial_{tt} \mathbf{A} - \Delta \mathbf{A} = \sigma_{2} \mathbf{J}, \label{eq:non-dim A lorenz} \\
    &\frac{1}{\kappa^2} \partial_{t} \phi + \nabla \cdot \mathbf{A} = 0, \label{eq:non-dim lorenz gauge}
\end{empheq}
where we have introduced the new parameters
\begin{equation}
    \label{eq:Normalized permittivity and permeability defs}
    \sigma_{1} = \frac{M \epsilon_{0}}{ Q^{2} T^{2} \bar{n} }, \quad \sigma_{2} = \frac{Q^2 L^2 \bar{n} \mu_{0}}{ M }.
\end{equation}
These are nothing more than normalized versions of the permittivity and permeability constants in the original equations.

\subsection{Particle Equations of Motion}
\label{app:Non-dim particles generalized momentum}

Starting with the position equation \eqref{eq:Position equation relativistic form}, we substitute the scales introduced at the beginning of this section and obtain
\begin{equation*}
    \frac{ L d \tilde{\mathbf{x}}_{i} }{T d \tilde{t} } = \frac{c^2 \left(P \tilde{\mathbf{P}}_{i} - QA_{0} \tilde{q}_{i}\tilde{\mathbf{A}}\right)}{\sqrt{c^2 \left(P \tilde{\mathbf{P}}_{i} - QA_{0} \tilde{\mathbf{A}}\right)^{2} + \left(m_{i} c^{2}\right)^{2} }}.
\end{equation*}
This equation can be simplified further by using the definition of $A_0$ from \eqref{eq:scales for the potentials} and noting that the scale for momentum is $P = MLT^{-1}$. Defining the normalized mass $\tilde{m}_{i} = m_{i}/M$ and charge $\tilde{q}_{i} = q_{i}/Q$, we obtain the non-dimensionalized position equation
\begin{equation*}
    \frac{ d \tilde{\mathbf{x}}_{i} }{d \tilde{t} } = \frac{\kappa^{2} \left(\tilde{\mathbf{P}}_{i} - \tilde{q}_{i}\tilde{\mathbf{A}}\right) }{\sqrt{\kappa^{2} \left(\tilde{\mathbf{P}}_{i} - \tilde{q}_{i}\tilde{\mathbf{A}}\right)^2 + \left( \tilde{m}_{i} \kappa^{2}\right)^{2} }},
\end{equation*}
where $\kappa = c/V$ is, again, the normalized speed of light.

Following an identical treatment for the generalized momentum equation \eqref{eq:Generalized momentum equation relativistic form} and appealing to the definition \eqref{eq:scales for the potentials}, after some simplifications, we obtain
\begin{equation*}
    \frac{ d \tilde{\mathbf{P}}_{i} }{d \tilde{t} } = -\tilde{q}_{i} \tilde{\nabla} \tilde{\phi} + \frac{\tilde{q}_{i} \kappa^2 \left( \tilde{\nabla} \tilde{\mathbf{A}} \right) \cdot \left(\tilde{\mathbf{P}}_{i} - \tilde{q}_{i}\tilde{\mathbf{A}}\right) }{\sqrt{\kappa^{2} \left(\tilde{\mathbf{P}}_{i} - \tilde{q}_{i}\tilde{\mathbf{A}}\right)^{2} + \left(\tilde{m}_{i} \kappa^{2}\right)^{2}}}.
\end{equation*}

Therefore, the non-dimensional form of the relativistic equations of motion is given by (dropping tildes)
\begin{empheq}[left=\empheqlbrace]{align*}
    \frac{ d \mathbf{x}_{i} }{d t} &= \frac{\kappa^{2} \left(\mathbf{P}_{i} - q_{i} \mathbf{A}\right)}{\sqrt{\kappa^{2} \left(\mathbf{P}_{i} - q_{i}\mathbf{A}\right)^2 + \left( m_{i} \kappa^{2}\right)^{2} }}, \\
    \frac{ d \mathbf{P}_{i} }{d t } &= -q_{i} \nabla \phi + \frac{q_{i} \kappa^2 \left( \nabla \mathbf{A} \right) \cdot \left(\mathbf{P}_{i} - q_{i}\mathbf{A}\right) }{\sqrt{\kappa^{2} \left(\mathbf{P}_{i} - q_{i}\mathbf{A}\right)^{2} + \left(m_{i} \kappa^{2}\right)^{2}}}.
\end{empheq}

Performing analogous manipulations in the non-relativistic limit leads to the system (again dropping tildes) 
\begin{empheq}[left=\empheqlbrace]{align*}
    \frac{ d \mathbf{x}_{i} }{d t} &= \frac{1}{m_{i}} \left(\mathbf{P}_{i} - q_{i} \mathbf{A}\right), \\
    \frac{ d \mathbf{P}_{i} }{d t } &= -q_{i} \nabla \phi + \frac{q_{i}}{m_{i}} \left( \nabla \mathbf{A} \right) \cdot \left(\mathbf{P}_{i} - q_{i}\mathbf{A}\right).
\end{empheq}

\section{Elements of the Integral Solution}

This section supplies additional details concerning the derivation and construction of methods to evaluate the integral solution used by the field solver proposed in this work. First, we discuss the recursive fast summation method, which computes the global convolution integral using an accumulation of local integrals. Then, we discuss the quadrature rules developed for the evaluation of these local integrals along with a sketch of the general development of such quadrature rules. We conclude with some brief comments on the application of boundary conditions in multi-dimensional problems.

\subsection{Fast Summation Method}
\label{app:Fast summation for 1-D}

In order to compute the inverse operators according to \eqref{eq:general inverse operator 1-D def 1}-\eqref{eq:general inverse operator 1-D def 3}, suppose we have discretized the one-dimensional computational domain $[a,b]$ into a mesh consisting of $N+1$ grid points:
\begin{equation*}
    a = x_0 < x_1 < \cdots < x_{N} = b,
\end{equation*}
with the spacing defined by
\begin{equation*}
    \Delta x_{j} = x_{j} - x_{j-1}, \quad j = 1, \cdots N.
\end{equation*}
If we directly evaluate the function $w(x)$ at each of the mesh points, according to \eqref{eq:general inverse operator 1-D def 2}, we obtain
\begin{equation*}
    w(x_{i}) = \frac{\alpha}{2} \int_{a}^{b} e^{-\alpha \lvert x_{i} - y \rvert} f(y) \,dy + A e^{-\alpha (x_{i} - a)} + B e^{-\alpha (b - x_{i})}, \quad i = 0, \cdots, N.
\end{equation*}
Since the evaluation of the integral term in the variable $y$ with quadrature requires $\mathcal{O}(N)$ operations, a direct approach requires a total of $\mathcal{O}(N^2)$ operations. The cost of evaluating these terms can be reduced from $\mathcal{O}(N^2)$ to $\mathcal{O}(N)$ by first introducing the operators
\begin{align}
    \mathcal{I}_{x}^{R}[f](x) \equiv \alpha \int_{a}^{x} e^{-\alpha (x - y)} f(y) \,dy , \label{eq:right convolution integral} \\
    \mathcal{I}_{x}^{L}[f](x) \equiv \alpha \int_{x}^{b} e^{-\alpha (y - x)} f(y) \,dy, \label{eq:left convolution integral}
\end{align}
so that the total integral over $[a,b]$ can be expressed as
\begin{equation}
    \label{eq:total convolutiion integral split}
    \mathcal{I}_{x}[f](x) = \frac{1}{2} \Big( \mathcal{I}_{x}^{R}[f](x) + \mathcal{I}_{x}^{L}[f](x) \Big).
\end{equation}

The task now relies on computing these integrals in an efficient manner. To develop a recursive expression, consider evaluating the right-moving convolution integral \eqref{eq:right convolution integral} at a grid point $x_{i}$. It follows that
\begin{align*}
    \mathcal{I}_{x}^{R}[f](x_{i}) &= \alpha \int_{a}^{x_{i}} e^{-\alpha (x_{i} - y)} f(y) \,dy, \\
                                  &= \alpha \int_{a}^{x_{i-1}} e^{-\alpha (x_{i} - y)} f(y) \,dy + \alpha \int_{x_{i-1}}^{x_{i}} e^{-\alpha (x_{i} - y)} f(y) \,dy, \\
                                  &= \alpha \int_{a}^{x_{i-1}} e^{-\alpha (x_{i} - x_{i-1} + x_{i-1} - y)} f(y) \,dy + \alpha \int_{x_{i-1}}^{x_{i}} e^{-\alpha (x_{i} - y)} f(y) \,dy, \\
                                 &= e^{-\alpha \Delta x_{i} } \left( \alpha \int_{a}^{x_{i-1}} e^{-\alpha (x_{i-1} - y)} f(y) \,dy \right) + \alpha \int_{x_{i-1}}^{x_{i}} e^{-\alpha (x_{i} - y)} f(y) \,dy, \\
                                 &\equiv e^{-\alpha \Delta x_{i} } \mathcal{I}_{x}^{R}[f](x_{i-1}) + \mathcal{J}_{x}^{R}[f](x_{i}).
\end{align*}
In the last line, we have introduced the local integral
\begin{equation}
    \label{eq:right local integral}
    \mathcal{J}_{x}^{R}[f](x_{i}) = \alpha \int_{x_{i-1}}^{x_{i}} e^{-\alpha (x_{i} - y)} f(y) \,dy.
\end{equation}
Through similar manipulations, one obtains the recursive expression for \eqref{eq:left convolution integral} given by
\begin{equation*}
    \mathcal{I}_{x}^{L}[f](x_{i}) = e^{-\alpha \Delta x_{i+1} } \mathcal{I}_{x}^{L}[f](x_{i+1}) + \mathcal{J}_{x}^{L}[f](x_{i}),
\end{equation*}
with
\begin{equation}
    \label{eq:left local integral}
    \mathcal{J}_{x}^{L}[f](x_{i}) = \alpha \int_{x_{i}}^{x_{i+1}} e^{-\alpha (y - x_{i})} f(y) \,dy.
\end{equation}
We see that the integrals can be expressed through a recursive weighting of an accumulated value plus an additional term that is localized in space. The recursive relations are initialized by setting 
$$\mathcal{I}_{x}^{R}[f](x_{0}) = 0, \quad \mathcal{I}_{x}^{L}[f](x_{N}) = 0, $$ which follows directly from the definitions \eqref{eq:right convolution integral} and \eqref{eq:left convolution integral}. Since the calculations of the local integrals \eqref{eq:right local integral} and \eqref{eq:left local integral} use quadrature over a collection of $M$ points, the cost of evaluating each integral is now of the form $\mathcal{O}(MN)$. Additionally, since the number of localized quadrature points $M$ is independent of the mesh size $N$, and we select $M \ll N$, the resulting approach scales as $\mathcal{O}(N)$. A similar argument is made for the other integral so that the final cost of evaluating their sum \eqref{eq:total convolutiion integral split} is also $\mathcal{O}(N)$.

\subsection{Approximating the Local Integrals}
\label{app:computing local integrals}

Here, we present the general process used to obtain quadrature rules for the local integrals defined by \eqref{eq:right local integral} and \eqref{eq:left local integral}, in the case of a uniform grid, i.e.,
\begin{equation*}
    \Delta x = x_{j} - x_{j-1}, \quad j = 1, \cdots, N.
\end{equation*}
Rather than use numerical quadrature rules, e.g., Gaussian quadrature or Newton-Cotes formulas, it was discovered that a certain form of analytical integration was required for stability \cite{causley2014method}. In this approach, the operand $f(x)$ is approximated by an interpolating function, which is then analytically integrated against the kernel. We provide a sketch of the approach to illustrate the idea. Specific details can be found in several papers, e.g., \cite{causley2017wave-propagation,christlieb2020NDAD}. 

First, it is helpful to transform the integrals \eqref{eq:right local integral} and \eqref{eq:left local integral} using a change of variable. Consider the integral \eqref{eq:right local integral} and let
\begin{equation*}
    y = (x_{j} - x_{j-1}) \tau + x_{j-1} \equiv \Delta x \tau + x_{j-1}, \quad \tau \in [0,1].
\end{equation*}
Then we can write
\begin{equation}
    \label{eq:right local integral transformed}
    \mathcal{J}_{x}^{R}[f](x_{i}) = \alpha \Delta x e^{-\alpha \Delta x} \int_{0}^{1} e^{\alpha \tau \Delta x } f(\tau \Delta x + x_{i-1}) \, d\tau.
\end{equation}

Next, we approximate $f(x)$ in \eqref{eq:right local integral transformed} using interpolation of a desired order of accuracy. As an example, suppose that we want to use linear interpolation with the data $\{f_{i-1},f_{i}\}$ using the basis $\{1,x - x_{i-1}\}$, which is shifted for convenience to cancel with the shift in \eqref{eq:right local integral transformed}. A direct calculation shows that the interpolating polynomial is
\begin{equation*}
    p(x) = f_{i-1} + \frac{f_{i} - f_{i-1}}{\Delta x} \left(x - x_{i-1}\right).
\end{equation*}
By replacing $f$ in \eqref{eq:right local integral transformed} with the above interpolant, and integrating the result analytically, we find that
\begin{align*}
    \mathcal{J}_{x}^{R}[f](x_{i}) &\approx \alpha \Delta x e^{-\alpha \Delta x} \int_{0}^{1} e^{\alpha \tau \Delta x } \Big( f_{i-1} + \left( f_{i} - f_{i-1} \right)\tau \Big) \, d\tau, \\
    &= w_{0} v_{i-1} + w_{1} v_{i},
\end{align*}
where the weights for integration are
\begin{align*}
    w_{0} &= \frac{1 - e^{-\alpha \Delta x} - \alpha \Delta x e^{-\alpha \Delta x} }{\alpha \Delta x}, \\
    w_{1} &= \frac{\left(\alpha \Delta x - 1\right) + e^{-\alpha \Delta x}}{\alpha \Delta x}. 
\end{align*}

Modifications of the above can be made to accommodate additional interpolation points, as well as techniques for shock capturing. In the latter case, methods have been devised following the idea of WENO reconstruction \cite{jiang1996efficient} to create quadrature methods that can address non-smooth features including shocks and cusps \cite{christlieb2020nonuniformHJE,christlieb2016weno}. In \cite{christlieb2020nonuniformHJE}, quadrature rules were developed using the exponential polynomial basis, which offers additional flexibility in capturing localized features through a ``shape" parameter in the basis functions. Such tools offer a promising approach to addressing problems with discontinuities in the material properties as well as more complex domains with non-smooth boundaries. Despite the notable differences in the type of approximating function used for the operand, the process is essentially identical to the example shown here. We also wish to point out that certain issues may arise when $\alpha \gg 1$ (i.e., $\Delta t \ll 1$). In such circumstances, when the weights are computed on-the-fly, the kernel function can be replaced with a Taylor expansion \cite{causley2014higher}. Otherwise, this results in a ``narrow" Green's function that is vastly under-resolved by the mesh, which causes wave phenomena to remain stagnant. Our experience has found this situation to be quite rare, but it is something to be aware of when a small CFL number is used in a simulation.

% BCs in multi-D
\subsection{Boundary Conditions in Multi-dimensional Problems}
\label{app:bcs in multi-D}

In this section we briefly discuss some of the issues concerning the application of boundary conditions for the multi-dimensional update given by \eqref{eq:BDF-1 2d semi-discrete} for the BDF-1 method. For convenience, the two-dimensional BDF-1 update is
\begin{equation*}
    \mathcal{L}_{x} \mathcal{L}_{y} u^{n+1} (\mathbf{x}) = 2 u^{n}(\mathbf{x}) -  u^{n-1} (\mathbf{x}) + \frac{1}{\alpha^2} S^{n+1}(\mathbf{x}), \quad \alpha := \frac{1}{c \Delta t}.
\end{equation*}

By inverting the operator, one direction at a time, using the techniques presented in \ref{subsec:inverting 1-D operators}, it follows that the solution is given by
\begin{equation*}
     u^{n+1} = \mathcal{L}_{x}^{-1} \mathcal{L}_{y}^{-1} \Big( 2 u^{n} - u^{n-1} + \frac{1}{\alpha^2} S^{n+1} \Big)(\mathbf{x}).
\end{equation*}
We wish to point out here that things are assumed to be smooth so that the ordering conventions used for operators are irrelevant. In other words, we can assume 
\begin{equation*}
    \mathcal{L}_{x} \mathcal{L}_{y} = \mathcal{L}_{y} \mathcal{L}_{x}.
\end{equation*}

\subsubsection{Sweeping Patterns in Multi-dimensional Problems}
\label{subsubsec:Sweep patterns in multi-D}

In the two-dimensional case, we need to construct terms of the form
\begin{equation*}
    \mathcal{L}_{y} \mathcal{L}_{x} w = f \implies w = \mathcal{L}_{x}^{-1} \mathcal{L}_{y}^{-1} \left( f \right),
\end{equation*}
with boundary data being prescribed for the variable $w$. The construction is performed over two steps. The first step inverts the $y$ operator, so we obtain
\begin{equation}
    \label{eq:First layer of sweeps}
    \mathcal{L}_{x} w = \mathcal{L}_{y}^{-1} \left( f \right).
\end{equation}
The first layer of sweeps given by equation \eqref{eq:First layer of sweeps} requires boundary data for the intermediate variable $\mathcal{L}_{x} w$ when we are only given boundary data for $w$. From the definition of $\mathcal{L}_{x}$, we note that
\begin{equation}
    \label{eq:Taylor argument for intermediate variables}
    \mathcal{L}_{x} w \equiv \left( \mathcal{I} - \frac{1}{\alpha^2} \partial_{xx} \right)w = w + \mathcal{O}\left( \frac{1}{\alpha^2} \right),
\end{equation}
% In other words, boundary conditions for $\mathcal{L}_{x} w$ can be approximated to second-order in time by those of $w$; however, unless we are dealing with outflow boundary conditions, we do not need to sweep along the boundary of the domain, so the approximation \eqref{eq:Taylor argument for intermediate variables} is not necessary. Proceeding further, the second step of the inversion process leads to the solution $w$
In other words, boundary conditions for $\mathcal{L}_{x} w$ can be approximated to second-order in time by those of $w$. Proceeding further, the second step of the inversion process leads to the solution $w$
\begin{equation}
    \label{eq:Second layer of sweeps}
    w = \mathcal{L}_{x}^{-1} \mathcal{L}_{y}^{-1} \left( f \right),
\end{equation}
which simply enforces the known boundary data on $w$. In the subsections that follow, we briefly summarize the changes associated with moving to multi-dimensional problems, including any changes necessitated by the proposed methods for calculating derivatives.

\subsubsection{Dirichlet Boundary Conditions}
\label{subsubsec:Dirichlet in multi-D}

In the case of Dirichlet boundary conditions, the values of the function are known along the boundary. Therefore, we only need to update the grid points corresponding to the interior of the domain. As mentioned earlier, rather than approximating the boundary conditions for the intermediate sweep, e.g., \eqref{eq:Taylor argument for intermediate variables}, we simply avoid sweeping along the boundary points of the domain, since these values are known. The direction corresponding to the intermediate sweep will now only use boundary data set by the solution, since the boundaries are left untouched. In the case of homogeneous Dirichlet conditions, the sweeps can be performed on the boundary with no effect. When sweeping over different directions for the derivatives, we note that along the boundary, the derivative information is not known. Therefore, the sweeps should extend all the way to the boundary. Otherwise the derivative will not be available there. In this case, the boundary data for the intermediate data can be approximated according to \eqref{eq:Taylor argument for intermediate variables}.

\subsubsection{Neumann Boundary Conditions}
\label{subsubsec:Neumann in multi-D}

The treatment of Neumann boundary conditions in multi-dimensional problems is identical to the procedure used to enforce Dirichlet boundary conditions discussed in the previous section for the case of Cartesian grids. In problems defined on complex geometries with embedded boundaries, the theory presented in \cite{Kreiss04neumann} suggests that dissipation is necessary to obtain stable numerical solutions \cite{causley2017wave-propagation} to the Neumann problem. This is not a problem for the BDF method, which is known to be dissipative. The treatment of Neumann boundary conditions for problems with curved boundaries was given in \cite{causley2017wave-propagation}. In such a setting, the normal direction along the curved boundary couples the sweeping directions of the interior grid points in a non-trivial manner. Authors in \cite{causley2017wave-propagation} devised an iterative technique that uses Hermite interpolation in the vicinity of the boundary to supply Dirichlet data at the ends of the lines over which sweeps are performed. Note that these caveats are also relevant to the proposed method for computing derivatives.

\subsubsection{Periodic Boundary Conditions}
\label{subsubsec:Periodic in multi-D}

Periodic boundary conditions in multi-dimensional problems can be enforced in a straightforward way by directly applying the one-dimensional approaches outlined in section \ref{subsubsec:BDF-1 periodic} along each dimension. No modifications are required for either the scheme or the proposed method for calculating derivatives.

\printbibliography

\end{document}